\documentclass[prx,amsmath,amssymb,floatfix,twocolumn,amsmath,superscriptaddress,nofootinbib,tighten]{revtex4-2}

\usepackage{graphicx}
\usepackage{dcolumn}
\usepackage{bm}
\usepackage{xcolor}
\usepackage{cancel}
\newcommand{\ket}[1]{| {#1} \rangle} 
\newcommand{\bra}[1]{\langle {#1} |} 
\newcommand{\braket}[2]{\langle {#1} \vphantom{#2} | {#2} \vphantom{#1} \rangle} 

\usepackage[utf8]{inputenc}
\usepackage[T1]{fontenc}
\usepackage[english,serbian]{babel}
\usepackage[OT1]{fontenc}
\usepackage{mathrsfs}
\usepackage{graphicx}
\usepackage{amssymb}
\usepackage{amsbsy}
\usepackage{latexsym}
\usepackage{mathtools}
\usepackage{titlesec}
\usepackage{amsmath}
\usepackage{amsthm}
\usepackage{color}
\usepackage[utf8]{inputenc}
\usepackage{lipsum}
\usepackage{fontenc}

\usepackage{chngcntr}

\usepackage{comment}

\def\de{\textrm{d}}

\renewcommand{\vec}[1]{\ensuremath{\mathbf{#1}}}
\def\dj{d\kern-0.4em\char"16\kern-0.1em}
\def \Dj {\mbox{\raise0.3ex\hbox{-}\kern-0.4em D}}

\usepackage{hyperref}

\usepackage{cleveref}

\begin{document}

\title{Symmetry-based theory of Dirac fermions on two-dimensional hyperbolic crystals:\\ Coupling to the spin connection }

\author{Ana Đorđević}

\affiliation{Faculty of Physics, University of Belgrade, Studentski Trg 12-16, 11000 Belgrade, Serbia}

\author{Marija Dimitrijevi\'c \'Ciri\'c}

\affiliation{Faculty of Physics, University of Belgrade, Studentski Trg 12-16, 11000 Belgrade, Serbia}

\author{Vladimir Juri\v{c}i\'c}\thanks{Corresponding author:vladimir.juricic@usm.cl}

\affiliation{ Departamento de F\'{i}sica, Universidad T\'{e}cnica Federico Santa Mar\'{i}a, Casilla 110, Valpara\'{i}so, Chile}

\selectlanguage{english}

\date{\today}

\begin{abstract}
Discrete fermionic and bosonic models for hyperbolic lattices have attracted significant attention across a range of fields since the experimental realization of hyperbolic lattices in metamaterial platforms, sparking the development of hyperbolic crystallography. However, a fundamental and experimentally consequential aspect remains unaddressed: fermions propagating in curved space inherently couple to the underlying geometry via the spin connection, as required by general covariance---a feature not yet incorporated in studies of hyperbolic crystals. Here, we introduce a symmetry-based framework for Dirac fermions on two-dimensional hyperbolic lattices, explicitly incorporating spin-curvature coupling via a discrete spin connection. Starting from the continuous symmetries of the Poincar\'e disk, we identify the isometry algebra and the principal-series spectral basis underlying the continuum Dirac theory. We show that, for massless Dirac fermions, this continuum theory predicts a finite density of states (DOS) at zero energy for any finite curvature in \(D\)-dimensional hyperbolic space with \(2\leq D\leq 4\), suggesting enhanced susceptibility of Dirac fermions to interaction-driven instabilities at weak coupling. We further derive the discrete translational and rotational symmetries of hyperbolic lattices with Schl\"afli symbols \(\{p,q\}\). Based on these symmetries, we construct a lattice spin-connection factor and implement it as a spin-dependent nearest-neighbor hopping. 
For finite open $\{10,3\}$ hyperbolic lattices, we implement the
geodesic Wilson-line phase in a nearest-neighbor tight-binding model.
Numerical calculations performed using the kernel polynomial method reveal a robust low-energy DOS
enhancement that persists across the studied system sizes at fixed
numerical resolution. At fixed lattice geometry, this behavior provides
qualitative lattice-level support for the continuum prediction of a
nonvanishing low-energy DOS.
Our results establish a symmetry-based framework for spin-curvature coupling on hyperbolic lattices, motivating future numerical studies of correlated Dirac phases and experimental exploration of spin-curvature effects in metamaterial platforms.
\end{abstract}
\maketitle

\tableofcontents

\selectlanguage{english}

\section{Introduction}

Negatively curved (hyperbolic) lattices have recently become directly
accessible in circuit-quantum-electrodynamics, photonic, and
metamaterial platforms, where suitably engineered hopping amplitudes,
refractive-index profiles, or effective dielectric constants emulate
the geometry of curved space in the laboratory
~\cite{NatureExperiment,
QuantumSimulationOfHyperbolicSpace,
DielectricConstant5,
DielectricConstant6,
CrystallographyOfHyperbolicLattices}. This experimental progress has driven the development of hyperbolic crystallography and renewed theoretical interest in quantum field theory on hyperbolic space more broadly, including its role in the AdS/CFT correspondence~\cite{Maldacena1997,Witten1998,Gubser1998} and in the connections between geometry, entanglement, and renormalization uncovered in holography and quantum information theory~\cite{Ryu_PRL2006,Ryu_JHEP2006,Swingle2012,Czech2015,Nishioka2018}.

One promising approach employs hyperbolic metamaterials, where spatial variations in the dielectric constant effectively encode  the effects of curvature~\cite{Leonhardt_2006,PhysRevA.78.043821,PhysRevLett.105.067402,DielectricConstant3,Chen_2010,DielectricConstant5,DielectricConstant6}. Other platforms for realizing the Dirac equation in curved space-time have been proposed, including ion traps~\cite{Sab_n_2017,Pedernales_2018}, optical waveguides~\cite{Koke_2016}, and optical lattices with position-dependent hopping and non-Abelian artificial gauge fields~\cite{Boada_2011}. Building on these developments, recent progress in quantum electrodynamics, superconducting microwave resonator networks~\cite{NatureExperiment}, and topoelectric circuits~\cite{Zhang2022,Lenggenhager2022,Zhang2023natcomm,Chen2023} has further expanded the capabilities for simulating lattice models and conformal field theories in hyperbolic space,  which also play a key role in the emerging field of hyperbolic crystallography~\cite{Maciejko2021,CrystallographyOfHyperbolicLattices,Cheng2022,Kienzle2022,Maciejko2022,Attar2022,Bzdusek2022}.

Hyperbolic crystals are regular tessellations of hyperbolic space generalizing the notion of the crystal in the Euclidean (flat) space to its negatively curved counterpart with a constant curvature where the tiling is realized by polygons of $p$ sides forming a network of sites (vertices) with the $q$ nearest neighbors. Such a lattice structure is commonly denoted by  Schl\"{a}fli symbols $\{p,q\}$, with the regular hyperbolic space tessellation achieved when $(p-2)(q-2)>4$ giving rise to a much richer structural landscape than in the Euclidean space where the condition  $(p-2)(q-2)=4$ yields only three possible regular tessellations corresponding to square $\{4,4\}$, triangular $\{3,6\}$, and honeycomb $\{6,3\}$ lattices. To embed the $\{p,q\}$ hyperbolic lattice into its continuum geometric setting, we employ the Poincar\'e disk as the reference geometry for our analysis.

The study of band structures in hyperbolic crystals has been developed within the framework of hyperbolic band theory~\cite{Maciejko2021,CrystallographyOfHyperbolicLattices,Maciejko2022,Cheng2022,Kienzle2022}, which extends the conventional Bloch theorem to negatively curved space. Notably, hyperbolic analogs of topological phases~\cite{THL-PRL2020,HPTB-PRL2022,CIHL-PRB2022,Zhang2022NatComm,Huang2024-NatComm,Chen2024,Sun-SciPost2024, HOTopology2, HOTopology3} have been explored, demonstrating that bulk topological invariants and protected edge modes remain well-defined in hyperbolic space. Moreover, the concept of topological band nodes has been generalized to negatively curved geometries \cite{Tummuru:2023wuq}, with a corresponding linear scaling of the density of states (DOS) found  for lattices where
$p/2$ is an odd integer~\cite{Roy_2024}. Beyond these topological aspects, a variety of condensed matter phenomena have been investigated in hyperbolic systems, including Hofstadter spectra \cite{Hofstadter1, Hofstadter2}, flat bands \cite{FlatBands1, FlatBands2, FlatBands3, FlatBands4}, strong correlation effects \cite{Roy_2024,StrongCorrelations1,StrongCorrelations2,StrongCorrelations3, Leong-Roy-PRB2026}, non-Hermitian phenomena \cite{Lv2022,hu2024-NHhyper,Shen-PRB2025}, and fracton physics \cite{Fractons1, Fractons2}.  \begingroup
Most existing tight-binding models on hyperbolic lattices employ a
single scalar amplitude per site and therefore do not include the
local-frame transformation law characteristic of spinor fields.
Their single-particle spectra may display Dirac-like features, but the
corresponding lattice degrees of freedom do not by themselves encode
the spin connection required by a continuum Dirac theory in curved
space.
\endgroup

\begingroup
Both scalar and spinor fields couple to the curved metric. Spinor
fields, however, require an additional spin connection associated
with covariance under local-frame rotations, whereas scalar fields
do not~\cite{Brill-Wheeler-1957}.
\endgroup More specifically, fermions couple to geometry through the spin connection tied to local Lorentz covariance, which fixes the covariant derivative to the universal form \(\nabla_\mu=\partial_\mu+\tfrac{1}{2}\omega_\mu{}^{bc}\Sigma_{bc}\), with the matrices $\Sigma_{bc}$ as the Lorentz generators, defined explicitly in Eq.~\eqref{Dirac_equation0}. This coupling is kinematic and, as such,  survives the nonrelativistic  limit, so slow (Schr\"odinger) fermions still ``feel'' the spin connection, which is in this case fixed by the non-relativistic Galilean invariance \cite{ObukhovDiracfermions, PhysRevD.107.045012}. Furthermore, in crystalline solids, lattice deformations define local frames \(e_a{}^{\mu}(x)\) yielding  a nontrivial spin connection \(\omega_\mu{}^{ab}(x)\).  Emergent Dirac quasiparticles  therefore couple via the same covariant derivative, with \(\omega_\mu\) term  symmetry-fixed and independent of microscopic details \cite{VOZMEDIANO2010109, DEJUAN2010625}. The same symmetry principle manifests in the nonrelativistic setting through Newton--Cartan effective field theory, where the covariant derivative for spin-\(s\) fields reads \(\nabla_\mu=\partial_\mu+i s\,\omega_\mu+iA_\mu+...\), with $A_\mu$ as the external U(1)
gauge field, which makes explicit why the spin-connection term is unavoidable and directly govern geometric responses of fractional quantum Hall fluids (e.g., Hall viscosity)~\cite{Son_2025, Andringa_2011, PhysRevD.91.045030}.

To the best of our knowledge, the coupling of an internal Dirac spinor degree of freedom to curvature through the spin connection has not yet been incorporated explicitly in tight-binding models of hyperbolic crystals. This coupling can generate a finite zero-energy DOS and thereby enhance the susceptibility of Dirac fermions to interaction-driven ordering at weak coupling. This expectation is supported by the scaling dimensions $[\rho]=D-1$ and $[R]=2$, which imply $\rho(0)\sim |R|^{(D-1)/2}$ in $D$ spatial dimensions. The resulting finite DOS may promote dynamical mass generation through spontaneous symmetry breaking. Further possibilities include curvature-driven band inversion, topological phase transitions on curved backgrounds, and geometric or gravitational-response phenomena in engineered electronic and classical-wave systems.

\subsection{Key results}

Motivated by these considerations, we systematically incorporate spin-curvature coupling on hyperbolic lattices. 
The symmetry analysis of the Poincaré disk~\cite{Kobayashi1972,Helgason2001,Ratcliffe2019} identifies the isometry algebra, its principal-series continuum basis, and its action on Dirac spinors, as summarized in Secs.~\ref{Irreducible_representations} and~\ref{section_spinors}; further representation-theory details are collected in App.~\ref{app:principal-series-details}. As we then show, massless Dirac fermions in this space couple to the geometry through the spin connection, yielding a finite density of states at zero energy for any finite curvature (Sec.~\ref{section_densityofstates}), with the form $\rho(0)\sim |R|^{(D-1)/2}$, consistent with general scaling arguments, which sheds light on the apparent universality of weak coupling instabilities of Dirac fermions coupled to the hyperbolic space~\cite{Inagaki-review-1997}.
To construct the discrete spin connection on an arbitrary hyperbolic lattice, we first derive the explicit rotational and translational symmetries fixed by the Schl\"{a}fli symbols $\{p,q\}$, given by Eqs.~\eqref{eq:discrete-rotation} and~\eqref{eq:discrete-translation}, respectively. \begingroup
Building on these discrete symmetries, we derive the lattice
spin-connection factor from the spinor transformation law and
implement it as a spin-dependent link factor. We then examine the
consequences of this construction in a nearest-neighbor tight-binding
model on finite $\{10,3\}$ hyperbolic graphs. This model should be
viewed as a minimal lattice realization of the geodesic
spin-connection link structure, rather than as a systematically
derived discretization of the full continuum Dirac operator.
Numerical calculations performed using the kernel polynomial method (KPM) yield a robust finite-resolution
zero-energy DOS enhancement, whereas the corresponding reference
model without the spin-connection link factor remains strongly
suppressed. The result is qualitatively consistent with the continuum
prediction, but its magnitude and detailed energy dependence are
lattice dependent; see
Figs.~\ref{fig:DOS_spin_connection_generations}--\ref{fig:saturation-vs-N}
and Table~\ref{tab:GC-floor}.
\endgroup

\subsection{Organization}

The paper is organized as follows. In Sec.~\ref{sectionII}, we study the geometry and symmetries of the Poincar\'{e} disk, summarize the principal-series continuum basis, and analyze continuum Dirac spinors and their density of states. In Section \ref{PointGroupSymmetry}, we obtain the discrete translation and rotational symmetries of the hyperbolic lattice embedded on the Poincar\'{e} disk. The explicit construction of the discrete spin connection and its tight-binding implementation on the $\{10,3\}$ lattice is carried out in Section \ref{LatticeDiracFermions}. We summarize our results in Section \ref{ConclusionsAndOutlook} and present technical details in Appendices.
\section{Geometry, symmetries, and Dirac spinors on the Poincar\'{e} disk}
\label{sectionII}

Understanding the properties of a hyperbolic lattice embedded in the Poincaré disk requires a careful examination of its symmetry properties~\cite{Kobayashi1972, Helgason2001, Ratcliffe2019}, which is crucial when constructing  spinors and the corresponding discrete spin connection therein. To this end, we start by introducing the Poincaré disk model, defined as $\mathbb{D} = \{z \in \mathbb{C} : |z| < l\}$, representing the hyperbolic plane and equipped with the metric
\begin{equation}\label{Poincare_metric}
\de s^2 = \frac{\de x^2 + \de y^2}{(1 - \frac{|z|^2}{l^2})^2},
\end{equation}
{yielding a constant scalar curvature $R=-8/l^2$}.
To uncover the isometries of the Poincaré disk, we take a geometric approach and identify the Killing vectors generating the coordinate transformations that preserve the metric in Eq.~\eqref{Poincare_metric}. These generators also determine the scalar Casimir and the principal-series spectral basis summarized in Sec.~\ref{Irreducible_representations}.
Furthermore,  as embedding spinors in the Poincaré disk can be  viewed as the long-wavelength limit of lattice spinors in a discrete hyperbolic lattice, it motivates us to explore the explicit form  of the spinors as the solutions of the Dirac equation on the Poincaré disk, as discussed in Sec.\ref{section_spinors}, and, in particular, to derive the corresponding DOS in Sec.\ref{section_densityofstates}.


\subsection{Killing vectors}
\label{sec:Killing vectors}

To set the stage, we first analyze  the symmetries of  Poincaré disk  by identifying its Killing vectors and their algebra,  which, at the same time, provides a more transparent geometric picture of its isometries. Using these results, we then determine the coordinate transformations generated by the Killing vectors and derive the geodesics, describing  the trajectories of a free particle in this space. To this end, we solve the Killing equation,
\begin{equation} \label{eq:Killing_equation}
\partial_\mu \xi_\nu + \partial_\nu \xi_\mu = 2\Gamma_{\mu\nu}^{\rho}\xi_{\rho},
\end{equation}
which also represents an integral of the geodesic equation. In Eq. \eqref{eq:Killing_equation}, $\xi_{\mu}$ represents the Killing vector and $\Gamma_{\mu\nu}^{\rho}$ is the Christoffel symbol. Now, invoking the form of the Christoffel symbols in Eqs.~\eqref{app-eq-conn-1}-\eqref{app-eq-conn-4} and  the metric of the Poincaré disk [Eq. \eqref{Poincare_metric}], we find that the Killing equation reduces to the Cauchy–Riemann conditions for the Killing vectors,
\begin{align} \label{eq:1}
\partial_x\xi^{x} &= \partial_y\xi^{y}, \\ \label{eq:2}
\partial_x\xi^{y} &= -\partial_y\xi^{x},
\end{align}
which are therefore equivalent to the Killing equations of the two-dimensional conformal group~\cite{MBlagojevic:2001}. Eqs.~\eqref{eq:1} and~\eqref{eq:2} then allow to express the  isometries of the Poincaré disk in the form of analytic and anti-analytic functions,
\begin{align} \label{eq:analytic_killing}
\xi &= \xi^x + i\xi^y, \\
\bar{\xi} &= \xi^x - i\xi^y,
\label{eq:antianalytic_killing}
\end{align}
where the analytic and anti-analytic  function are, respectively, given by $\xi = a_{-1} + a_0z + a_1z^2$ and  $\bar{\xi} = \bar{a}_{-1} + \bar{a}_0\bar{z} + \bar{a}_1\bar{z}^2$, with $z=x+i y$. Substituting this \emph{ansatz} into Eq. \eqref{eq:Killing_equation} yields  three Killing vectors,
\begin{align} \label{eq:Killing_vector1}
\xi_1 &= \left(-l^2 + x^2-y^2\right)\partial_x + 2xy\partial_y, \\ \label{eq:Killing_vector2}
\xi_2 &= -y\partial_x + x\partial_y, \\ \label{eq:Killing_vector3}
\xi_3 &= 2xy\partial_x + \left(-l^2- x^2+y^2\right)\partial_y.
\end{align}
Therefore, Killing vector $\xi_2$ represents the generator of rotations in the plane, i.e.,  the angular momentum. The other two Killing vectors $\xi_1$ and $\xi_3$ can be interpreted as a combination of generators of translations and special conformal transformations. Furthermore, the obtained  Killing vectors correspond to the generators of the Witt algebra of the conformal group, specifically $\{L_{\pm 1}, L_0, \bar{L}_{\pm 1}, \bar{L}_0\}$, with the identifications,
\begin{align}
\xi_1 &= L_{-1} - L_1 + \bar{L}_{-1} - \bar{L}_1, \\
\xi_2 &= i(-L_0 + \bar{L}_0), \\
\xi_3 &= i[(L_{-1} + L_1) - (\bar{L}_{-1} + \bar{L}_1)].
\end{align}
As such, they satisfy the Lie algebra
\begin{align}
\label{eq:Killing_algebra1}
[\xi_1,\xi_2] &= \xi_3,\\
\label{eq:Killing_algebra2}
[\xi_1,\xi_3] &= 4l^2\xi_2,\\
\label{eq:Killing_algebra3}
[\xi_2,\xi_3] &= \xi_1.
\end{align}
which is isomorphic to the algebra ${\rm sl}(2,\mathbb{R})$, generating the symmetry of  Poincaré disk. As a consequence,  the corresponding finite coordinate transformations take the form
\begin{equation} \label{eq:coordinate_transformation}
z' = e^{(\bar{a}\frac{z^2}{l} + i\theta z - al)\partial}z,
\end{equation}
with $a \in \mathbb{C}$ and $\theta\in\mathbb{R}$ being the parameters of the transformation. The isometries of the disk $\mathbb{D}$ correspond to conformal automorphisms,
\begin{equation} \label{eq:Mobius_transformation}
z' = l^2e^{i\tilde{\theta}}\frac{z-c}{l^2-\bar{c}z},
\end{equation}
as explicitly demonstrated in App.~ \ref{AppendixA}. As expected, these transformations are isomorphic to the group PSL(2,$\mathbb{R}$) of Möbius transformations on the upper half-plane~\cite{QuantumSimulationOfHyperbolicSpace}.

\par
\begin{figure}[t!]
\centering
\includegraphics[width=4cm, height=4cm]{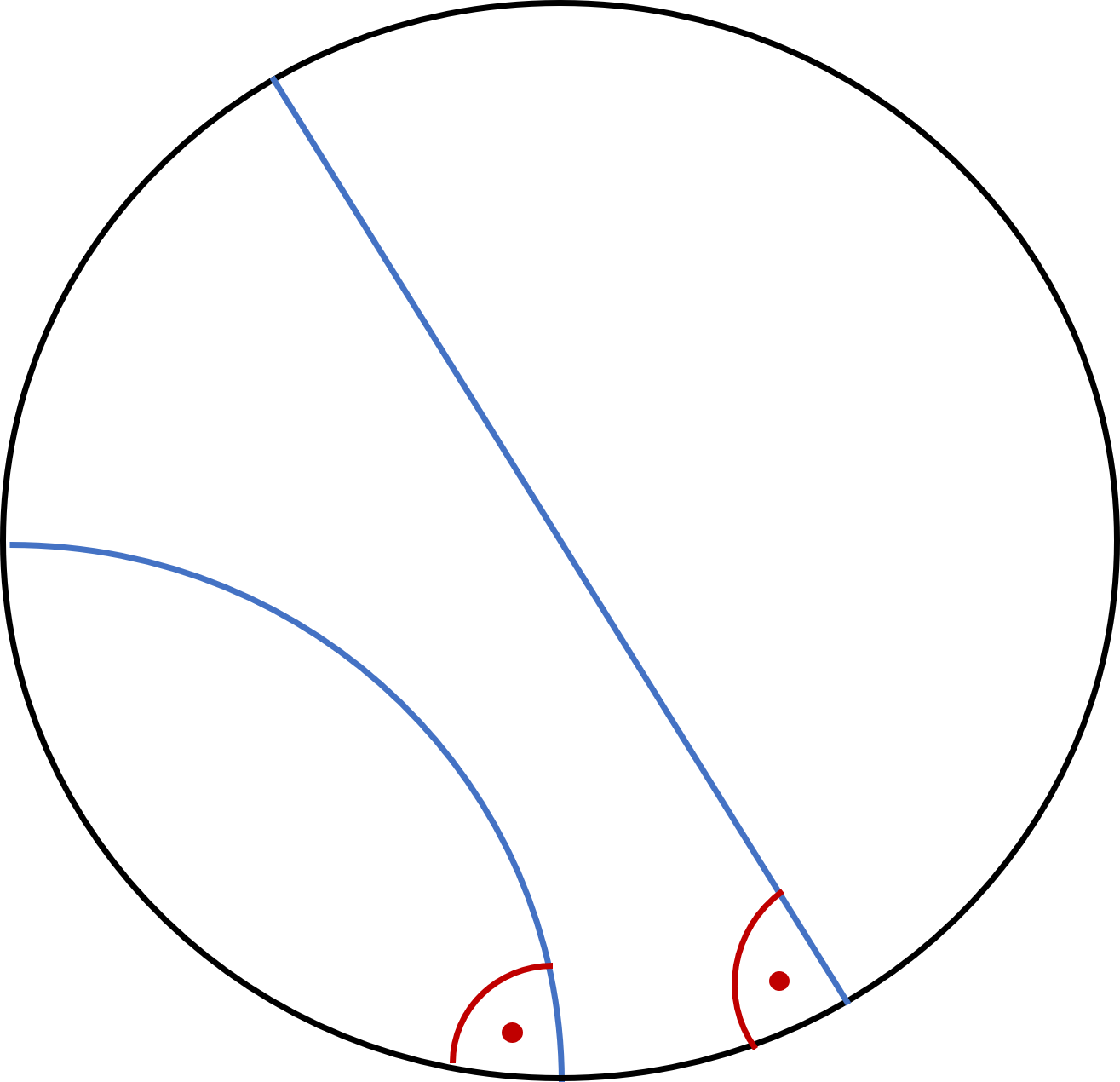}
  \caption{Geodesics on the Poincar\' e disk. First class of geodesics given by the straight line passes through the center of the disk, and intersects the edges at the angle of $\pi/2$, as given by Eq.~\eqref{geodesic1}. Second class of geodesics is represented by circles  that do not pass through the center and are orthogonal to the edges, as given by Eq.~\eqref{geodesic2}. }\label{fig1}
\end{figure}

 We use these Killing vectors and conserved quantities to find the equations of  geodesic curves in this space,
\begin{align}
g_{\mu\nu}\frac{\de x^{\mu}}{\de\tau}\frac{\de x^{\nu}}{\de\tau}&=C_4 ,\\
\label{conserved_killing}
  \frac{\de x^{\mu}}{\de\tau}\xi_\mu&=C,
\end{align}
where $C$ and $C_4$ are constants, and $\tau$ as the geodesic parameter, which ultimately yield the following geodesic curves:
\begin{align}
\label{geodesic1}
  y&=Dx, \\
  \label{geodesic2}
(x-x_0)^2+(y-y_0)^2&=x_0^2+y_0^2-l^2,
\end{align}
where $D$, $x_0$ and $y_0$ are real constants, and correspond to two distinct classes of geodesics. First, the geodesic line in  Eq.~\eqref{geodesic1} describes a particle moving along a straight line crossing the center of the disk $\mathbb{D}$ (Fig.~\ref{fig1}), while the ones in Eq.~\eqref{geodesic2} imply that the particle can also follow a circular trajectory orthogonal to the edge circle of the disk lying at infinity, as shown in Fig. \ref{fig1}. See  App.~\ref{AppendixA}  for details.
Finally, the distance between two points on the Poincar\'{e} disk is
\begin{equation}
\label{hyperbolic_distance}
d(z,z')=\frac{l}{2}\,\, \mathrm{arcosh}\left(1+\frac{2l^2|z-z'|^2}{(l^2-|z|^2)(l^2-|z'|^2)}\right).
\end{equation}
The corresponding scalar Casimir and the continuum spectral basis are summarized next; the full ladder-operator construction and explicit coordinate-space eigenfunctions are deferred to App.~\ref{app:principal-series-details}.

\subsection{Continuum spectral basis}
\label{Irreducible_representations}

\begingroup
The orientation-preserving isometry group of the Poincar\'e disk is
\begin{equation}
{\rm PSL}(2,\mathbb R)\simeq {\rm PSU}(1,1).
\end{equation}
For scalar fields, the quadratic Casimir associated with this symmetry is the Laplace--Beltrami operator,
\begin{equation}
\label{eq:main-casimir-laplacian}
\xi^2=-\frac{l^2}{4}\Delta_g,
\qquad
\Delta_g=
\left(1-\frac{|z|^2}{l^2}\right)^2
(\partial_x^2+\partial_y^2).
\end{equation}
The continuum spectral resolution relevant to square-integrable fields on the hyperbolic plane is furnished by the principal series,
\begin{equation}
\label{eq:main-principal-series}
\xi^2\,\Phi_{s,m}
=
\left(\frac14+s^2\right)\Phi_{s,m},
\qquad s\in\mathbb R,
\end{equation}
with the angular quantum number taking the values appropriate to the chosen single-valuedness condition. These generalized eigenstates form the complete Plancherel basis on the Poincar\'e disk~\cite{kitaev2018,bargmann1947}. The ladder-operator construction, the relation to the other unitary series, and the explicit coordinate-space eigenfunctions are given in App.~\ref{app:principal-series-details}.
\endgroup

\subsection{Spinors on the Poincaré disk}
\label{section_spinors}
In this section, we investigate the properties of spinors in the long-wavelength limit as a preliminary step toward the construction of the discrete spin connection and the embedding of spinor fields on a hyperbolic lattice. To this end, we introduce an \emph{expanded Poincar\'{e} disk} geometry, where the addition of a temporal coordinate extends the Poincar\'{e} disk metric to the form
\begin{equation}
\label{expanded_metric}
  G(t,x,y) =
  \begin{pmatrix}
  -1 & 0 & 0 \\
  0 & \frac{1}{\left(1-\frac{x^2+y^2}{l^2}\right)^2} & 0 \\
  0 & 0 & \frac{1}{\left(1-\frac{x^2+y^2}{l^2}\right)^2}
  \end{pmatrix}.
\end{equation}
Since the time coordinate is incorporated as a flat direction, this modification does not introduce additional nonzero Christoffel symbols in the geometry.

The dynamics of a massive fermion in this background are governed by the Dirac equation in natural coordinates, including the spin connection $\omega_{\mu}{}^{ab}$,
\begin{equation}
\label{Dirac_equation0}
  \left[e^{\mu}{}_a \gamma^a \left(\partial_{\mu} + \frac{1}{2}\omega_{\mu}{}^{cb}\Sigma_{cb}\right) - m\right]\Psi(x) = 0,
\end{equation}
\begingroup
where \(e^{\mu}{}_a\) is the inverse vielbein and
\(\Sigma_{cb}=\frac14[\gamma_c,\gamma_b]\) denotes the Lorentz generator
with lowered local-frame indices. Greek indices
\(\mu\in\{t,x,y\}\) label coordinates on the Poincar\'e disk, whereas
Latin indices \(a,b,c\in\{0,1,2\}\) refer to the local Minkowski frame.
\endgroup

By solving the stationary Dirac equation~\eqref{Dirac_equation0}, we obtain explicit spinor solutions on the hyperbolic disk, as detailed in Appendix~\ref{app:Dirac-curved-space}. These results are employed  for the calculation of physical observables, such as the density of states via the corresponding Green's function, discussed in the subsequent section.

We now determine the symmetry generators of the Dirac field on the Poincar\'{e} disk. To this end, we focus on stationary solutions of the Dirac equation, restricting the Greek indices to spatial coordinates. These generators  satisfy the algebraic structure defined by the Killing vectors, as given in Eqs.~\eqref{eq:Killing_algebra1}--\eqref{eq:Killing_algebra3}.

The infinitesimal coordinate transformation $x \to x' = x + \delta x$ takes the form
\begin{equation}
\label{eq:infinitesimal_coordinate_transformation}
  \delta x^{\mu} = w^{\mu}{}_{\nu} x^{\nu} + 2(c \cdot x)x^{\mu} - c^{\mu}(x^2 + l^2),
\end{equation}
where $w^{\mu\nu} = \theta \varepsilon^{\mu\nu}$, with $\varepsilon^{\mu\nu}$ the Levi-Civita tensor. This, in turn,  motivates  to represent  a general variation of the spinor field's form  under this transformation as follows:
\begin{equation}
\label{eq:ansatz_transformation_spinor}
  \delta_0 \Psi(x) = \left(\frac{1}{2} w^{\mu\nu} M_{\mu\nu} + c^{\mu} K_{\mu}\right) \Psi(x),
\end{equation}
where the coordinate and matrix parts of the generator read
\begin{align}
\label{eq:ansatz_Mmunu}
  M_{\mu\nu} &= M_{\mu\nu}{}^{\rho} \partial_{\rho} + M_{\mu\nu}^{(0)}, \\
  \label{eq:ansatz_Kmu}
  K_{\mu}  &= K_{\mu}{}^{\rho} \partial_{\rho} + K_{\mu}^{(0)}.
\end{align}
The explicit form of the spinor transformation is fixed by requiring invariance of the Dirac equation; see Appendix~\ref{app:symmetry_dirac_equation} for details. The resulting symmetry generators are
\begin{align}
\label{eq:operator-Kx}
  K_x &= \left(-l^2 + x^2 - y^2\right)\partial_x + 2 x y \partial_y + i y \sigma_z, \\
  M  &= x\partial_y - y\partial_x + i \frac{\sigma_z}{2}, \label{eq:operator-M} \\
  K_y &= 2 x y \partial_x + \left(-l^2 - x^2 + y^2\right)\partial_y - i x \sigma_z,
  \label{eq:operator-Ky}
\end{align}
where $\sigma_z$ is the Pauli matrix and the coordinate parts correspond precisely to the Killing vectors in Eqs.~\eqref{eq:Killing_vector1}--\eqref{eq:Killing_vector3}. Note that the matrix part of generators is also coordinate dependent.

\begin{figure*}[t!]
  \centering
  \includegraphics[width=.9\linewidth]{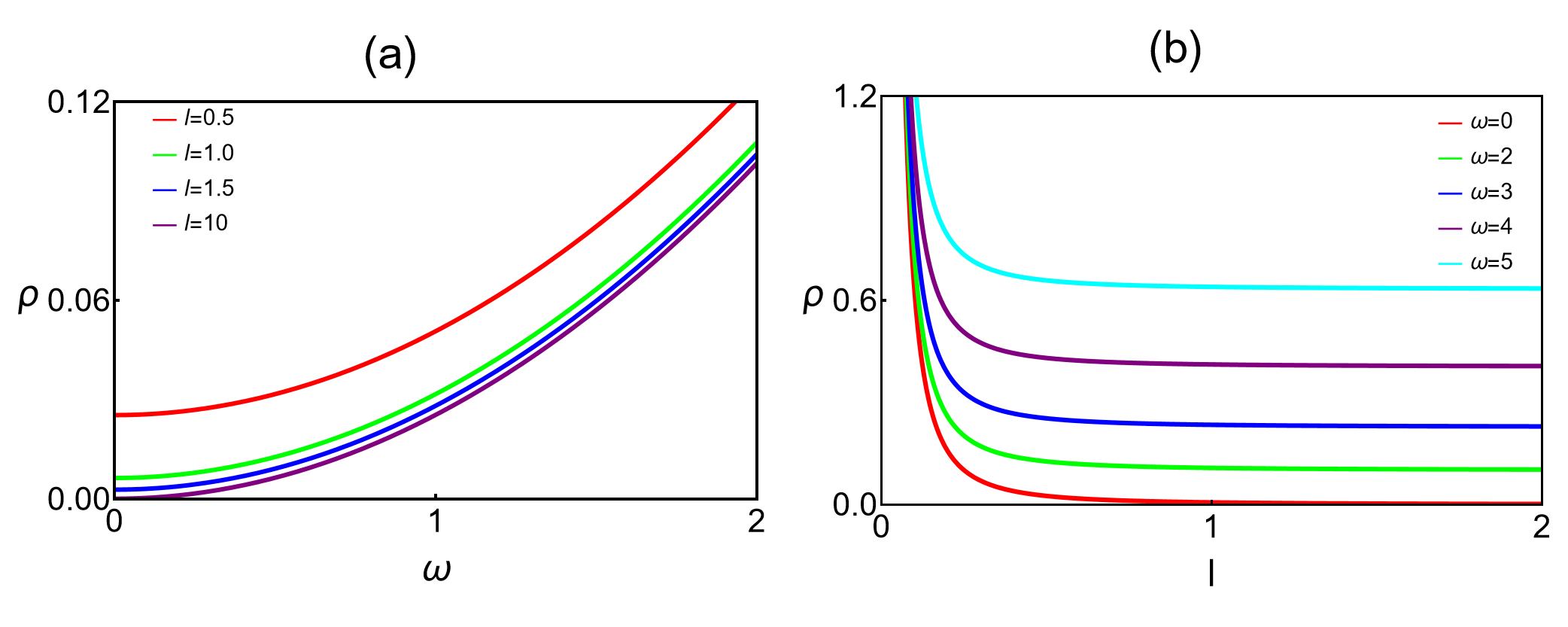}
  \caption{The density of states (DOS) for Dirac fermions in three-dimensional hyperbolic  space, as given by Eq.~\eqref{eq:3D-DOS}. (a) The DOS as a function of energy for several values of the radius of the hyperbolic space, $l$. We observe that  the zero-energy DOS is non-vanishing, but tends to zero as $l$ increases. (b) The DOS as a function of the radius, $l$, for several values of the energy.  }
  \label{fig:3D_rho(w,l)}
\end{figure*}
\begingroup
The associated dimensionless Casimir operator is
\begin{equation}
  \Xi^2 = \xi^2
  +\left(1-\frac{x^2+y^2}{l^2}\right)
  \frac{i\sigma_z}{2}M ,
\end{equation}
with \(M\) given in Eq.~\eqref{eq:operator-M}. Its eigenvalue equation,
\begin{equation}
  \Xi^2\Psi(x)=\kappa\Psi(x),
\end{equation}
is satisfied by the Dirac eigenfunctions derived in
App.~\ref{app:Dirac-curved-space}, with
\begin{equation}
  \kappa
  =
  \frac14\left[l^2\left(E^2-m^2\right)+1\right],
\end{equation}
in the units used here. This dimensionless combination is the one that
appears in the radial Dirac equation, Eq.~\eqref{R_Dirac_eq}.
These symmetry generators provide the starting point for the lattice
spinor transformations discussed below.
\endgroup

\subsection{Density of states}
\label{section_densityofstates}
Here we calculate the density of states (DOS) of Dirac fermions in $D-$dimensional hyperbolic space with $2\leq D\leq4$, to show that the coupling to the spin connection introduces a finite density of states at zero energy. We find the DOS from the corresponding  Green's function $G(x,y)$ in the real space
\begin{align}
\label{eq:greenfunction_def}
  \left[e^{\mu}{}_a\gamma^a\left(\partial_{\mu}+\frac{1}{2}\omega_{\mu}{}^{cb}\Sigma_{cb}\right)-m\right]&G(x,y)\nonumber\\
  &=i\delta(x-y),
\end{align}
\begingroup where \(x\) and \(y\) denote real-space coordinates and \(\delta(x-y)\) is the Dirac delta function.\endgroup We also define an auxiliary function $S(x,y)$ as
\begin{equation}
  \left[e^{\mu}{}_a\gamma^a\left(\partial_{\mu}+\frac{1}{2}\omega_{\mu}{}^{cb}\Sigma_{cb}\right)+m\right]S(x,y)=G(x,y),
\end{equation}
which we substitute into Eq.~\eqref{eq:greenfunction_def}. We introduce the spectral parameter \(s^2=E^2-m^2\).
\begingroup
For the DOS results discussed below, we specialize to massless Dirac fermions,
\(m=0\), so that \(s=|E|\) and the analytic continuation is directly in the
real energy. The formulas for the Green's function are first written
in terms of \(s\) and then evaluated in this massless limit.
\endgroup
An ansatz for the auxiliary function contains a matrix part \(U(x,y)\) and a function of the distance between two different points in space, with \(U(x,x)=1\). The  Green's function in $D$-dimensional hyperbolic space then takes the form~\cite{Camporesi:1992tm},
\begin{widetext}
\begin{align}
\label{greens_function}
  G(x,y;s)=&-i\frac{l^{1-D}}{(4\pi)^{\frac{D}{2}}}\frac{\Gamma(\frac{D}{2}+isl)\Gamma(isl+1)}{\Gamma(2isl+1)}\left[\cosh\left({\frac{d}{2l}}\right)\right]^{1-D-sl}\times \nonumber\\
  &\bigg{[}U(x,y)\cosh\left(\frac{d}{2l}\right){}_2F_1\left(\frac{D}{2}+isl,isl;2isl+1;\cosh^{-2}\left({\frac{d}{2l}}\right)\right)+\nonumber\\
  &+\gamma_in^iU(x,y)\sinh\left(\frac{d}{2l}\right){}_2F_1\left(\frac{D}{2}+isl,isl+1;2isl+1;\cosh^{-2}\left(\frac{d}{2l}\right)\right)\bigg{]}.
\end{align}
\end{widetext}
The DOS in $D$-dimensional hyperbolic space is found from the Green's function as
\begingroup
\begin{equation}
  \rho_D(\omega)= -\frac{1}{\pi}\lim_{\delta\rightarrow 0^+}\mathrm{Im}\mathrm{Tr}\!\left[\lim_{y\to x}G(x,y;s\rightarrow \omega+i\delta)\right],
\end{equation}
\endgroup
where analytical continuation in energy is performed. It should be noted that, in the Matsubara (imaginary-time) formalism, the energy parameter is purely imaginary, taking the form $s \to i s$, with $s \in \mathbb{R}$. To relate the Matsubara Green's function to physical observables defined for the real frequency, one performs analytic continuation by substituting $i s \to \omega + i\delta$, where $\omega$ is the real frequency and $\delta>0$ is  infinitesimal.


\begin{figure*}[t!]
  \centering
  \includegraphics[width=.9\linewidth]{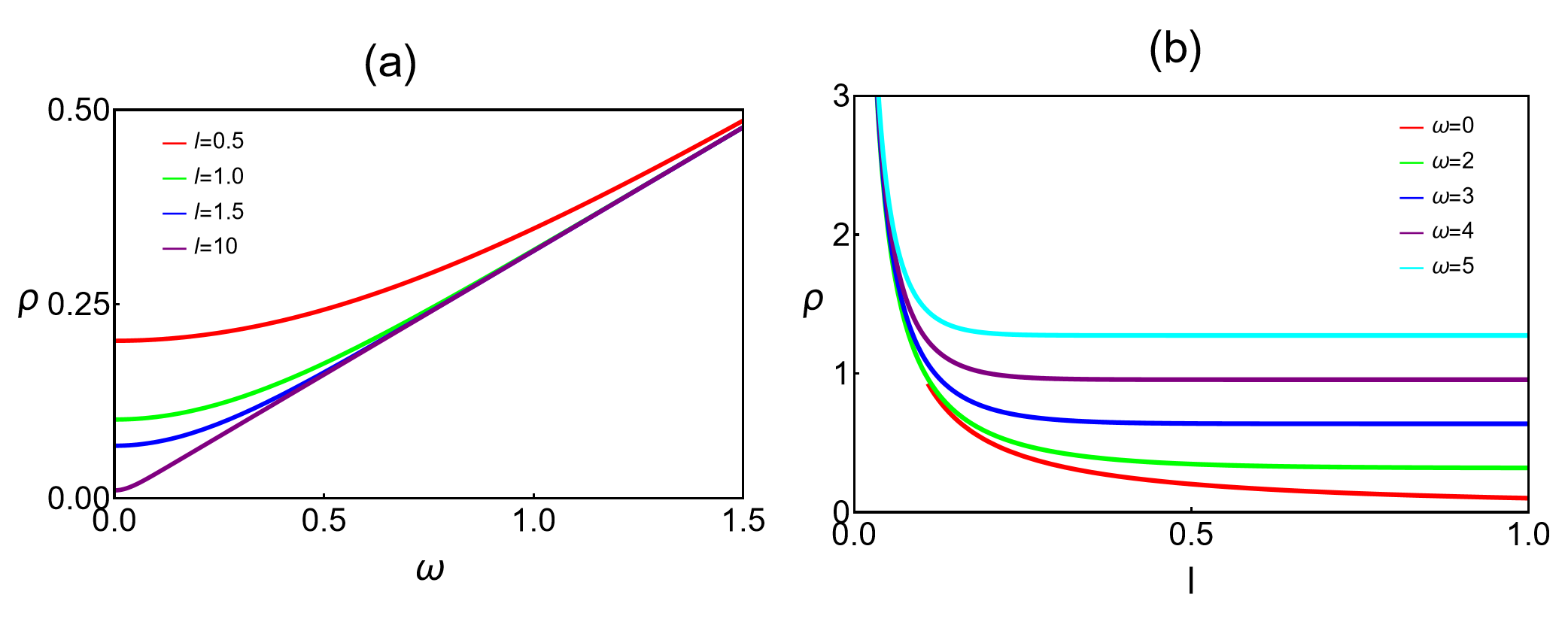}
  \caption{The density of states (DOS) for Dirac fermions  in the two-dimensional hyperbolic space, as given by Eq.~\eqref{eq:2D-DOS}. (a) The DOS as a function of energy for several values of the radius of the hyperbolic space, $l$.  We observe  that $\rho(\omega=0)$ tends to zero as $l$ increases, while for any finite $l$, the zero-energy DOS, $\rho(\omega=0)$, is finite. Notice that in the regime $\omega\gg l^{-1}$, the DOS approaches the linear scaling with energy, $\rho(\omega)\sim \omega$, consistent with the flat-space result. (b) The DOS as a function of the radius $l$, $\rho(l)$, for different energies. Notice that for $\omega\to0$, the DOS is finite, for  any finite radius, $l$ (red  curve), while for any non-zero energy, the DOS approaches  a finite value as $l$ increases.}
  \label{fig:2D_rho(w,l)}
\end{figure*}


\par We first focus on  the calculation of the  DOS in  two-dimensional space, $D=2$. After taking the trace of Eq. (\ref{greens_function}),  we find  a  divergence which we treat by employing  the dimensional regularization, with $\varepsilon=2-D$. Taking the limit $\delta\to0$ first, and then $D\to2$ (equivalently, $\varepsilon\to0$),
\begin{align}
  \rho_2(\omega,l)=-\frac{4}{\pi}\lim_{\varepsilon\rightarrow 0}\lim_{\delta\rightarrow 0} &\mathrm{Im}\frac{l^{\varepsilon-1}}{(4\pi)^{1-\frac{\varepsilon}{2}}} \nonumber\\
  &\times\frac{\Gamma(1-\frac{\varepsilon}{2}-il\omega+\delta l)\Gamma(\frac{\varepsilon}{2})}{\Gamma(-il\omega+\delta l+\frac{\varepsilon}{2})}.
\end{align}

We then obtain
\begin{equation}
  \rho_2(\omega,l)=-\frac{1}{\pi^2}\left(2\omega \mathrm{Im}[\psi(-il\omega)]+\frac{1}{l}\right),
  \label{eq:2D-DOS}
\end{equation}
where $\psi(z)$ is the digamma function, defined as $\psi(z)=d\ln\Gamma(z)/dz$. The plot of the DOS in the two-dimensional space is shown in Fig.~\ref{fig:2D_rho(w,l)}. The flat space limit ($l\to\infty$) of the density of state function in two-dimensional space is given by
\begin{equation}
\rho_2(\omega,l\to\infty)=\frac{\omega}{\pi},
\end{equation}
The normalization used here corresponds to a four-component Dirac fermion with Fermi velocity set to unity; a single irreducible two-component Dirac spinor contributes one half of this DOS. Importantly, in the zero-energy limit,  $\omega\rightarrow 0$, the DOS is non-vanishing and it scales with the scalar curvature as $\rho(0,R)\sim|R|^{1/2}$, with the explicit form given by
\begin{equation}
  \rho_2(0,R)=\frac{1}{\pi^2\sqrt{8}}|R|^{1/2},
  \label{eq:2D-DOS-zero-energy}
\end{equation}
For finite curvature, the expansion around zero energy further reads
\begin{equation}
\rho_2(\omega,R)=
  \rho_2(0,R)
  \left[
  1+\frac{8\pi^2}{3|R|}\,\omega^2
  +O\!\left(\frac{\omega^4}{|R|^2}\right)
  \right],
\label{eq:2D-DOS-correction}
\end{equation}
showing that the curvature-induced finite DOS is followed by a leading quadratic energy correction.
This is also in agreement with general scaling argument implying that the scaling dimension of the DOS $[\rho]=D-1$, in units of momentum (inverse length), with the dynamic exponent $z=1$, as it should be for a (pseudo-)relativistic system.
Notice that this result agrees with the form obtained in Ref.~\cite{gorbar2008gap} using a  different approach.

\par For completeness, we also consider  three-dimensional hyperbolic space,  $D=3$, yielding the DOS in the form
\begin{equation}
\rho_3(\omega,l)=\frac{\omega^2}{\pi^2}+\frac{1}{(2\pi l)^2},
\label{eq:3D-DOS}
\end{equation}
which therefore shows a separable form in the energy and the curvature. The effect of the curvature is to simply add a linear-in-curvature shift to the usual form of the DOS for 3D massless Dirac fermions in flat space, see
also Figure \ref{fig:3D_rho(w,l)}. Remarkably, in the zero-energy limit $\omega\rightarrow 0$, the DOS is non-zero and linearly scales with the curvature,
\begin{equation}
\rho_3(\omega=0,R)=\frac{|R|}{8(2\pi)^2}.
\end{equation}
On the other hand, in the limit $l\to \infty$ ($R\to 0$), we recover the well known  result in the flat space,
\begin{equation}
\label{eq:rho3D_l=infty}
\rho_3(\omega,l\to\infty)=\frac{\omega^2}{\pi^2}.
\end{equation}
Similarly, in $D=4$, we find
\begin{equation}
\rho_4(\omega,l)=-\frac{\omega}{4\pi^3l^2}(1+l^2\omega^2){\rm Im} \left[\psi(2-i l \omega)+\psi(-1-i l \omega)\right],
\end{equation}
also featuring  finite zero-energy DOS given by
\begin{equation}
\rho_4(\omega=0,R)=\frac{|R|^{3/2}}{64\pi^3\sqrt{2}}.
\end{equation}
Furthermore, the flat-space limit of the DOS is
\begin{equation}
\rho_4(\omega,l\to\infty)=\frac{\omega^3}{4\pi^2},
\end{equation}
in agreement with the result obtained by calculating the DOS of Dirac fermions directly in $D=4$ flat space.
\begin{figure}[t!]
  \centering
  \includegraphics[width=0.7\linewidth]{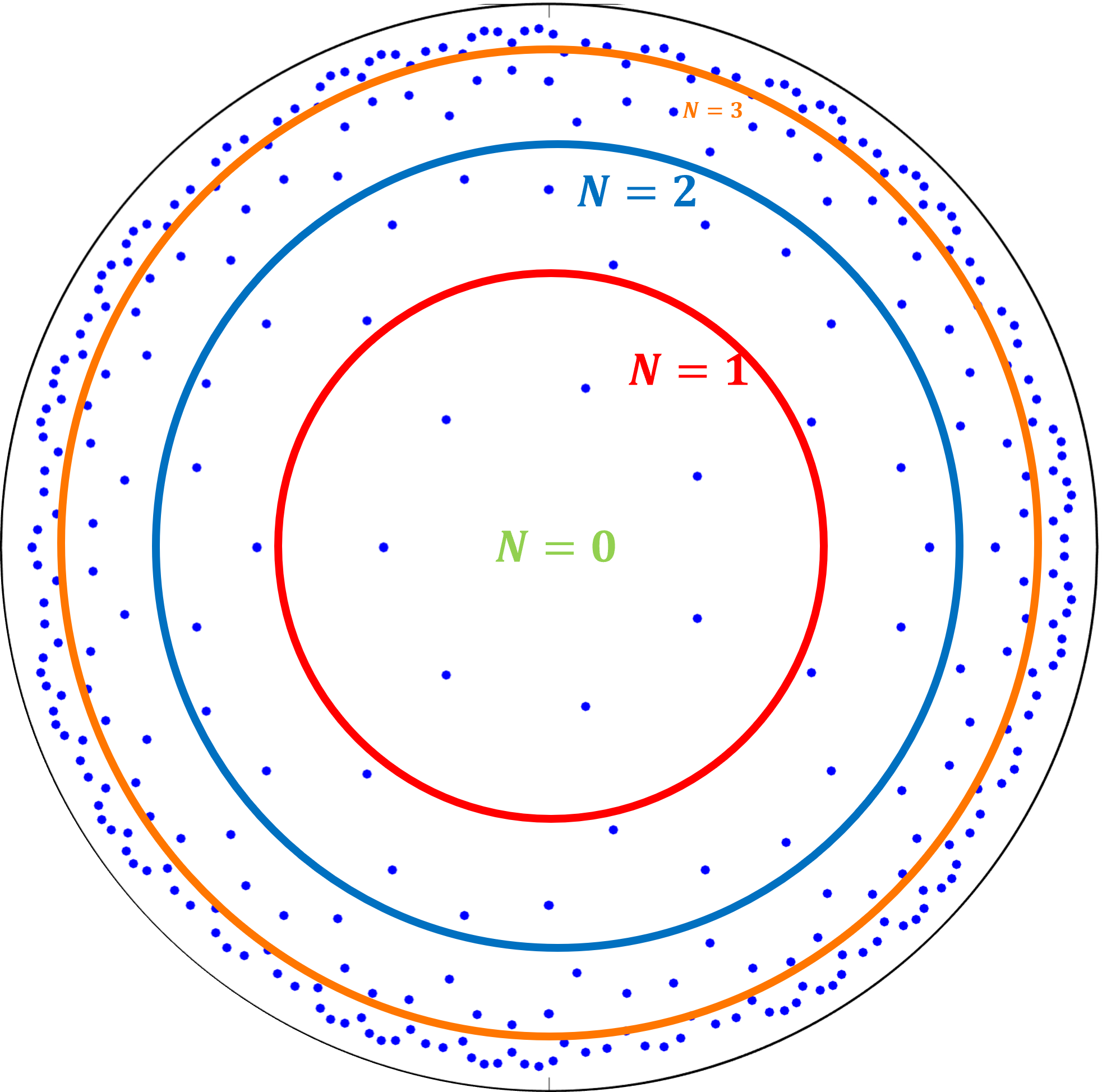}
  \caption{Schematic representation of the $\{7,3\}$ hyperbolic lattice, illustrating the hierarchical organization of polygons by generation for a tessellation of the Poincaré disk with Schl\"afli symbol $\{p,q\}$. The central polygon is generation $g=0$. The red, blue, and orange circles mark generations $g=1$, $2$, and $3$, respectively. Subsequent generations are identified analogously.}
  \label{fig:generations}
\end{figure}

\section{Discretization of Poincar\'{e} disk}
\label{PointGroupSymmetry}
Motivated by this nontrivial behavior of the Dirac fermions coupled to the spin connection in the hyperbolic space, we now consider discretization of the Poincaré disk. To this end, in this Section we obtain the discrete symmetries of a hyperbolic lattice starting from their continuum counterparts on the Poincar\'{e} disk, and thus consider the point group symmetry of the unit cell, in Sec. \ref{Point_group_symmetry}, and the translations, in Sec. \ref{translations_section}. For convenience, we place the center of a polygon at the origin of the tessellated Poincar\'{e} disk and systematically organize the polygons into successive generations, as illustrated in Fig.~\ref{fig:generations}. From this section on, we fix the radius of the Poincaré disk to unity ($l = 1$), unless stated otherwise.

\subsection{Point group symmetry}
\label{Point_group_symmetry}

\par We begin by constructing the cyclic point group corresponding to the Möbius transformation given in Eq.~\eqref{eq:Mobius_transformation}. This construction is then used to derive the explicit form of discrete rotational symmetries on the tessellated Poincaré disk.

To determine the structure of the point group operations, we recall that repeated application of a group element eventually yields the identity transformation; the number of applications required is referred to as the order of the element. Specifically, we consider the symmetry transformation corresponding to the application of the transformation
$p$ times to a given point, as follows:
\begin{align}
  \label{Mtrans_1} z^{(1)}&=e^{i\Tilde{\theta}}\frac{z-c}{1-\Bar{c}z}, \\
  z^{(2)}&=e^{i\Tilde{\theta}}\frac{z^{(1)}-c}{1-\Bar{c}z^{(1)}}, \\
  &\hspace{2mm}\vdots, \\
  \label{Mtrans_p} z^{(p)}&=e^{i\Tilde{\theta}}\frac{z^{(p-1)}-c}{1-\Bar{c}z^{(p-1)}}.
\end{align}
We  can thus identify parameters $\theta_p$ and $c_p$ after every transformation as:
\begin{align}
  \label{theta_p}
e^{i\theta_p}&=e^{i\Tilde{\theta}}\frac{e^{i\theta_{p-1}}+c\Bar{c}_{p-1}}{1+\Bar{c}c_{p-1}e^{i\theta_{p-1}}}, \\
  \label{c_p}
  c_p&=\frac{c_{p-1}e^{i\theta_{p-1}}+c}{e^{i\theta_{p-1}}+c\Bar{c}_{p-1}}, \quad p \in \mathbb{N}.
\end{align}
\begin{figure*}[h]
  \centering
  \includegraphics[width=0.8\textwidth, height=11cm]{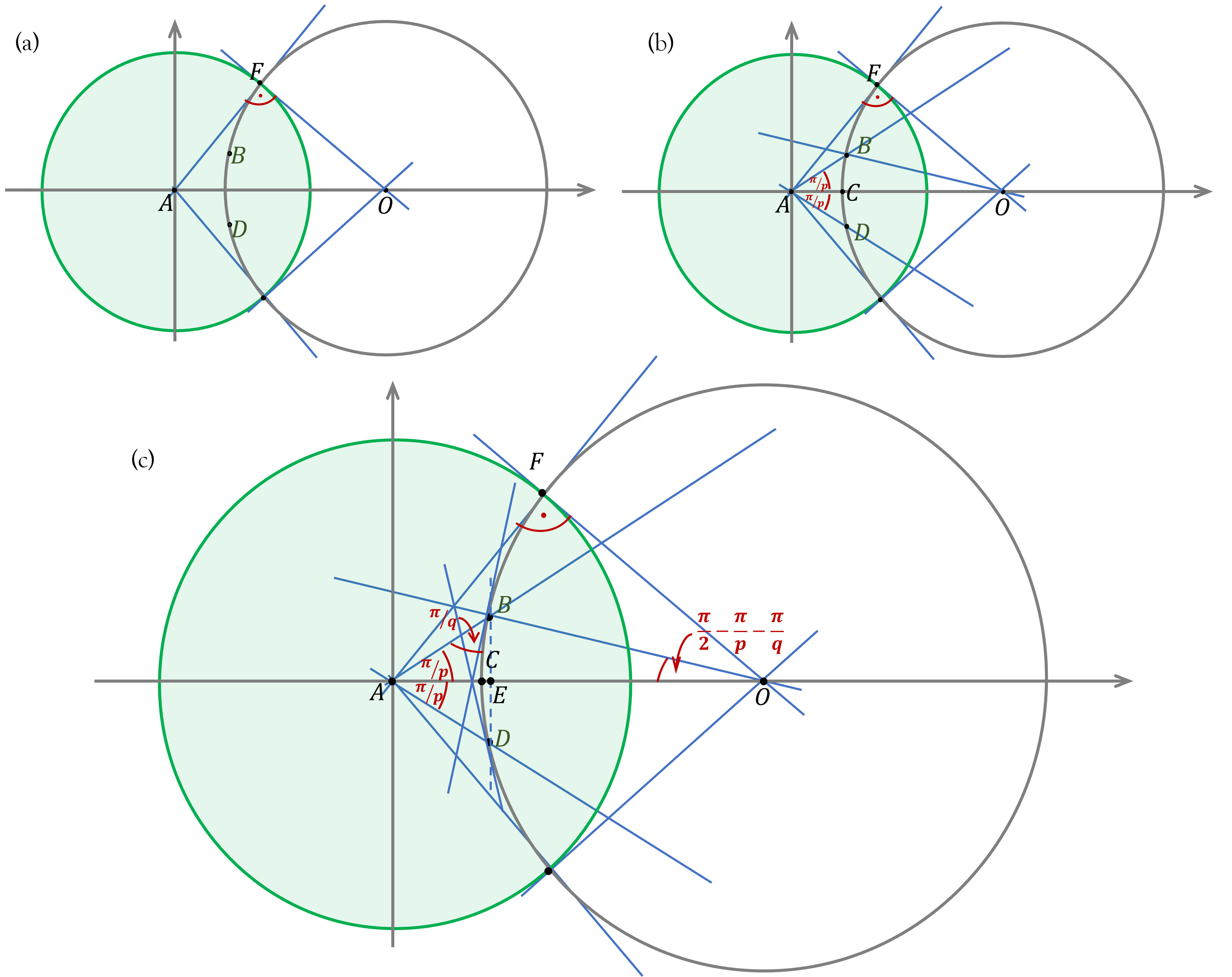}
  \caption{Geometric construction within the Poincar'{e} disk, illustrating the main  quantities for defining the hyperbolic $\{p, q\}$ lattice. The disk is centered at point $A(0,0)$, and its geodesic is represented as a circular arc of radius $R$ centered at $O(\kappa+R, 0)$. (a) Geodesics in the Poincar'{e} disk correspond to circles orthogonal to the disk boundary. At point $F$, a right triangle $\triangle AFO$ can be constructed. Points $B$ and $D$ indicate two vertices of a regular $p$-gon centered at $A$ and lying on the same geodesic. (b) We denote the Euclidean distance from the center of the disk to the nearest points of the hyperbolic lattice $r_0$, i.e. $AB=AD=r_0$. Point $C$ marks the intersection of the $x$-axis, passing through $A$ and $O$, with the geodesic circle, such that $AC = \kappa$.  As we deal with regular $p$-gons on the $\{p,q\}$ hyperbolic lattice, an important angle in the construction is given as $\angle BAC=\angle CAD=\frac{\pi}{p}$. (c) Another important  angle is $\angle ABD=\angle ADB=\frac{\pi}{q}$.  The shortest Euclidean distance from point $B$ ($D$) to the line $AO$ is indicated by $BE$ ($DE$), depicted as a dotted line.}\label{fig2}
\end{figure*}

Let us assume  that after applying  the transformation $p$ times, we obtain $z^{(p)}=z$, that corresponds to the unit (trivial) element of the discrete group. This then yields the conditions on the parameters,
\begin{equation}
  e^{i\theta_p}=1, \quad c_p=0,
\end{equation}
which result in the constraints on the arbitrary parameters,
\begin{equation}
  c_{p-k}e^{i\theta_{p-k}}=-c_k.
\end{equation}
Finally, the condition imposed on the parameters of continuum group symmetry  to yield the point group is given by:
  \begin{equation}
\label{condition_on_parameters}
  1+\cos\tilde{\theta}=(1-|c|^2)\left(1+\cos\frac{2n\pi}{p}\right),
\end{equation}
where $n \in \{1,...,p-1\}$.
We connect these parameters to those introduced in Eq. \eqref{eq:coordinate_transformation} for the continuous Möbius transformation, as follows:
\begin{align}
  \frac{1+\cos\Tilde{\theta}}{1-|c|^2}&=1+\cosh(\sqrt{-\theta^2+4|a|^2}) \nonumber\\
  \Longrightarrow & \cosh(\sqrt{-\theta^2+4|a|^2})=\cos \frac{2n\pi}{p}, \quad \theta\ge 2|a| \nonumber\\
  \Longrightarrow & \sqrt{\theta^2-4|a|^2}=\pm \frac{2n\pi}{p}+2\pi k, \quad k\in \mathbb{Z}.
\end{align}
In this way, we constructed the simplest discrete group for our problem, with one generator of order $p$, namely the cyclic group $C_p$. For the transformation to have exact order $p$, one must additionally require that $n$ and $p$ are coprime. See also App.~\ref{AppendixB} for the technical details.

\par Next, we  determine the rotations around an arbitrary point $z_o$ of the tessellated Poincaré disk in which the center of a polygon is positioned  at the center of the disk and employ polygon generations, as illustrated in Fig.~\ref{fig:generations}.  To this end, we will use arbitrary $l$ and perform the transformation that brings an arbitrary point to the origin,
\begin{equation}
  z_o'=l^2e^{i\Tilde{\theta}}\frac{z_o-c}{l^2-\Bar{c}z_o},
\end{equation}
such that $z_o'=0$,
which in turn implies the form of  the parameters given by
\begin{equation}
  z_o=c, \quad  \tilde{\theta}=2n\pi,
\end{equation}
where $n\in\mathbb{Z}$.
We then choose parameter $\tilde{\theta}=0$, i.e. $n=0$. The rotations  are defined as
\begin{equation}
  z_R'=e^{i\alpha}z',
\end{equation}
where the primed coordinates denote the transformed coordinates of the site, while the inverse transformation reads
\begin{equation}
  z_R=l^2\frac{z_R'+z_o}{l^2+\Bar{z_o}z_R'}.
\end{equation}

Finally, the discrete rotations on the lattice embedded in the Poincaré disk take the form
\begin{equation}
  z_R=\frac{l^2e^{i\alpha}-|z_o|^2}{l^2-|z_o|^2e^{i\alpha}}\,\,\frac{z-z_ol^2\frac{e^{i\alpha}-1}{l^2e^{i\alpha}-|z_o|^2}}{1-\Bar{z_o}\frac{1-e^{i\alpha}}{l^2-|z_o|^2e^{i\alpha}}z},
  \label{eq:discrete-rotation}
\end{equation}
where the parameters of the rotation, as defined in Eq.~\eqref{eq:Mobius_transformation}, are given by
\begin{align}
  c&=z_o l^2 \frac{e^{i\alpha}-1}{l^2e^{i\alpha}-|z_o|^2}, \\
e^{i\tilde{\theta}}&=\frac{l^2e^{i\alpha}-|z_o|^2}{l^2-|z_o|^2e^{i\alpha}}.
\end{align}
The choice of Schläfli symbols, which characterize the tessellation of the Poincaré disk, therefore uniquely determines the rotation parameter $\alpha$ and, consequently, fully specifies the discrete rotational symmetries of the lattice.

\subsection{Translations}
\label{translations_section}

In the previous subsection, we derived the conditions under which the continuum symmetry of space reduces to the discrete symmetry of a lattice, as captured by Eq.~\eqref{condition_on_parameters}. We now turn to the construction of discrete transformations corresponding to translations inherited from flat space.

To proceed, we first determine the distance from the center of the Poincar\'e disk to the nearest site of the $\{p,q\}$ hyperbolic lattice. This distance is obtained by an elementary Euclidean construction involving the corresponding geodesic circles. As illustrated in Fig.~\ref{fig2}, when the origin coincides with the disk center, these geodesics appear as circles orthogonal to the boundary of the disk, a property that will play a crucial role in our analysis.


We introduce the following notation for the relevant distances, as depicted in Fig.~\ref{fig2},
\begin{align}
  AC &= \kappa, \quad OC = OF = OB = R, \label{eq:parameter_kappaR} \\
  AB &= AD = r_0, \label{eq:parameter_r0}
\end{align}
which satisfy the relations
\begin{align}
  (\kappa+R)^2 &= l^2+R^2, \label{Pythagorean_theorem} \\
  r_0\sin\frac{\pi}{p} &= R\sin\left(\frac{\pi}{2} - \frac{\pi}{p} - \frac{\pi}{q}\right), \label{B_to_line} \\
  \kappa+R &= r_0\cos\frac{\pi}{p} + R\cos\left(\frac{\pi}{2} - \frac{\pi}{p} - \frac{\pi}{q}\right). \label{AO_line}
\end{align}
Here, Eq.~\eqref{Pythagorean_theorem} is the Pythagorean relation for the triangle $\triangle AFO$, Eq.~\eqref{B_to_line} gives the shortest distance from $B$ to the line $AO$, and Eq.~\eqref{AO_line} specifies the distance between $A$ and $O$.

Solving these equations yields the exact distance from the center to the nearest site, which depends only on the Schl\"{a}fli symbols:
\begin{equation}
  r_0 = l\sqrt{\frac{\cos\left(\frac{\pi}{p} + \frac{\pi}{q}\right)}{\cos\left(\frac{\pi}{p} - \frac{\pi}{q}\right)}}.
  \label{eq:central_distance_r0}
\end{equation}
Similarly, we obtain the distance from the center $A$ to the nearest circular geodesic at point $C$ as\begin{equation}
\label{eq:kappa}
  \kappa = l\frac{\cos\frac{\pi}{q} - \sin\frac{\pi}{p}}{\sqrt{\cos(\frac{\pi}{p}+\frac{\pi}{q})\cos(\frac{\pi}{p}-\frac{\pi}{q})}}.
\end{equation}
Since we are after the form of  translations, we now compute the corresponding hyperbolic distance,
\begin{equation}
  d_{\kappa} = \frac{l}{2} \mathrm{arcosh}\left(\frac{\cos\frac{\pi}{q}}{\sin\frac{\pi}{p}}\right).
\end{equation}
with the hyperbolic distance between two polygon centers being $2d_{\kappa}$. Furthermore,  the Euclidean distance $d_E$ between these centers, obtained from Eq.~(\ref{hyperbolic_distance}), is  given by
\begin{equation}
\label{eq:Euclidian_distance_dE}
  d_E = l\sqrt{1 - \frac{\sin^2\frac{\pi}{p}}{\cos^2\frac{\pi}{q}}}.
\end{equation}

\begingroup
To write the translation in dimensionless form, we introduce
\begin{equation}
  w\equiv\frac{z}{l},
  \qquad
  \delta_E\equiv\frac{d_E}{l}.
\end{equation}
A translation from the origin toward direction
$\phi_m=2\pi m/p$ then acts as
\begin{equation}
  w'=t_m w
  =\frac{w+\delta_Ee^{i\phi_m}}
  {1+\delta_Ee^{-i\phi_m}w},
  \qquad z'=lw',
  \label{eq:traslation_transformation}
\end{equation}
where $m\in\{0,\ldots,p-1\}$. A normalized matrix representative is
\begin{equation}
\label{eq:translation_matrix}
  t_m=\frac{1}{\sqrt{1-\delta_E^2}}
  \begin{pmatrix}
  1 & \delta_Ee^{i\phi_m} \\
  \delta_Ee^{-i\phi_m} & 1
  \end{pmatrix}.
\end{equation}

Proceeding along a fixed direction requires repeated application of this transformation. Its $n$th power is
\begin{widetext}
\begin{equation}
  t_m^n = \frac{1}{2(1-\delta_E^2)^{n/2}}
  \begin{pmatrix}
  (1+\delta_E)^n+(1-\delta_E)^n & e^{i\phi_m}\!\left[(1+\delta_E)^n-(1-\delta_E)^n\right] \\
  e^{-i\phi_m}\!\left[(1+\delta_E)^n-(1-\delta_E)^n\right] & (1+\delta_E)^n+(1-\delta_E)^n
  \end{pmatrix}.
  \label{eq:discrete-translation}
\end{equation}
\end{widetext}

Finally, the corresponding Lie-group parameters introduced in Eq.~\eqref{eq:coordinate_transformation} are
\begin{align}
  a &= -\mathrm{arctanh}(\delta_E)e^{i\phi_m}, \\
  \bar a &= -\mathrm{arctanh}(\delta_E)e^{-i\phi_m}, \\
  \tilde\theta &= 0.
\end{align}
These parameters will be used in the next section to construct the discrete spin connection.
\endgroup We emphasize that, unlike ordinary translations in Euclidean space, the transformations \(t_m\) inherited from the M\"obius action generally do not commute. This noncommutativity, together with the conditions under which commuting translations may be recovered, is analyzed in App.~\ref{AppendixC}.

\begin{figure*}[t]
  \centering
\includegraphics[width=0.95\textwidth]{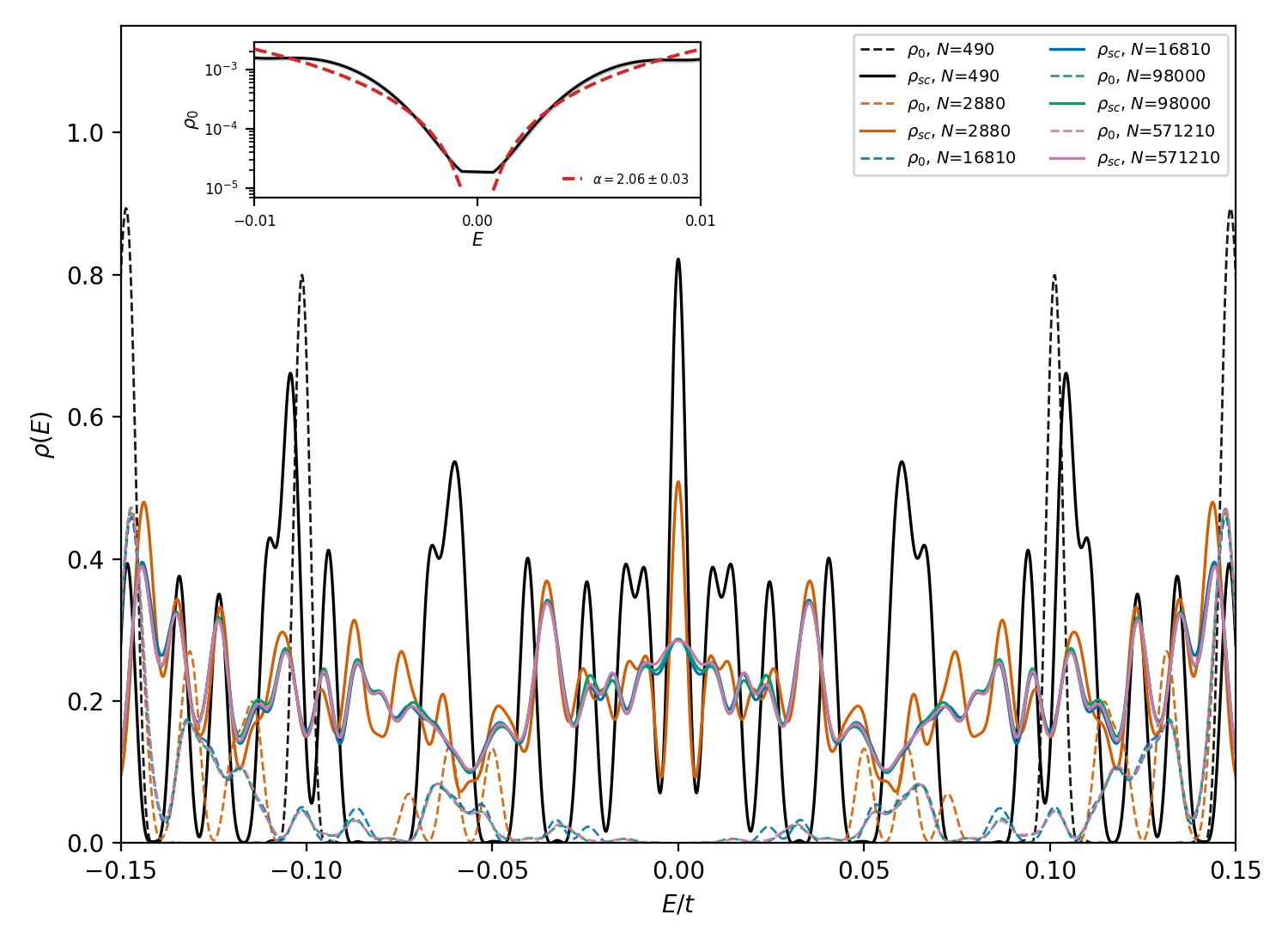}
  \caption{Density of states (DOS) with
($\rho_{\rm sc}$, solid) and without ($\rho_0$, dashed) the spin
connection for open $\{10,3\}$ lattices with generations $2-6$
($N=490$, $2880$, $16810$, $98000$, and $571210$), evaluated at $t=1$ and
$\ell=1$ using kernel-polynomial method (KPM) with $M=4096$ Chebyshev moments and
$R=150,72,36,12,12$ stochastic vectors, respectively.
Both $H_{\rm ref}=H_0\otimes\mathbb{I}_2$ and $H_{\rm sc}$ obey
chiral symmetry,
$\Gamma_{\rm ch}H\Gamma_{\rm ch}=-H$, with
$\Gamma_{\rm ch}=\tau_z\otimes\mathbb{I}_2$, and the curves show the
symmetrized estimator
$\rho_{\rm sym}(E)=[\rho(E)+\rho(-E)]/2$.
The residual asymmetry of the unsymmetrized estimator decreases from
$6.4\%$ to $0.9\%$ of the peak height over the displayed size range.
Inset: low-energy reference DOS for the largest lattice, recomputed
with $R=48$ stochastic vectors and shown on a logarithmic vertical
scale. A joint fit to both signs of the energy over
$7\times10^{-4}\lesssim |E|<10^{-2}$ yields
$\rho_0(E)=C|E|^\alpha$ with $\alpha=2.06\pm0.03$ (red dashed line),
consistent with Ref.~\cite{Leong-Roy-PRB2026}.}
\label{fig:DOS_spin_connection_generations}
\end{figure*}

\section{Lattice implementation of the discrete spin connection}
\label{LatticeDiracFermions}

\begingroup
In this section, we construct a minimal spinor tight-binding model on
a hyperbolic lattice that incorporates the parallel-transport factor
associated with the continuum spin connection. The construction
preserves the local $\mathrm{Spin}(2)$ link structure and the
corresponding plaquette holonomy. It is not intended as a unique or
complete lattice discretization of the continuum Dirac operator;
rather, it isolates the spectral consequences of introducing the
geodesic spin-connection factor into a simple nearest-neighbor model.
The starting point is the transformation law of the continuum Dirac
spinor generated by
Eqs.~\eqref{eq:operator-Kx}--\eqref{eq:operator-Ky}.
\endgroup In the coordinate representation, we write
\begin{equation}
  \ket{\Psi}=\sum_{\sigma}\int dz\, \Psi_{\sigma}(z)\ket{z,\sigma},
\end{equation}
\begingroup
Here \(\sigma=\pm\) labels an internal two-component spinor degree of freedom
at each lattice site, i.e., the eigenvalue of \(\sigma_z\). This degree of
freedom is distinct from the sublattice pseudospin associated with the
bipartite \(\{10,3\}\) graph, so the lattice Hilbert space has the product
structure \(\mathcal H_{\rm lattice}\otimes\mathbb C^2_{\sigma}\).
Depending on the experimental platform, the \(\sigma\) sector may represent
physical spin or a synthetic polarization/orbital doublet. The spin
connection acts in this internal sector and gives opposite link phases to
the two components.
\endgroup
Using the Baker--Campbell--Hausdorff formula, the symmetry transformation of the spinor can be written as (see App.~\ref{App:DerivationEq105} for the details of the derivation)
\begin{equation}
  \label{eq:infinitesimal-Dirac-transf}
  \Psi_{\sigma}'(z')=
  e^{i f_{\gamma}(z',z)\sigma_z/2}\,e^{\partial_{\xi}}\Psi_{\sigma}(z),
\end{equation}
with
\begin{equation}
\label{Eq:Symf(z,z)}
  f_{\gamma}(z',z)=\theta-\frac{i}{z_2-z_1}
  \left(
  z_2\log\frac{z'-z_2}{z-z_2}
  -
  z_1\log\frac{z'-z_1}{z-z_1}
  \right),
\end{equation}
and
\begin{equation}
  \xi=
  \frac{1}{\sqrt{-\theta^2+4|a|^2}}
  \log{\frac{z-z_2}{z-z_1}}.
\end{equation}
Here \(z_1\) and \(z_2\) are defined in Eqs.~\eqref{eq:z1} and~\eqref{eq:z2}. Equation~\eqref{eq:infinitesimal-Dirac-transf} separates the transformation into two parts. The operator \(e^{\partial_\xi}\) implements the coordinate transformation \(z\mapsto z'=\gamma^{-1}z\), thereby generating hoppings between neighboring lattice sites.
The factor \(e^{if_{\gamma}(z',z)
\sigma_z/2}\) describes the spinor transformation induced by the
finite M\"obius transformation \(\gamma^{-1}\). This finite-isometry
spin rotation is not, in general, identical to Levi--Civita parallel
transport along the geodesic bond connecting two neighboring lattice
sites~\cite{Helgason2001}; see App.~\ref{App:Paralell_transport_equation_solving} for the
parallel transport along such a bond. To implement the lattice spin
connection, we therefore use the geodesic Wilson line, which directly
represents parallel transport along the bond.

\subsection{Tight-binding implementation and density of states}

We now implement this spin-connection factor in a tight-binding model on the \(\{10,3\}\) hyperbolic lattice. As a reference, we first consider the spinless nearest-neighbor model
\begin{equation}
  H_0=-t\sum_{\langle i,j\rangle}
  \left(c_i^\dagger c_j+\mathrm{h.c.}\right),
\end{equation}
\begingroup
For a direct comparison in the same \(2N\)-dimensional spinor
Hilbert space, we define the reference Hamiltonian as
\(H_{\rm ref}\equiv H_0\otimes\mathbb{I}_2\),
\endgroup
\begingroup
where each undirected nearest-neighbor bond \(\langle i,j\rangle\) is included once. The same KPM procedure used for the spin-connection Hamiltonian is applied to \(H_0\). The dashed curves in
Fig.~\ref{fig:DOS_spin_connection_generations} show that the
fixed-resolution reference DOS is strongly suppressed near \(E=0\),
consistent with the previously reported nodal low-energy structure of
the nearest-neighbor \(\{10,3\}\)
model~\cite{Roy_2024,Leong-Roy-PRB2026}.
\endgroup

\begin{figure*}[t]
  \centering
  \includegraphics[width=0.8\textwidth]{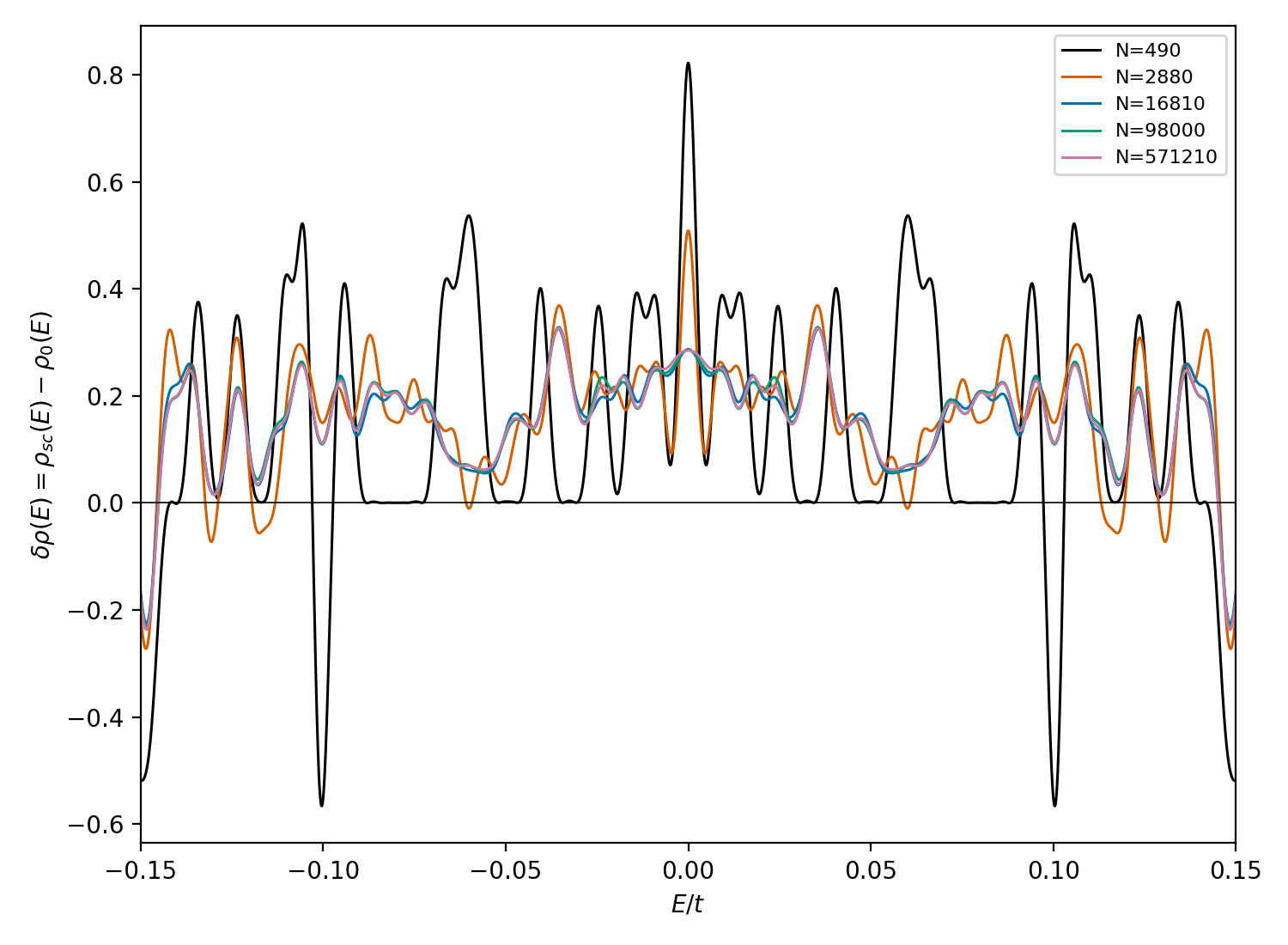}
  \caption{Difference $\delta\rho(E)=\rho_{\rm sc}(E)-\rho_0(E)$ for generations 2 through 6 (as defined in Fig.~\ref{fig:generations}) of the $\{10,3\}$ lattice. Both terms are calculated with $M=4096$ Chebyshev moments, as in Fig.~\ref{fig:DOS_spin_connection_generations}, using the same energy-symmetrized estimators, $\rho_{\rm sym}(E)=[\rho(E)+\rho(-E)]/2$. The positive low-energy enhancement persists across all five sizes and is nearly size independent from $N=16810$ through $N=571210$, with larger values at the two smallest sizes. See also Table~\ref{tab:GC-floor}.}
  \label{fig:DeltaDOS_spin_connection_generations}
\end{figure*}

We next incorporate the spinor structure. The matrix part of Eq.~\eqref{eq:infinitesimal-Dirac-transf} acts differently on the two spin components, producing opposite spin-dependent link factors. In the coordinate basis \(\{\ket{z,\sigma}\}\), this gives
\begin{widetext}
\begin{equation}\label{eq:Hsc-TB}
  H_{\rm sc}=
  -t\sum_{z}
  \begin{pmatrix}
  \ket{\gamma^{-1}z,+} & \ket{\gamma^{-1}z,-}
  \end{pmatrix}
  \begin{pmatrix}
  e^{\frac{i}{2}f^{\rm W}_{ij}} & 0 \\
  0 & e^{-\frac{i}{2}f^{\rm W}_{ij}}
  \end{pmatrix}
  \begin{pmatrix}
  \bra{z,+} \\
  \bra{z,-}
  \end{pmatrix}
  +\mathrm{h.c.},
\end{equation}
where the neighboring site is \(z'=\gamma^{-1}z\), with \(\gamma\) determined by the coordinate part of the transformation. \begingroup Now,
the spin-connection Hamiltonian can be written as
\begin{align}
  H_{\rm sc}=-t\sum_{z}
  \Big[
  e^{if^{\rm W}_{ij}/2}
  \ket{\gamma^{-1}z,+}\bra{z,+}
  +
  e^{-if^{\rm W}_{ij}/2}
  \ket{\gamma^{-1}z,-}\bra{z,-}
  \Big]+\mathrm{h.c.},
\end{align}
\end{widetext}
\endgroup
with \(z'=\gamma^{-1}z\). Here $i$ and $j$ label the sites at $z$ and $z'$, respectively, and $f^{\rm W}_{ij}$ denotes the geodesic Wilson-line phase defined below, rather than the finite-isometry phase $f_{\gamma}$. \begingroup
In the $\sigma_z$ basis the Hamiltonian decomposes as
\begin{equation}
  H_{\rm sc}=H_+\oplus H_-,\qquad H_-=H_+^*,
\end{equation}
so the two spin blocks have identical spectra.
\endgroup Each block is equivalent to a $U(1)$ Peierls model with the opposite plaquette flux, which is fixed geometrically by spin parallel transport and by the Gauss--Bonnet curvature enclosed by each plaquette. Thus, the discrete spin connection enters as a spin-dependent hopping factor attached to each nearest-neighbor link. The sum contains one orientation of each undirected bond; the explicit Hermitian-conjugate term implements the reverse hopping and enforces \(\mathcal U_{ji}=\mathcal U_{ij}^{\dagger}\), where the link phase is
\begin{equation}
\mathcal{U}_{ij}=e^{i\frac{ f^{\rm W}_{ij}}{2}\sigma_z},
\end{equation}
with $f^{\rm W}_{ij}=\int_{\gamma_{ij}}dx^\mu\,\omega_\mu(x)$. \begingroup
Under a local $\mathrm{Spin}(2)$ frame rotation
$g_i=e^{i\alpha_i\sigma_z/2}$,
\begin{equation}
  \Psi_i\to g_i\Psi_i,
  \qquad
  \mathcal U_{ij}\to g_i\mathcal U_{ij}g_j^{-1},
\end{equation}
which leaves the hopping term gauge covariant and makes closed-loop holonomies gauge invariant.
\endgroup \begingroup
For the finite-link Hamiltonian, we therefore define the lattice spin connection directly through the geodesic Wilson line. This provides a gauge-covariant link variable for each nearest-neighbor bond.
\endgroup
In the numerical calculations, the phase entering the Hamiltonian~\eqref{eq:Hsc-TB} is evaluated from the exact geodesic Wilson line in Eq.~\eqref{eq:exact-wilson-line}, thereby preserving the correct finite-bond holonomy of the tessellation.

\begin{table*}
\begin{center}
\begin{tabular}{|c|c|c|c|c|}
\hline
Gen.\ of polygons & $N$ & $\rho_0(0)$ & $\rho_{\mathrm{sc}}(0)$ & $\delta\rho(0)$ \\
\hline
2 & 490  &  $(3.93\pm0.11)\times10^{-7}$ &  $0.822\pm0.039$ &  $0.822\pm0.039$ \\
\hline
3 & 2880  &  $(4.12\pm0.13)\times10^{-7}$ &  $0.509\pm0.023$ &  $0.509\pm0.023$ \\
\hline
4 & 16810  &  $(7.21\pm0.21)\times10^{-7}$ &  $0.2870\pm0.0089$ &  $0.2870\pm0.0089$ \\
\hline
5 & 98000  &  $(8.23\pm0.66)\times10^{-6}$ &  $0.2858\pm0.0057$ &  $0.2858\pm0.0057$ \\
\hline
6 & 571210 &  $(1.419\pm0.079)\times10^{-5}$ &  $0.2846\pm0.0016$ &  $0.2846\pm0.0016$ \\
\hline
\end{tabular}
\end{center}
\caption{Zero-energy density of states (DOS) for the
$\{10,3\}$ lattice at $\ell=1$, obtained with the kernel polynomial
method and normalized per single-particle state.
Here $\delta\rho(0)=\rho_{sc}(0)-\rho_0(0)$. We use $M=4096$ Chebyshev
moments, 6001 energy points, and $R=150,72,36,12,12$ stochastic
vectors used in simulations in the lattices from
generation 2--6. The values correspond to the fixed numerical resolution
used in Figs.~\ref{fig:DOS_spin_connection_generations}--\ref{fig:saturation-vs-N}.
Uncertainties quoted are standard errors of the mean, $\sigma/\sqrt{R}$,
where $\sigma$ is the sample standard deviation of the $R$ independent
stochastic-trace estimates at each system size. These error bars quantify only stochastic-trace sampling and do not include systematic effects associated with finite system size, finite Chebyshev order, Jackson broadening, or spectral rescaling. Since $\rho_0(0)$ is negligible on the displayed scale, the uncertainty of $\delta\rho(0)$ coincides with that of $\rho_{\rm sc}(0)$ to the precision shown.}
\label{tab:GC-floor}
\end{table*}

\begin{figure*}[t!]
\centering
\includegraphics[width=0.7\linewidth]{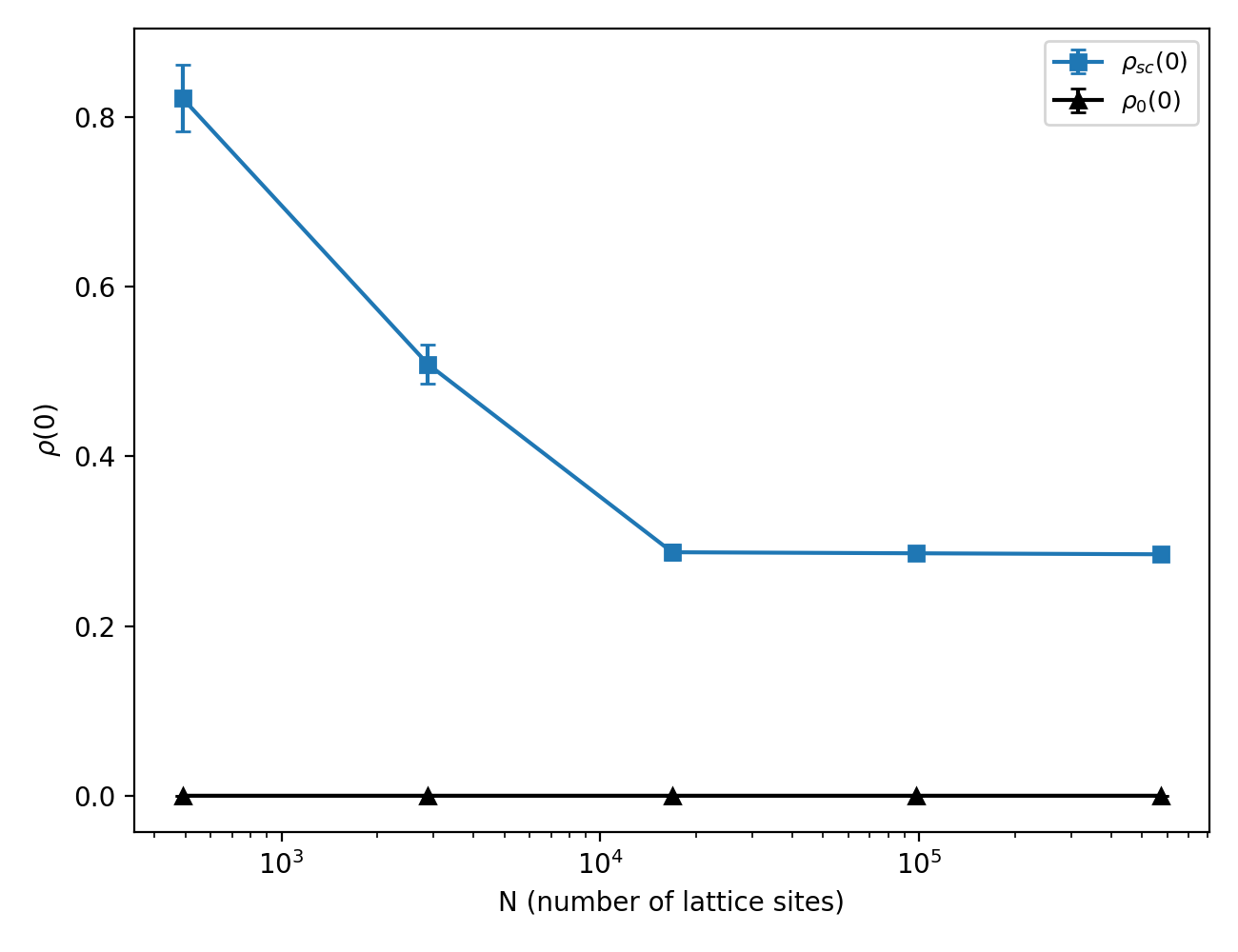}
\caption{Zero-energy density of states with ($\rho_{\rm sc}(0)$) and without ($\rho_0(0)$) the spin connection as a function of system size for open $\{10,3\}$ lattices with generations 2--6 and $l=1$, evaluated with $M=4096$ Chebyshev moments. Error bars show the stochastic standard error of the mean, $\sigma/\sqrt{R}$, estimated from the $R$ independent stochastic-trace samples at each system size (Table~\ref{tab:GC-floor}); they do not include finite-size or finite-$M$ systematic effects. The three largest systems are mutually compatible at this stochastic-error level, while the two smallest systems yield the larger values $0.822\pm0.039$ and $0.509\pm0.023$. The reference value remains at or below $(1.42\pm0.08)\times10^{-5}$. Lines connecting the points are guides to the eye.}
\label{fig:saturation-vs-N}
\end{figure*}

\begingroup
We implement the finite-link spin connection through the geodesic Wilson line, which provides the gauge-covariant parallel-transport operator associated with the continuum spin connection~\cite{MBlagojevic:2001}. For an oriented nearest-neighbor bond \(i\to j\), we define
\begin{equation}
\mathcal U_{ij}
=
\mathcal P\exp\!\left[
\int_{x_i}^{x_j}dx^\mu\,\Omega_\mu(x)
\right],
\qquad
\Omega_\mu
=
\frac{1}{2}\omega_\mu{}^{ab}\Sigma_{ab}.
\end{equation}
For convenience, we adopt the notation

\begin{equation}
f^{\rm W}_{ij}
=
\int_{\gamma_{ij}} dx^\mu\,\omega_\mu(x),
\label{eq:exact-wilson-line}
\end{equation}
where \(\gamma_{ij}\) is the geodesic connecting \(z_i\) and \(z_j\). For the chosen Poincar\'e-disk vielbeins, see Eq.~\eqref{AppEq:Vielbein}, the spin connection is given by
\begin{equation}
\omega
=
\frac{2(y\,dx-x\,dy)}{l^2-|z|^2}
=
i\,\frac{\bar z\,dz-z\,d\bar z}{l^2-|z|^2}.
\end{equation}
The integral in Eq.~\eqref{eq:exact-wilson-line} can be evaluated exactly, following the App. \ref{App:ContinuumLimitDiscreteSpinConnection} derivation, as
\begin{align}
 f^{\rm W}_{ij}&=-i\log\left(\frac{l^2-z_i\bar{z_j}}{l^2-\bar{z_i}z_j}\right),
\label{eq:closed-form-geodesic-phase}\\
e^{if^{\rm W}_{ij}}
&=
\frac{l^2-z_i \bar z_j}
  {l^2-\bar z_i z_j}.
\label{eq:closed-form-geodesic-phase-exp}
\end{align}
Here \(\log\) denotes the principal logarithm, thus
\(f^{\rm W}_{ij}\) is taken on the principal branch. The numerical code evaluates Eq.~\eqref{eq:closed-form-geodesic-phase} directly, without path discretization or numerical quadrature.

For bonds short compared with the curvature radius the expansion,
\begin{equation}
\mathcal U_{ij}\Psi(x_j)
=
\Psi(x_i)
+
\delta x_{ij}^{\mu}
\bigl(\partial_\mu+\Omega_\mu\bigr)\Psi(x_i)
+
O(\delta x_{ij}^{2}),
\end{equation}
recovers the continuum covariant derivative
\begin{equation}
D_\mu
=
\partial_\mu
+
\frac{1}{2}\omega_\mu{}^{ab}\Sigma_{ab}.
\end{equation}
At finite bond length, the same construction also reproduces the Gauss--Bonnet holonomy, equal to $4\pi/3$ for the $\{10,3\}$, around every elementary plaquette exactly, as shown in App.~\ref{App:ContinuumLimitDiscreteSpinConnection}. The numerical results in Figs.~\ref{fig:DOS_spin_connection_generations}--\ref{fig:DeltaDOS_spin_connection_generations} use this Wilson-line implementation.
\endgroup
\begingroup
We evaluate both \(H_0\) and \(H_{\rm sc}\) with the KPM~\cite{KPM} for every system size reported. Figures~\ref{fig:DOS_spin_connection_generations}--\ref{fig:saturation-vs-N} and Table~\ref{tab:GC-floor} use outer generations $2-6$ (\(N=490,2880,16810,98000, 571210\)). All results shown here use \(M=4096\) Chebyshev moments, Jackson damping, 6001 energy points, and an adaptive rescaling fixed by the estimated spectral radius, cross-checked via Lanczos tridiagonalization. The stochastic trace uses \(R=150,72,36,12,12\) random-phase vectors for the five sizes, respectively.

The random vectors are normalized so that the DOS is per single-particle Hilbert-space state,
\begin{equation}
  \int dE\,\rho(E)=1.
\end{equation}
Thus, for the \(2N\)-dimensional spinor Hamiltonian, the DOS per lattice site would be \(2\rho_{\rm sc}\). The trace is taken over the complete open graph, including boundary sites. The values quoted as \(\rho(0)\) are therefore Jackson-broadened, fixed-\(M\) full-lattice values rather than exact delta-function DOS values.

Figure~\ref{fig:DOS_spin_connection_generations} compares the DOS with and without the spin connection, while Fig.~\ref{fig:DeltaDOS_spin_connection_generations} shows
\begin{equation}
  \delta\rho(E)=\rho_{\rm sc}(E)-\rho_0(E)
\end{equation}
demonstrating that the geodesic Wilson-line phases generate a robust low-energy enhancement. This supports the qualitative continuum statement that curvature can remove the vanishing Dirac DOS, but the coarse fixed \(\{10,3\}\) tessellation cannot  test the detailed form of the continuum low-energy DOS in Eq.~\eqref{eq:2D-DOS-correction}.

Table~\ref{tab:GC-floor} lists the values of zero-energy DOS for five sizes.  At \(M=4096\), the three largest systems yield \(\rho_{\rm sc}(0)=0.2846\)--\(0.2870\) and are mutually compatible at the level of their stochastic standard errors. The two smallest systems give the larger values \(0.822\pm0.039\) and \(0.509\pm0.023\), reflecting stronger finite-size and discrete-spectrum effects. The reference value remains at or below \((1.42\pm0.08)\times10^{-5}\) across all five sizes. \begingroup
Thus, at the fixed resolution $M=4096$, the reference DOS remains
strongly suppressed, whereas the spin-connection model retains
a robust finite low-energy spectral weight.
Independent spatially resolved checks
(App.~\ref{App:SpatialBoundaryCheck}) show that this enhancement is
not confined to the boundary. We emphasize that $H_{\rm ref}=H_0\otimes\mathbb{I}_2$ and
$H_{\rm sc}$ act on the same finite graph and the same $2N$-dimensional
Hilbert space; their only difference is the geodesic spin-connection
link factor. The observed enhancement therefore directly establishes the
spectral effect of this link structure within the chosen
tight-binding model. Its agreement with the continuum Dirac result is
qualitative: no \(M\to\infty\) extrapolation is attempted, and the
fixed \(\{10,3\}\) tessellation does not provide independently tunable
lattice spacing and curvature radius. It therefore does not permit a
controlled continuum extrapolation or a test of the continuum
curvature law \(\rho(0)\propto |R|^{1/2}\)
[Eq.~\eqref{eq:2D-DOS-zero-energy}]. Such a test would require a
genuine continuum-refinement sequence together with matched velocity,
area, and degeneracy normalizations, which we leave for the future.
\endgroup
\endgroup




\section{Conclusions and outlook}
\label{ConclusionsAndOutlook}

\begingroup
In this work, we have developed a symmetry-based framework for
incorporating spin-curvature coupling into lattice models on
hyperbolic crystals. Starting from the isometries of the
Poincar\'e disk and their action on continuum Dirac spinors, we
derived the corresponding lattice spin-connection factor and
implemented it as a geodesic Wilson line on nearest-neighbor bonds.
For massless continuum Dirac fermions, the covariant Dirac theory on hyperbolic space, which necessarily includes the spin connection, yields a finite zero-energy DOS scaling as
\(\rho(0)\sim |R|^{(D-1)/2}\) for spatial dimensions
\(2\leq D\leq4\). We also determined the explicit discrete
translational and rotational symmetries of the hyperbolic lattice,
which are fixed by the Schl\"afli symbols \(\{p,q\}\).
\endgroup

\par\begingroup
Building on these symmetries, we constructed a minimal two-component
tight-binding model that incorporates the geodesic Wilson-line factor
of the continuum spin connection. Using the KPM with
\(M=4096\) Chebyshev moments for all five system sizes, we find a
robust finite-resolution zero-energy DOS enhancement that remains
close to \(0.28\) from generations \(4\) through \(6\),
while the reference value without the spin-connection link factor
remains strongly suppressed ($\lesssim1.5\times10^{-5}$). Spatially resolved checks
(App.~\ref{App:SpatialBoundaryCheck}) confirm that the enhancement is
not boundary dominated. The resulting DOS enhancement provides
lattice-level evidence that spin-connection link phases can
qualitatively reproduce one of the central low-energy signatures
predicted by the continuum Dirac theory. Its magnitude and detailed
energy dependence remain lattice dependent, and the fixed
\(\{10,3\}\) tessellation cannot test the continuum
curvature law \(\rho(0)\propto |R|^{1/2}\)
[Eq.~\eqref{eq:2D-DOS-zero-energy}].
\endgroup Hyperbolic tilings with
\(p~({\rm mod}~4)=2\), \(p>6\), and \(q=3\) have been reported to
display Dirac-like nodal low-energy structure in the absence of the
spin-connection link factor~\cite{Roy_2024,Leong-Roy-PRB2026},
making this family a natural setting in which to test the present
construction. Within this family, however, one additional discrete selection rule applies: the elementary plaquette holonomy of a $\{p,3\}$ tessellation is $(p-6)\pi/3$, and whenever this is an exact multiple of $4\pi$ -- i.e., $p\equiv6\pmod{12}$, such as $p=18,30,\dots$ -- the SU(2) holonomy $\exp[i(p-6)\pi\sigma_z/6]$ is exactly the identity, so, on a simply connected open patch, the link field is gauge equivalent to the identity and has no spectral effect (App.~\ref{App:ContinuumLimitDiscreteSpinConnection}). The $\{10,3\}$ lattice used above is unaffected, since $10\equiv10\pmod{12}$. For instance, our results should motivate numerical studies of the lattice models of interacting Dirac fermions incorporating the discrete spin connection in quantum Monte Carlo simulations, as has been carried out for short-range interacting Dirac fermions in flat space~\cite{Sorella_1992,Assaad-Herbut-2013,Sorella-2016}. Our findings should further motivate a reconsideration of the corresponding field-theoretical description of such phase transitions of Dirac fermions in terms of Gross-Neveu-Yukawa quantum-critical theories in the curved space~\cite{Inagaki-review-1997}.

\par It is important to emphasize that the spin-curvature coupling introduced here for lattice models of Dirac fermions opens new avenues for the exploration of topological states in curved geometries. A particularly compelling direction for future work is the investigation of curvature-induced topological phase transitions—specifically, whether the spin-curvature coupling proposed in this study can give rise to emergent topological phases, analogous to the role played by spin-orbit coupling in conventional quantum materials~\cite{kane-mele-2005,BHZ-2006}. An additional set of challenges lies in characterizing the appropriate topological invariants in this context, such as determining the precise form of the Chern number in the presence of a spin connection, and analyzing the resultant topological phase transitions. We note that initial progress in this direction has recently been made for spinless quasiparticles on hyperbolic space~\cite{Sun-SciPost2024}; extending such analyses to systems with spin-curvature coupling represents an interesting open problem.

\par Our approach can be straightforwardly extended to the Poincaré ring representing a continuum embedding of time-constant slices of Bañados-Teitelboim-Zaneli (BTZ) black hole in $(2+1)D$ AdS space~\cite{BTZ}, which should reveal tessellations in this space and open the frontier studying the band structures therein. This may ultimately yield new insights into the effects of a black hole on the fermionic matter. In particular, such studies may also motivate simulations of BTZ black hole coupled to the fermions, and exploration of the effects of ensuing topological bands, interaction effects, etc.

\par Finally, we point out that the experimental realization of our theoretical proposal appears to be well within reach, given recent advances in the engineering of spinless tight-binding models on hyperbolic lattices. Notably, such models have been implemented in topolectric circuits~\cite{Lenggenhager2022,Zhang2023natcomm} and photonic resonator lattices~\cite{Huang2024-NatComm}. A particularly promising platform for realizing our proposal is silicon photonics based on coupled optical ring resonators. Hyperbolic photonic topological insulators have now been {experimentally} demonstrated on a \(\{6,4\}\) hyperbolic lattice using this architecture, with controllable connectivity and tunable complex hopping phases engineered via link rings~\cite{Huang2024-NatComm}.  This phase tunability is precisely what our scheme requires: the present two-dimensional spin connection is a diagonal \(\mathrm{Spin}(2)\simeq U(1)\) subgroup represented by opposite Abelian phases for the two spinor components. More general \(\mathrm{SU}(2)\) gauge structures for light have already been realized in coupled-resonator platforms~\cite{Cheng2023_NonAbelianSU2Photons,Yang2024_NonAbelianLightSound}, so the simpler spin-dependent link structure required here is within reach of current photonic technology.

\section*{Data availability}
The data and software code for generating the figures presented in the main text and appendices are available at \cite{GitHub-Code}.

\section*{Acknowledgments}
We would like to thank Stefan Đorđevi\'{c}, Ilija Buri\'{c} and Bitan Roy for fruitful discussions and useful comments. A.Đ. and M.D.\'{C} acknowledge the funding provided by the Faculty of Physics, University of Belgrade, through the grant
number 451-03-136/2025-03/200162 by the Ministry of
Science, Technological Development and Innovations of
the Republic of Serbia. A.Đ. acknowledges the support by the Science Fund of the Republic of Serbia, grant number TF C1389-YF, Towards a Holographic Description
of Noncommutative Spacetime: Insights from ChernSimons Gravity, Black Holes and Quantum Information
Theory - HINT. V. J. acknowledges the support by the Swedish Research Council Grant No. VR 2019-04735 and  Fondecyt (Chile) Grant No. 1230933.

\appendix
\setcounter{figure}{0}
\renewcommand{\thefigure}{A\arabic{figure}}
\section*{Appendix}

\begingroup
\section{Details of the principal-series representations}
\label{app:principal-series-details}
For completeness, we summarize the standard principal-series
representations of ${\rm PSL}(2,\mathbb R)$ and its covers in the
normalization used in the main text~\cite{bargmann1947,kitaev2018}.
In particular, the eigenvalues of the compact generator form the
unbroken sequence $m=\mu+n$, with $n\in\mathbb Z$.
The double covers ${\rm SU}(1,1)\simeq{\rm SL}(2,\mathbb R)$ act naturally on spinors, while the universal cover is denoted by $\widetilde{{\rm SU}(1,1)}$. To cast the Killing-vector algebra in dimensionless form, we define
\begin{equation}
J_0=-i\xi_2,
\qquad
J_1=-\frac{i}{2l}\xi_1,
\qquad
J_2=-\frac{i}{2l}\xi_3.
\end{equation}
Equations~\eqref{eq:Killing_algebra1}--\eqref{eq:Killing_algebra3} then imply
\begin{equation}
[J_0,J_1]=iJ_2,
\qquad
[J_0,J_2]=-iJ_1,
\qquad
[J_1,J_2]=-iJ_0,
\end{equation}
which is the ${\rm su}(1,1)$ algebra. Introducing $J_\pm=J_1\pm iJ_2$ gives
\begin{align}
\label{KillingVsAdjoint_operators}
[J_0,J_\pm]&=\pm J_\pm,\\
[J_+,J_-]&=-2J_0.
\end{align}

It is useful to distinguish the standard algebraic Casimir from the positive scalar Casimir used in the main text:
\begin{align}
\xi^2_{\rm alg}&=J_0^2-J_1^2-J_2^2=-\xi^2, \nonumber\\
\xi^2
&=\xi_2^2-\frac{\xi_1^2+\xi_3^2}{4l^2}
=-\frac{l^2}{4}\Delta_g.
\end{align}
We choose simultaneous eigenstates
\begin{align}
\label{Kasimir&Kiling_eigenbasis1}
J_0\ket{q,m}&=m\ket{q,m},\\
\label{Kasimir&Kiling_eigenbasis2}
\xi^2\ket{q,m}&=q\ket{q,m},
\end{align}
with real $q$. The ladder operators act as
\begin{align}
\label{eq:eta+}
J_+\ket{q,m}
&=\sqrt{m(m+1)+q}\,\ket{q,m+1},\\
\label{eq:eta-}
J_-\ket{q,m}
&=\sqrt{m(m-1)+q}\,\ket{q,m-1}.
\end{align}
For the principal series, $q=1/4+s^2$, and the two squared ladder
amplitudes reduce to
\[
m(m\mp1)+q=\left(m\mp\frac12\right)^2+s^2.
\]
They are therefore positive for generic $s$, so the ladder operators
connect the complete set $m\in\mu+\mathbb Z$, without a forbidden
interval in $m$~\cite{bargmann1947,kitaev2018}. The special endpoint
$s=0$, $\mu=1/2$ corresponds to the reducible limit-of-discrete-series
case. The unitary dual also contains the discrete and complementary series, which are not required for the continuum spectral resolution used in the main text. Nontrivial finite-dimensional representations of ${\rm SU}(1,1)$ are nonunitary.

In coordinate space, the Casimir eigenvalue problem reads
\begin{equation}
-\frac{\hbar^2}{2m^*}\Delta_g\Phi_{q,m}(r,\varphi)
=
\frac{4q\hbar^2}{2m^*l^2}\Phi_{q,m}(r,\varphi),
\end{equation}
with  $\Phi_{q,m}(r,\varphi)=\braket{r,\varphi}{q,m}$ as the coordinate representation of the symmetry adapted basis. Consequently, we can interpret $\frac{4q\hbar^2}{2m^{*}l^2}$ as energy, where $m^{*}$ is the effective mass. Moreover, we can write the eigenvalue $q$ as a function of the momentum eigenvalue $k$, $4q=(kl)^2$.
By simultaneously solving Eqs.~(\ref{Kasimir&Kiling_eigenbasis1}) and (\ref{Kasimir&Kiling_eigenbasis2}) in the real space, we obtain the following form of the corresponding eigenfunctions (see App.~\ref{app:Dirac-curved-space} for details)
\begin{widetext}
\begin{equation}
\label{eq:principal-eigenstates}
\Phi_{q,m}(r,\varphi)
=
\frac{e^{im\varphi}}{\sqrt{2\pi}}
\left(1-\frac{r^2}{l^2}\right)^\lambda
\left[
A_m(k)r^m
{}_2F_1\!\left(\lambda,\lambda+m;m+1;\frac{r^2}{l^2}\right)
+
B_m(k)r^{-m}
{}_2F_1\!\left(\lambda,\lambda-m;1-m;\frac{r^2}{l^2}\right)
\right].
\end{equation}
\end{widetext}
Here ${}_2F_1$ is the hypergeometric function, $4q=(kl)^2$, and
\begin{equation}
\lambda=\frac12+is,
\quad
q=\lambda(1-\lambda)=\frac14+s^2,
\quad
s=\frac12\sqrt{(kl)^2-1}.
\end{equation}
The coefficients $A_m(k)$ and $B_m(k)$ are fixed by normalization. These principal-series eigenfunctions provide the Plancherel spectral resolution of square-integrable functions on the Poincar\'e disk~\cite{kitaev2018,bargmann1947} and vanish as $r\to l$.

A discrete subgroup of M\"obius transformations acting properly discontinuously on the disk is a Fuchsian group~\cite{ActaMath:1993}. This structure underlies the lattice discretization in Sec.~\ref{PointGroupSymmetry} and the spin-connection construction in Sec.~\ref{LatticeDiracFermions}.
\endgroup

\section{Coordinate transformation and geodesics}
\label{AppendixA}

In this section, we will explicitly derive Killing vectors, coordinate transformations and also give explicit calculation of equation of motion for a free particle.
\par To find an explicit form of Eq.~\eqref{eq:Killing_equation} for the Killing vectors, we first obtain  the Christoffel symbols
\begin{align}
  \Gamma_{xx}^y&=-\frac{1}{l^2}\frac{2y}{1-\frac{x^2+y^2}{l^2}},\label{app-eq-conn-1} \\
  \Gamma_{yy}^x&=-\frac{1}{l^2}\frac{2x}{1-\frac{x^2+y^2}{l^2}}, \label{app-eq-conn-2}\\
  \Gamma_{xy}^y&=\frac{1}{l^2}\frac{2x}{1-\frac{x^2+y^2}{l^2}},\label{app-eq-conn-3} \\
  \Gamma_{yx}^x&=\frac{1}{l^2}\frac{2y}{1-\frac{x^2+y^2}{l^2}}.
  \label{app-eq-conn-4}
\end{align}
Eq. \eqref{eq:Killing_equation} then explicitly reads as
\begin{align}
\label{CR-1}
  \partial_x \xi_x=&\frac{1}{l^2}\frac{2x}{1-\frac{x^2+y^2}{l^2}}\xi_x-\frac{1}{l^2}\frac{2y}{1-\frac{x^2+y^2}{l^2}}\xi_y ,\\
  \label{CR-2}
  \partial_y \xi_y=& -\frac{1}{l^2}\frac{2x}{1-\frac{x^2+y^2}{l^2}}\xi_x+\frac{1}{l^2}\frac{2y}{1-\frac{x^2+y^2}{l^2}}\xi_y, \\
\partial_x\xi_y+\partial_y\xi_x&=\frac{1}{l^2}\frac{4y}{1-\frac{x^2+y^2}{l^2}}\xi_x+\frac{1}{l^2}\frac{4x}{1-\frac{x^2+y^2}{l^2}}\xi_y.
\end{align}
By raising the indices of the components of Killing vector and summing Eq. \eqref{CR-1} and Eq. \eqref{CR-2}  we arrive at the Cauchy-Riemann conditions (Eqs. \eqref{eq:1} and \eqref{eq:2}). Conformal Killing equation in two dimensions also turns to Cauchy-Riemann conditions, so further steps are analogous to that of solving conformal Killing equation. As mentioned in the main text, we can write symmetry transformations as a form of the analytic (Eq. \eqref{eq:analytic_killing}) and anti-analytic function (Eq. \eqref{eq:antianalytic_killing}). When applying this ansatz in Eq. \eqref{CR-1}, we get conditions on parameters as
\begin{align}
  a_0+\bar{a}_0&=0, \\
  a_1+\bar{a}_{-1}&=0.
\end{align}
We conclude that the parameter $a_0$ is purely imaginary and can be written as $a_0=i\theta$, $\theta\in \mathbb{R}$. Furthermore, we write parameter $a_{-1}=-a$, so parameter $a_1$ becomes $a_1=\bar{a}$. These parameters will be used when finding the coordinate transformation generated by the Killing vectors. It is straightforward to derive vectors given by Eqs. \eqref{eq:Killing_vector1}, \eqref{eq:Killing_vector2} and \eqref{eq:Killing_vector3}.
\par We now proceed to determine the coordinate transformations. We consider the transformation of $z=x+iy$ coordinate on the disk, as for $\Bar{z}=x-iy$ coordinate can be done similarly. Starting from the algebra of Killing vectors, we construct the group element by exponentiation, as given by Eq. \eqref{eq:coordinate_transformation}. Since the exponential function is analytical, we can expand the expression as a Taylor series to obtain,
\begin{equation}
  z'=\frac{z\frac{z_2-z_1e^{\sqrt{-\theta^2+4|a|^2}}}{z_2e^{\sqrt{-\theta^2+4|a|^2}}-z_1}-z_1z_2\frac{1-e^{\sqrt{-\theta^2+4|a|^2}}}{{z_2 e^{\sqrt{-\theta^2+4|a|^2}}-z_1}}}{{z\frac{1-e^{\sqrt{-\theta^2+4|a|^2}}}{z_2 e^{\sqrt{-\theta^2+4|a|^2}}-z_1}}+1},
\end{equation}
where $z_1$ and $z_2$ read
\begin{align}
\label{eq:z1}
  z_1&=\frac{-i\theta-\sqrt{-\theta^2+4|a|^2}}{2\Bar{a}},\\
  \label{eq:z2}
  z_2&=\frac{-i\theta+\sqrt{-\theta^2+4|a|^2}}{2\Bar{a}}.
\end{align}
To obtain the M\"{o}bius transformation (Eq. \eqref{eq:Mobius_transformation}) as the isometry of the Poincar\'e disk, we will redefine our parameters as
\begin{align}
  e^{i\tilde{\theta}}&= \frac{z_2-z_1e^{\sqrt{-\theta^2+4|a|^2}}}{z_2e^{\sqrt{-\theta^2+4|a|^2}}-z_1},\label{eqapp:mob-1}\\
  c&=z_1z_2\frac{1-e^{\sqrt{-\theta^2+4|a|^2}}}{z_2-z_1e^{\sqrt{-\theta^2+4|a|^2}}}.
  \label{eqapp:mob-2}
\end{align}
\par We derived the M\"{o}bius transformation using the fact $z_1\neq z_2$. We now turn to the case $z_1=z_2$, which implies $\theta^2=4|a|^2$, with the M\"{o}bius parameters given as
\begin{align}
  &e^{i\tilde{\theta}}=\frac{2+i\theta}{2-i\theta}, \\
  &c=\frac{\theta^2}{2\bar{a}(i\theta+2)}.
\end{align}
\par Let us now find the geodesic equation from conserved quantities together with the corresponding Killing vectors (Eq.~\eqref{conserved_killing}). Accordingly, we assign the constant $C$ from Eq.~(\ref{conserved_killing}) three distinct values, denoted $C_1$, $C_2$, and $C_3$, each corresponding to the equations associated with the Killing vectors $\xi_1$, $\xi_2$, and $\xi_3$, respectively. The explicit forms of the conserved quantities and the equations for the Killing vectors are given by
\begin{align}
\label{c1}
  C_1=&\frac{1}{(1-x^2-y^2)^2}\left[\left(\frac{\de x}{\de\tau}\right)^2+\left(\frac{\de y}{\de\tau}\right)^2\right] \\
  \label{c2}
  C_2=& \frac{\de x}{\de\tau} \frac{1}{(1-x^2-y^2)^2}\left(-y+x\frac{\de y}{\de x}\right)\\
  \label{c3}
  C_3=& \frac{\de x}{\de \tau}\frac{1}{(1-x^2-y^2)^2} \left(2xy + \frac{\de y}{\de x}\left(-1-x^2+y^2\right)\right) \\
  \label{c4}
  C_4=& \left(\frac{\de x}{\de\tau}\right)^2\frac{1}{(1-x^2-y^2)^2}\left(1+\left(\frac{\de y}{\de x}\right)^2\right) .
\end{align}
We introduce another constant $D_1^2=\frac{C_4}{C_2^2}$ to get a solvable form,
\begin{equation}
  x\frac{dy}{dx}-y=\frac{1}{D_1}(1-x^2-y^2)\sqrt{1+\left(\frac{\de y}{\de x}\right)^2}.
\end{equation}
Using this in the geodesic equation,
\begin{equation}
\label{geodesic_equation}
  \frac{\de^2x^{\mu}}{\de\tau^2}+\Gamma_{\rho\sigma}^{\mu}\frac{\de x^{\rho}}{\de\tau}\frac{\de x^{\sigma}}{\de\tau}=0,
\end{equation}
 in the limit $D_1\rightarrow +\infty$ we obtain geodesics given by the Eq. (\ref{geodesic1}). For an arbitrary constant $D_1$ we obtain geodesics given by the Eq. (\ref{geodesic2}).

\section{Point group discretization}
\label{AppendixB}

Given a space-time manifold endowed with symmetry described by a Lie group, the discretization of the background via lattice construction necessitates a systematic procedure for connecting  the continuous symmetries of the underlying geometry with the discrete symmetries inherent to the lattice. This correspondence imposes specific constraints on the parameters characterizing the Lie group.

In this appendix, we provide a detailed derivation of the results presented in Sec.~\ref{PointGroupSymmetry}. The transformation specified in Eq.~\eqref{eq:Mobius_transformation} defines  the action of a group element on a point within the disk, and thus serves as a representative element of the discrete symmetry group, contingent upon the imposition of appropriate conditions. As established in point group theory, repeated application of a group element yields the identity transformation. Here, we focus on the construction of rotational analogues, paralleling those found in flat space. While point group symmetries in Euclidean space can also encompass reflections, the presence of reflection symmetry in the hyperbolic setting is highly sensitive to the choice of Schl\"{a}fli symbols $\{p, q\}$. Since our aim is to accommodate arbitrary $\{p, q\}$ tessellations, our analysis is restricted to rotational symmetries at this stage.

Let $\theta_p$ denote the angle associated with $p$ successive applications of the transformation in Eq.~\eqref{Mtrans_p}, and let $\tilde{\theta}$ correspond to the angle for a single application, as in Eq.~\eqref{Mtrans_1}. We introduce the notation
\begin{equation}
  a_p = e^{i\theta_p}, \qquad a = e^{i\tilde{\theta}},
\end{equation}
and  then obtain a recurrence relation given by the Eq. \eqref{theta_p}, which can be rewritten as
\begin{equation}
  a_p=a\frac{a_{p-1}+c\Bar{c}_{p-1}}{1+\Bar{c}c_{p-1}a_{p-1}}.
\end{equation}
The same can be done with the equations for $c_p$. Now we define a new series as
\begin{equation}
  b_p\equiv a_pc_p,
\end{equation}
with the condition $b_1=ac$. These series can be also written as a recurrence relation,
\begin{equation}
  b_p=a\frac{b_{p-1}+c}{1+\Bar{c}b_{p-1}}.
\end{equation}
In order to solve this recurrence relation, we define another two set of series as $b_p=\frac{m_p}{n_p}$, so the relation becomes:
\begin{equation}
  \frac{m_p}{n_p}=a\frac{m_{p-1}+cn_{p-1}}{n_{p-1}+\bar{c}m_{p-1}},
\end{equation}
where we can equate separately the numerator and the denominator from the left and right-hand side of the equation. This yields the difference equation
\begin{equation}
  m_{p+1}-(a+1)m_p+a(1-|c|^2)m_{p-1}=0,
\end{equation}
which can be solved by introducing the parameter $\lambda$ as
\begin{equation}
  \lambda^2-(a+1)\lambda+a(1-|c|^2)=0.
\end{equation}
The solution to this quadratic equation can be readily found
\begin{equation}
  \lambda_{1/2}=\frac{a+1\pm\sqrt{(a-1)^2+4a|c|^2}}{2}.
\end{equation}
Now the general solution for the elements of  two defined series':
\begin{align}
  m_p&=A\lambda_1^p+B\lambda_2^p, \\
  n_p&=\frac{1}{ac}\left[A\lambda_1^p(\lambda_1-a)+B\lambda_2^p(\lambda_2-a)\right],
\end{align}
or the previously defined series' element:
\begin{equation}
  b_p=ac\frac{\lambda_1^p-\lambda_2^p}{(\lambda_1-a)\lambda_1^p-(\lambda_2-a)\lambda_2^p},
\end{equation}
where $A$ and $B$ are arbitrary constants.
\par Since the $p$-time applied transformation is equivalent to the unit element of the group, the element of the series becomes zero, i.e. $b_p=0$. This implies that $\lambda_1^p=\lambda_2^p$, and we obtain
\begin{equation}
  \left(\frac{\lambda_1}{\lambda_2}\right)^p=1 \Longleftrightarrow \frac{\lambda_1}{\lambda_2}=e^{i\frac{2n\pi}{p}}, \quad n\in\mathbb{Z},
\end{equation}
which implies
\begin{align}
\label{b12}
  \lambda_1^2& =\lambda_1\lambda_2e^{i\frac{2n\pi}{p}}=a(1-|c|^2)e^{i\frac{2n\pi}{p}}, \\
  \label{b13}
  \lambda_1^2+\lambda_2^2&=2a(1-|c|^2)\cos\left(\frac{2n\pi}{p}\right).
\end{align}
By further transforming Eqs. \eqref{b12} and \eqref{b13}, the condition on Lie group parameters given by Eq. \eqref{condition_on_parameters} can be readily obtained.

\section{Noncommutativity of translations}
\label{AppendixC}
As discussed in Sec.~\ref{translations_section}, the translation operators in curved space-time generally do not commute, which  distinguishes them from their flat space counterparts. In this section, we derive the explicit conditions under which the translation operators commute.

To elucidate this, consider the standard quantization procedure analogous to that leading to Bloch's theorem. We begin by introducing a basis for the Hilbert space $\mathcal{H}$, denoted by $\{\ket{z} \mid z \in \text{solution}\}$. The translation operators are then defined as
\begin{equation}
  T_m \overset{\mathrm{def.}}{=} \sum_z \ket{z}\bra{t_m z}.
\end{equation}
Assuming the flat-space approach remains valid, the corresponding eigenvalue problem for these operators can be formulated as
\begin{equation}
  T_m \ket{\lambda_1, \ldots, \lambda_{p/2}} = e^{i\lambda_m} \ket{\lambda_1, \ldots, \lambda_{p/2}},
\end{equation}
where the eigenvectors can be expressed as linear combinations of the basis vectors in  Hilbert space,
\begin{equation}
  \ket{\lambda_1,...,\lambda_{p/2}} =\sum_z a_{\lambda_1...\lambda_{p/2}}(z)\ket{z}.
\end{equation}
Our definition of the translation operator leads to the following equation:
\begin{equation}
  T_m\ket{z}=\ket{t_m^{-1}z}.
\end{equation}
The condition on the eigenvalue $\lambda_m$ is given by:
\begin{equation}
  a_{\lambda_1...\lambda_{p/2}}(t_mz)=e^{i\lambda_m} a_{\lambda_1...\lambda_{p/2}}(z).
\end{equation}
Up to this point, our analysis has not encountered any specific obstacles arising from the curvature of the underlying space. However, in order to determine the eigenvalues $\lambda_m$, it is necessary to derive the commutation relations between two arbitrary translation operators:
\begin{align}
  [T_n, T_m] &= \left[\sum_z \ket{z}\bra{t_n z}, \sum_{z'} \ket{z'}\bra{t_m z'}\right] \\
  &= \sum_z \left(\ket{z}\bra{t_m t_n z} - \ket{z}\bra{t_n t_m z}\right).
\end{align}
We therefore compute  the commutator of two translation transformations acting on a coordinate $z$:
\begin{equation}
  [t_n, t_m]z = t_n t_m z - t_m t_n z,
\end{equation}
where $t_{n}$ denote the transformations defined in Eq.~\eqref{eq:traslation_transformation}. This expression can be recast as a rational function, $[t_n, t_m]z = \frac{G}{D}$. Commutativity of the translation transformations is achieved if $G=0$ and $D \neq 0$, with  explicit forms of $G$ and $D$  given by
\begin{widetext}
\begin{align}
  G = 4 \Big\{ & x^3 \cos\frac{\pi}{p}(n-m) \left[\sin\frac{2\pi}{p}(n-m) \sin\frac{\pi}{p}(n+m)(z^2-1) + i \cos\frac{\pi}{p}(n+m)\sin\frac{2\pi}{p}(n-m)(z^2+1)\right] \\
  & + i z x^2 \sin\frac{2\pi}{p}(n-m)\left(1 + x^2 \cos\frac{2\pi}{p}(n-m)\right) \Big\}, \\
  D = \Big( 1 &+ 2 z x e^{-i\frac{\pi}{p}(n+m)} \cos\frac{\pi}{p}(n-m) \Big)^2 + x^2 \left[x^2 + 2 \cos\frac{2\pi}{p}(n-m)\left(1 + 2 z x e^{i\frac{\pi}{p}(n+m)}\right)\right].
\end{align}
\end{widetext}

From these results, it follows that the condition for commutativity of two translation operators reads
\begin{equation}
  \label{commutativity_condition}
  n - m = \frac{p k}{2}, \quad n, m \in \{0, \ldots, p-1\},\ k \in \mathbb{Z}.
\end{equation}
This condition is satisfied provided that  $n = m$ and  $t_n$ and $t_m$ are inverse to each other. Commutativity of translation operators, implying one-dimensional irreducible representations of pure translations, is essential for the form of the  Bloch's theorem in flat space. For further details on the generalization of the Bloch's theorem to hyperbolic crystals, see Refs.~\cite{CrystallographyOfHyperbolicLattices, Maciejko2022}.
\section{Dirac equation in hyperbolic space}
\label{app:Dirac-curved-space}
Here we will analyze spinor fields on expanded Poincar\'{e} disk defined by the metric in Eq. \eqref{expanded_metric}. We recall the Dirac equation \eqref{Dirac_equation0} and choose local $\gamma$ matrices as Pauli matrices:
\begin{align}
\label{gamma_matrix_z}
  \gamma^0 &=i\sigma_z = \begin{pmatrix}
  i & 0 \\
  0 & -i
  \end{pmatrix},\\
  \label{gamma_matrix_x}
  \gamma^1 &=\sigma_x = \begin{pmatrix}
  0 & 1 \\
  1 & 0
  \end{pmatrix},\\
  \label{gamma_matrix_y}
  \gamma^2 &=\sigma_y = \begin{pmatrix}
  0 & -i \\
  i & 0
  \end{pmatrix},
\end{align}
with $\{\gamma_a,\gamma_b\}=2\eta_{ab}$.
\begingroup
With the Minkowski metric \(\eta_{ab}={\rm diag}(-1,1,1)\), lowering the
time-like index gives \(\gamma_0=-\gamma^0\), while
\(\gamma_1=\gamma^1\) and \(\gamma_2=\gamma^2\). Therefore, for
\(\Sigma_{ab}=\frac14[\gamma_a,\gamma_b]\),
\begin{align}
  \Sigma_{01}
  &=\frac14[\gamma_0,\gamma_1]
  =\frac12\gamma_2,\\
  \Sigma_{02}
  &=\frac14[\gamma_0,\gamma_2]
  =-\frac12\gamma_1,\\
  \Sigma_{12}
  &=\frac14[\gamma_1,\gamma_2]
  =-\frac12\gamma_0
  =\frac12\gamma^0.
\end{align}
\endgroup
Next, we introduce vielbeins to satisfy $G_{\mu\nu}=e_{\mu}{}^ae_{\nu}{}^b\eta_{ab}$, where $\eta_{ab}$ is metric of the Minkowski spacetime. Thus, vielbeins are not uniquely defined. With that in mind, we  choose them in the form
\begin{equation}
\label{AppEq:Vielbein}
  e_t{}^0=1, \quad
  e_x{}^1=e_y{}^2=\frac{1}{1-\frac{x^2+y^2}{l^2}}.
\end{equation}
Now, the components of the spin connection read \cite{MBlagojevic:2001}:
\begin{equation}
  \omega_{\mu}{}^a{}_b=e_{\nu}{}^ae^{\lambda}{}_b\Gamma_{\mu\lambda}^{\nu}-e^{\lambda}{}_b\partial_{\mu}e_{\lambda}{}^a.
\end{equation}
For our choice of vielbeins, we obtain the following non-zero spin connection coefficients,
\begin{align}
  \omega_x{}^{1}{}_{2}&=\frac{2y}{l^2}\frac{1}{1-\frac{x^2+y^2}{l^2}}, \\
  \omega_y{}^{1}{}_{2}&=-\frac{2x}{l^2}\frac{1}{1-\frac{x^2+y^2}{l^2}}.
\end{align}
\par The Dirac equation (\ref{Dirac_equation0}) can now be rewritten as,
\begin{widetext}
  \begin{equation}
\label{Dirac_equation}
  [i\sigma_z\partial_t+\left(1-\frac{x^2+y^2}{l^2}\right)(\sigma_x\partial_x+\sigma_y\partial_y)+\frac{1}{l^2}(y\sigma_y+x\sigma_x)-m]\Psi=0.
\end{equation}
\end{widetext}

We change coordinates corresponding to the metric of Ref.~\cite{HyperbolicEigenFunction_Camporesi} which leads to the form of the Dirac equation therein. To this end, we use the form of the Poincar\'{e} disk metric in polar coordinates,
\begin{equation}
\mathrm{d}s^2=\frac{\mathrm{d}r^2+r^2\mathrm{d}\varphi^2}{\left(1-\frac{r^2}{l^2}\right)^2}
\end{equation}
and then introduce the coordinate transformation as $r=l\tanh(\chi/2)$ and $\varphi=\eta$. Under this change of variables, the metric becomes
\begin{equation}
  \mathrm{d}s^2=\frac{l^2}{4}\left(\mathrm{d}\chi^2+\sinh^2{\chi}\,\mathrm{d}\eta^2\right).
\end{equation}
Hence, the Poincaré disk metric is equivalent to the standard hyperbolic metric up to an overall multiplicative factor of $\frac{l^2}{4}$, which rescales the curvature while preserving its constant negative character. Using this result, it can be shown that the form of the Green's function in Eq.~\eqref{greens_function} can be obtained from the eigenstates of the Dirac operator, given by Eq.~\eqref{eq:principal-eigenstates}, as demonstrated in Ref.~\cite{HyperbolicEigenFunction_Camporesi}.

\par Our goal now is to find the stationary solution of Eq. \eqref{Dirac_equation} and the corresponding wavefunctions for fermions. An ansatz for stationary solution of the Dirac equation (\ref{Dirac_equation}) can be written as,
\begin{equation}
\label{ansatz}
  \Psi(t,x,y)=\begin{pmatrix}
  \phi(x,y) \\ \chi(x,y)
  \end{pmatrix} e^{-iEt},
\end{equation}
where $E$ is energy-like quantity. Implementing the ansatz to Eq. \eqref{Dirac_equation}, we obtain
\begin{widetext}
\begin{align}
  E\phi+\left(1-\frac{x^2+y^2}{l^2}\right)(\partial_x-i\partial_y)\chi+\frac{1}{l^2}(x-iy)\chi-m\phi &=0, \\
  -E\chi+\left(1-\frac{x^2+y^2}{l^2}\right)(\partial_x+i\partial_y)\phi+\frac{1}{l^2}(x+iy)\phi-m\chi &=0.
\end{align}
\end{widetext}
It is more convenient to solve these equations in polar coordinates with $\phi=R(r)F(\varphi)$ and use symmetry based solution for $F(\varphi)=e^{in\varphi}$ to  find
\begin{widetext}
  \begin{equation}
  \label{R_Dirac_eq}
  R''(r)+\frac{1}{r}R'(r)+\left[\frac{E^2-m^2+\frac{1}{l^2}}{\left(1-\frac{r^2}{l^2}\right)^2}+\frac{2n+1}{l^2\left(1-\frac{r^2}{l^2}\right)}-\frac{n^2}{r^2}\right]R(r)=0.
\end{equation}
\end{widetext}
The dimensional analysis implies  that the third term in Eq. \eqref{R_Dirac_eq} corresponds to a momentum-like quantity $\frac{l}{\hbar}\sqrt{\frac{E^2}{c^2}-m^2c^2}=\frac{l}{\hbar c}\sqrt{p^2c^2}=lk$, where $p=\hbar k$.

\par Now, we can write the solution for the free massive fermion on the Poincar\'e disk:
\begin{widetext}
  \begin{align}
  \psi_{n,k}^{<}(r,\varphi)&=A_n(k)e^{in\varphi}r^{-n}\begin{pmatrix}
  \left(1-\frac{r^2}{l^2}\right)^{\frac{1}{2}(1-ilk)}\,_2F_1\left(-\frac{i}{2}lk+1,-\frac{i}{2}lk-n;1-n;\frac{r^2}{l^2}\right) \\
  \left(1-\frac{r^2}{l^2}\right)^{\frac{1}{2}(1+ikl)}\,_2F_1\left(\frac{i}{2}kl,\frac{i}{2}kl-n+1;1-n;\frac{r^2}{l^2}\right)
  \end{pmatrix}, n<0 \\
  \psi_{n,k}^{\ge}(r,\varphi)&= B_n(k)e^{in\varphi}r^{n}\begin{pmatrix}
  \left(1-\frac{r^2}{l^2}\right)^{\frac{1}{2}(1-ilk)}\,_2F_1\left(-\frac{i}{2}lk,-\frac{i}{2}lk+n+1;1+n;\frac{r^2}{l^2}\right) \\
  \left(1-\frac{r^2}{l^2}\right)^{\frac{1}{2}(1+ikl)}\,_2F_1\left(\frac{i}{2}kl+1,\frac{i}{2}kl+n;n+1;\frac{r^2}{l^2}\right)
  \end{pmatrix}, n\ge 0
  \end{align}
\end{widetext}
where $A_n(k)$ and $B_n(k)$ are normalization constants dependent on the curvature radius $l$ and the quantum numbers $n$ and $k$.When employing a Euclidean time coordinate, the appearance of the imaginary unit in terms such as $kl$ is avoided.

\par Eq. \eqref{R_Dirac_eq} corresponds to the first component of the eigenvalue problem for the Casimir operator in the coordinate representation, as found in Sec. \ref{section_spinors}. Similar steps can be performed for the second component. Consequently, the obtained spinor components denote the symmetry-adapted basis.

Notice that as we approach  the edge of the Poincar\'{e} disk, $r\rightarrow l$, these wave functions approach zero. A general solution of the  Dirac equation can be written in the symmetry-adapted basis as
\begin{align}
\label{eq.app:result_Dirac}
  \Psi(t,r,\varphi)=\int \de k\sum_n \bigg{(}&\alpha_n(k)\psi_{n,k}e^{-i E_kt} \nonumber\\
  &+\beta_n(k)\psi_{n,k}^{*}e^{i E_kt}\bigg{)},
\end{align}
where $\alpha_n(k)$ and $\beta_n(k)$ are the functions of the momentum-like quantity and the angular momentum.

We also note that, in curved space, casting the Dirac equation as a Hermitian Hamiltonian eigenvalue problem requires some care, since the spin connection contributes nontrivial terms that may, in principle, spoil Hermiticity. In the present case, however, the Hamiltonian obtained from Eq.~\eqref{Dirac_equation} is Hermitian, as we show below. Multiplying the Dirac equation by \(\sigma_z\), and writing \(p_j=-i\partial_j\), \(j=x,y\), we obtain
\begin{widetext}
\begin{equation}
H\Psi =
\left[
\left(1-\frac{x^2+y^2}{l^2}\right)(\sigma_y p_x-\sigma_x p_y)
+\frac{i}{l^2}(x\sigma_y-y\sigma_x)
+m\sigma_z
\right]\Psi
=E\Psi .
\end{equation}
\end{widetext}
Let \(f(x,y)=1-(x^2+y^2)/l^2\) and \(A=\sigma_y p_x-\sigma_x p_y\). Then
\[
H=fA+\frac{i}{l^2}(x\sigma_y-y\sigma_x)+m\sigma_z .
\]
Its Hermitian conjugate is
\[
H^\dagger=A f-\frac{i}{l^2}(x\sigma_y-y\sigma_x)+m\sigma_z .
\]
Using
\[
[A,f]=\frac{2i}{l^2}(x\sigma_y-y\sigma_x),
\]
we find
\[
\begin{aligned}
H^\dagger
&=fA+[A,f]-\frac{i}{l^2}(x\sigma_y-y\sigma_x)+m\sigma_z \\
&=fA+\frac{i}{l^2}(x\sigma_y-y\sigma_x)+m\sigma_z
=H,
\end{aligned}
\]
which demonstrates  Hermiticity of the Hamiltonian.

\subsection{Dirac equation in flat space}
\label{AppendixE}

\par For completeness, we here consider the flat-space limit of the Dirac equation  in the hyperbolic space.  Since  the scalar curvature of the expanded Poincar\'{e} disk is given by
\begin{equation}
\label{scalar_curvature}
  R=-\frac{8}{l^2}.
\end{equation}
 the curvature vanishes in the limit $l\rightarrow \infty$. Therefore, we can use this to find the flat space form  of our equations and compare our results (Eq. \eqref{eq.app:result_Dirac}) with well-known results in flat space.
\par First, let us consider the metric \eqref{expanded_metric} we introduced in Sec. \ref{section_spinors}:
\begin{equation}
  \de s^2=-c^2\de t^2+\frac{\de x^2+\de y^2}{\left(1-\frac{x^2+y^2}
  {l^2}\right)^2},
\end{equation}
when $l\rightarrow \infty$, we obtain  the ordinary three dimensional Minkowski space.
\par The Dirac equation \eqref{Dirac_equation}, in the limit $l\rightarrow+\infty$ can be written in the form
\begin{equation}
  [i\sigma_z\partial_t+(\sigma_x\partial_x+\sigma_y\partial_y)-m]\Psi=0.
\end{equation}
With the choice of $\gamma$ matrices [Eqs.~\eqref{gamma_matrix_z}-\eqref{gamma_matrix_y}], we find
\begin{equation}
  (\gamma^{\mu}\partial_{\mu}-m)\Psi=0,
\end{equation}
which represents the Dirac equation in flat space.
Let us now solve it by employing the same stationary ansatz \eqref{ansatz} yielding
\begin{align}
  (E+m)\phi+(\partial_x-i\partial_y)\chi&=0, \\
  -(E-m)\chi+(\partial_x+i\partial_y)\phi&=0.
\end{align}
Separation of the variables for $\phi=R(r)F(\varphi)$ yields  the solution for angular part as $F(\varphi)=e^{in\varphi}$, while the solution for function $R(r)$ is given by the Bessel function $J_n(kr)$, where $k=\sqrt{E^2-m^2}$,. Therefore, the solution for $\phi$ reads
\begin{equation}
  \phi(r,\varphi)=C J_n(kr)e^{in\varphi},
\end{equation}
where $C$ is the normalization constant. One can show that the hypergeometric function can be reduced to the Bessel function in this limit $l\rightarrow +\infty$.

\section{Symmetries  of the Dirac equation on the Poincar\'{e} disk}
\label{app:symmetry_dirac_equation}

In this section, we present the detailed derivation of the transformation generators for the spinor field, as introduced in Sec.~\ref{section_spinors}. The infinitesimal coordinate transformation $z' = z + \delta z$ given in Eq.~\eqref{eq:Mobius_transformation} takes the form
\begin{equation}
  z' = z + i\theta z + \bar{c} z^2 - c.
\end{equation}
Expressing $c = c_1 + i c_2$ and $z = x + i y$, the corresponding transformations for the real coordinates $x$ and $y$ are
\begin{align}
  x' &= x - \theta y + c_1(-1 + x^2 - y^2) + 2 c_2 x y, \\
  y' &= y + \theta x + c_2(-1 - x^2 + y^2) + 2 c_1 x y.
\end{align}
These expressions directly generalize the infinitesimal coordinate transformation given in Eq.~\eqref{eq:infinitesimal_coordinate_transformation}. To determine the corresponding transformation generators for the Dirac field, we consider the covariance of the Dirac equation under these transformations:
\begin{equation}
\label{app:eq:primed_Dirac_equation}
  \left[\gamma^{\prime\mu}(x^{\prime}) D_{\mu}^{\prime}(x^{\prime}) - m \right]\Psi^{\prime}(x^{\prime}) = 0,
\end{equation}
where $D_{\mu}$ denotes the covariant derivative, defined as $D_{\mu} = \partial_{\mu} + \omega_{\mu} = \partial_{\mu} + \frac{1}{2} \omega_{\mu}{}^{ab} \Sigma_{ab}$, and $\gamma^{\mu}(x) = e^{\mu}{}_a(x) \gamma^a$. In Eq.~\eqref{app:eq:primed_Dirac_equation}, Greek indices refer to both spatial and temporal coordinates; however, in the following, we focus on stationary solutions of the Dirac equation and accordingly restrict Greek indices to spatial components. The transformation law for the metric is given by the variation
\begin{equation}
  \delta_0 g_{\mu\nu}(x) = 0.
\end{equation}

\par We now express the total variation of the metric tensor as
\begin{equation}
  \delta g_{\mu\nu}(x) = \delta x^{\rho}\, \partial_{\rho} \left(e_{\mu}{}^{a}(x)\, e_{\nu}{}^{b}(x)\right) \eta_{ab},
\end{equation}
where $\eta_{ab}$ is the identity matrix, as we consider only spatial coordinates, and $e_{\mu}{}^{a\prime}(x) = e_{\mu}{}^{a}(x)$.

From this, the transformation laws for the spin connection and the $\gamma$-matrices follow as
\begin{align}
  \omega^{\prime}_{\mu}(x) &= \omega_{\mu}(x), \\
  \gamma^{\prime\mu}(x) &= \gamma^{\mu}(x).
\end{align}
The infinitesimal transformation of the covariant derivative can thus be written as
\begin{align}
  D_{\mu}^{\prime}(x^{\prime}) &= \partial_{\mu}^{\prime} + \omega_{\mu}^{\prime}(x^{\prime}) \\
  &= \frac{\partial x^{\rho}}{\partial x^{\prime\mu}} \partial_{\rho} + \omega_{\mu}(x) + \delta x^{\rho} \partial_{\rho} \omega_{\mu}(x),
\end{align}
and the $\gamma$-matrices transform as
\begin{equation}
  \gamma^{\prime\mu}(x^{\prime}) = \gamma^{\mu}(x) + \delta x^{\rho} \partial_{\rho} \gamma^{\mu}(x).
\end{equation}
Taking into account the transformation from Eq.~\ref{eq:ansatz_transformation_spinor}, the spinor field transforms as
\begin{align}
  \Psi'(x') = \left[ 1 + \frac{1}{2} w^{\mu\nu} M_{\mu\nu} + c^{\mu} K_{\mu} \right] \Psi(x)
  + \delta x^{\mu} \partial_{\mu} \Psi(x).
\end{align}

\par Now we introduce the ansatz for symmetry generators as given in Eqs. \eqref{eq:ansatz_Mmunu} and \eqref{eq:ansatz_Kmu}. Noting that the rotation parameter can be expressed as $w^{\mu\nu} = \theta \varepsilon^{\mu\nu}$, we introduce the shorthand notation $M^{(0)} \equiv \varepsilon^{\mu\nu} M_{\mu\nu}^{(0)}$ and $M^{\rho} \equiv \varepsilon^{\mu\nu} M_{\mu\nu}{}^{\rho}$. Applying these transformations to Eq.~\eqref{app:eq:primed_Dirac_equation}, we obtain the following algebra of the generators:
\begin{align}
\label{app:eq:coordinate_M}
  [\gamma^{\rho},M^{\mu}]&=0, \\
\label{app:eq:matrix_M}
  [\gamma^{\rho},M^{(0)}]&=M^{\mu}(\partial_{\mu}\gamma^{\rho})-\gamma^{\mu}(\partial_{\mu}M^{\rho}), \\
\label{app:eq:coordinate_K}
  [\gamma^{\rho},K_{\mu}{}^{\nu}]&=0, \\
\label{app:eq:matrix_K}
  [\gamma^{\rho},K_{\mu}^{(0)}] &=K_{\mu}{}^{\nu}(\partial_{\nu}\gamma^{\rho})-\gamma^{\nu}(\partial_{\nu}K_{\mu}{}^{\rho}).
\end{align}
Eqs.~\eqref{app:eq:coordinate_M} and \eqref{app:eq:coordinate_K} confirm that $M^{\mu}$ and $K_{\mu}{}^{\nu}$ act purely as coordinate representations, as assumed. The Killing vectors derived in Eqs.~\eqref{eq:Killing_vector1}, \eqref{eq:Killing_vector2}, and \eqref{eq:Killing_vector3} correspond directly to these coordinate representations of the generators. Incorporating this identification into Eqs.~\eqref{app:eq:matrix_M} and \eqref{app:eq:matrix_K} yields the explicit forms of the matrix components of the transformation generators:
\begin{align}
  K_x^{(0)} &= i y \sigma_z, \\
  M  &= i \frac{\sigma_z}{2}, \\
  K_y^{(0)} &= -i x \sigma_z.
\end{align}

It can then be readily checked that the obtained generators satisfy the algebra of the Killing vectors from Eqs. \eqref{eq:Killing_algebra1}-\eqref{eq:Killing_algebra3}.

{
\section{Derivation of the spinor transformation law on the lattice}
\label{App:DerivationEq105}

In this appendix, we derive Eq.~\eqref{eq:infinitesimal-Dirac-transf}, which is used in Sec.~\ref{LatticeDiracFermions} to identify the discrete spin-connection factor entering the tight-binding model.

We start from the continuum form of the symmetry generator acting on a Dirac spinor on the Poincar\'e disk. In complex coordinates, the coordinate part of the generator can be written as
\begin{equation}
  V(z)\partial_z,
  \,\, {\rm with}  \,\,
  V(z)=\bar a z^2+i\theta z-a ,
\end{equation}
while the spinor part is proportional to \(\sigma_z\). Thus, the full generator takes the form
\begin{equation}
  \mathcal{D}
  =
  V(z)\partial_z
  +\frac{i}{2}\big(\theta-i\bar a z\big)\sigma_z .
\end{equation}
The first term generates the M\"obius transformation of the coordinate \(z\), whereas the second term acts on the spinor index.

For \(z_1\neq z_2\), we factorize
\begin{equation}
  V(z)=\bar a (z-z_1)(z-z_2),
\end{equation}
where we used $z_1$ and $z_2$ defined in Eqs. \eqref{eq:z1} and \eqref{eq:z2}. We then introduce a coordinate \(\xi\) along the orbit generated by \(V(z)\partial_z\), defined by
\begin{equation}
  \de\xi=\frac{\de z}{V(z)} .
\end{equation}
Using the above factorization, this gives
\begin{equation}
  \xi
  =
  \frac{1}{\sqrt{-\theta^2+4|a|^2}}
  \log\frac{z-z_2}{z-z_1},
\end{equation}
since
\begin{equation}
  \bar a (z_2-z_1)=\sqrt{-\theta^2+4|a|^2}.
\end{equation}
Therefore,
\begin{equation}
  V(z)\partial_z=\partial_\xi .
\end{equation}
In the \(\xi\) coordinate, the generator becomes
\begin{equation}
  \mathcal{D}
  =
  \partial_\xi
  +
  \frac{i}{2}\big(\theta-i\bar a z(\xi)\big)\sigma_z .
\end{equation}

Exponentiating this first-order differential operator separates the coordinate translation from the spinor rotation. Since the matrix part is always proportional to \(\sigma_z\), it commutes with itself along the orbit, and the finite transformation can be written as
\begin{equation}
  e^{\mathcal{D}}\Psi_\sigma(z)
  =
  \exp\left[
  \frac{i\sigma_z}{2}
  \int_{\xi}^{\xi+1}\de \xi'\,
  \big(\theta-i\bar a z(\xi')\big)
  \right]
  e^{\partial_\xi}\Psi_\sigma(z).
\end{equation}
The operator \(e^{\partial_\xi}\) translates the coordinate along the orbit and therefore implements
\begin{equation}
  z\mapsto z'=\gamma^{-1}z ,
\end{equation}
where we introduce the notation $\gamma^{-1}$ as the coordinate transformation. We thus define
\begin{equation}
  f_{\gamma}(z)
  =
  \int_{\xi}^{\xi+1}d\xi'\,
  \big(\theta-i\bar a z(\xi')\big).
\end{equation}
Using \(d\xi=dz/V(z)\), this integral becomes
\begin{align}
  f_{\gamma}(z)
  &=
  \theta
  -i\bar a\int_z^{z'}\frac{u\,\de u}{V(u)}
  \nonumber\\
  &=
  \theta
  -i\int_z^{z'}\frac{u\,\de u}{(u-z_1)(u-z_2)} .
  \label{AppEq:G12}
\end{align}
The integrand can be decomposed as
\begin{equation}
  \frac{u}{(u-z_1)(u-z_2)}
  =
  \frac{1}{z_2-z_1}
  \left(
  \frac{z_2}{u-z_2}
  -
  \frac{z_1}{u-z_1}
  \right).
\end{equation}
Therefore,
\begin{align}
  f_{\gamma}(z',z)
  &=
  \theta
  -\frac{i}{z_2-z_1}
  \left[
  z_2\log\frac{z'-z_2}{z-z_2}
  -
  z_1\log\frac{z'-z_1}{z-z_1}
  \right].
\end{align}
Substituting this result into the finite transformation gives
\begin{equation}
  \Psi'_\sigma(z')
  =
  e^{i f_{\gamma}(z',z)\sigma_z/2}\,
  e^{\partial_\xi}\Psi_\sigma(z),
\end{equation}
which is Eq.~\eqref{eq:infinitesimal-Dirac-transf}.  \begingroup The first exponential describes the local spin-frame rotation generated by the finite isometry, while the second exponential implements the coordinate transformation between neighboring lattice sites. The degenerate case \(z_1=z_2\) can also be obtained taking the integral in this case Eq.~\eqref{AppEq:G12},
\begin{equation}
  f_{\gamma}(z',z)=\ln \frac{z'-z_1}{z-z_1}+\frac{z_1(z'-z)}{(z'-z_1)(z-z_1)}.
\end{equation}
\endgroup
}

\begingroup
\section{Parallel transport method for discrete spin connection}
\label{App:Paralell_transport_equation_solving}

\par We now outline an alternative method for deriving the discrete spin connection based on directly solving the parallel transport equation. As established in Sec.\ref{sec:Killing vectors}, the Poincar\'{e} disk admits two types of geodesics: straight lines passing through the disk's center [Eq.\eqref{geodesic1}] and orthogonal circles [Eq.~\eqref{geodesic2}]. Parallel transport along the straight-line geodesics can be readily analyzed, analogously to the Euclidean case but with the normalization condition imposed by the Poincar\'{e} disk metric. In contrast, for geodesics corresponding to orthogonal circles, the parallel transport equation must be solved explicitly, as detailed below. To simplify the calculation, we solve parallel transport equation along the geodesics in polar coordinates which read
\begin{align}
\label{parallel_transport_equation1}
  \frac{\de V^{r}}{\de t}+\Gamma_{\varphi\varphi}^{r}\frac{\de \varphi}{\de t}V^{\varphi}+\Gamma_{rr}^{r}\frac{\de r}{\de t}V^r&=0, \\
  \label{parallel_transport_equation2}
\frac{\de V^{\varphi}}{\de t}+\Gamma_{r\varphi}^{\varphi}\left(\frac{\de r}{\de t}V^{\varphi}+\frac{\de\varphi}{\de t}V^{r}\right)&=0,
\end{align}
with Christoffel symbols also in polar coordinates. Non-zero Christoffel symbols obtained in the chosen model of hyperbolic space are given as
\begin{align}
  \Gamma_{\varphi\varphi}^{r}  &= -r\,\frac{r^2+1}{1-r^2}, \\
  \Gamma_{r\varphi}^{\varphi}  &= \frac{1}{r}\frac{r^2+1}{1-r^2}, \\
  \Gamma_{rr}^{r}  &= \frac{2r}{1-r^2},
\end{align}
where we have set $l=1$. We proceed by simultaneously solving the parallel transport equations, Eqs.~(\ref{parallel_transport_equation1}) and~(\ref{parallel_transport_equation2}), and recasting them in a form that can be transformed into the equation of a circle. To facilitate this, we multiply each equation by an appropriate factor as follows:
\begin{align}
\label{app:p_transport1}
  \frac{\de V^r}{\de r} - r\frac{r^2+1}{1-r^2} \frac{\de \varphi}{\de r} V^{\varphi} + \frac{2r}{1-r^2} V^r &= 0
  \Big/ \cdot (1 - r^2)^{-1} \\
\label{app:p_transport2}
  \frac{\de V^{\varphi}}{\de r} + \frac{1}{r}\frac{r^2+1}{1-r^2} V^{\varphi} + \frac{1}{r}\frac{r^2+1}{1-r^2} \frac{\de \varphi}{\de r} V^r &= 0
  \Big/ \cdot \frac{r}{1 - r^2}
\end{align}
This procedure reveals a symmetry between the equations, motivating the introduction of the variables:
\begin{align}
\label{zeta_phi}
  \zeta^{\varphi} &\equiv \frac{r}{1 - r^2} V^{\varphi}, \\
\label{zeta_r}
  \zeta^{r}  &\equiv \frac{1}{1 - r^2} V^r.
\end{align}
\par After implementing expressions (\ref{zeta_phi}) and (\ref{zeta_r}), and by summing the equations of parallel transport (\ref{app:p_transport1}) and (\ref{app:p_transport2}), we obtain
\begin{equation}
  \frac{\de}{\de r}[(\zeta^{\varphi})^2+(\zeta^r)^2]=0 \quad \Longrightarrow \quad (\zeta^{\varphi})^2+(\zeta^r)^2=a^2,
\end{equation}
where $a$ is an arbitrary constant. Solutions for defined vectors (Eqs. \eqref{zeta_phi} and \eqref{zeta_r}) are now given by $\zeta^{\varphi}=a\cos\theta$ and $\zeta^r=a\sin\theta$. The differential equations for parallel transport (Eqs. \eqref{parallel_transport_equation1} and \eqref{parallel_transport_equation2}) can be rewritten as
\begin{equation}
  \frac{\mathrm{d}\theta}{\mathrm{d}r}-\left(\frac{r^2+1}{1-r^2}\right)\frac{\mathrm{d}\varphi}{\mathrm{d}r}=0,
\end{equation}
where we calculate $\frac{\mathrm{d}\varphi}{\mathrm{d}r}$ from the geodesic Eq.~\eqref{geodesic2}. We subsequently find
\begin{equation}
\label{eq:dtheta/dr}
  \frac{\de \theta}{\de r}=-\frac{\cot(\chi-\varphi)}{r}.
\end{equation}
Here, there is the ambiguity of the sign of $\theta$ depending on the angle $\chi$, which is defined as
\begin{equation}
  \cos\chi=\frac{x_0}{\sqrt{x_0^2+y_0^2}}, \quad \sin\chi=\frac{y_0}{\sqrt{x_0^2+y_0^2}}.
\end{equation}
Taking this into account, we can now integrate the Eq. \eqref{eq:dtheta/dr}, yielding the form of angle $\theta$ as
\begin{widetext}
  \begin{equation}
  \label{eq:theta_solved}
  \theta=\mp\frac{1}{2}\left[\arcsin\frac{\frac{1}{r^2}+1-2(x_0^2+y_0^2)}{2\sqrt{(x_0^2+y_0^2)(x_0^2+y_0^2-1)}} -\arcsin\frac{r^2+1-2(x_0^2+y_0^2)}{2\sqrt{(x_0^2+y_0^2)(x_0^2+y_0^2-1)}}
  \right]+\theta_0,
  \end{equation}
where $\theta_0$ is an integration constant, $x_0$ and $y_0$ are the coordinates of the center of the corresponding geodesic curve.  From this we can readily write the vectors on the arbitrary site given by
\begin{align}
\label{parallel_transport_vector1}
  V^r&=\pm a\frac{1-r^2}{2r\sqrt{R^2+1}}[\sqrt{4r^2(R^2+1)-(r^2+1)^2}\cos\theta_0
  +(r^2+1)\sin\theta_0],\\  \label{parallel_transport_vector2}
  V^{\varphi}&=\pm a\frac{1-r^2}{2r^2\sqrt{R^2+1}}[(r^2+1)\cos\theta_0
  -\sqrt{4r^2(R^2+1)-(r^2+1)^2}\sin\theta_0],
\end{align}
where $a$ and $\theta_0$ are integration constants, which determine the initial conditions of the vector being parallel transported, and $R$ is the radius of the arbitrary geodesic curve from Eq. \eqref{geodesic2}. The sign ambiguity comes from solving the parallel transport equation, Eq.~\eqref{eq:dtheta/dr}. At last, having found the components of the parallel transported vectors, we express them in Cartesian coordinates.

\par By utilizing the result for the parallel transport equations Eq.~\eqref{parallel_transport_vector1} and Eq.~\eqref{parallel_transport_vector2} in Cartesian coordinates, for site $B$ in Fig. \ref{fig2}, for example, we find
  \begin{equation}
  \vec{V_B}=\frac{a(1-r_0^2)}{2r_0\sqrt{R^2+1}}\begin{pmatrix}
  \sqrt{4r_0^2(R^2+1)-(r_0^2+1)^2}\cos(\theta_0-\frac{\pi}{p})+(r_0^2+1)\sin(\theta_0-\frac{\pi}{p}) \\
  -\sqrt{4r_0^2(R^2+1-(r_0^2+1)^2}\sin(\theta_0-\frac{\pi}{p})+(r_0^2+1)\cos(\theta_0-\frac{\pi}{p})
  \end{pmatrix} .
  \end{equation}
\end{widetext}
\endgroup
{
\section{Continuum limit of the discrete spin connection}
\label{App:ContinuumLimitDiscreteSpinConnection}

In this appendix, we show explicitly how the discrete spin-connection factor used in Sec.~\ref{LatticeDiracFermions} reduces to the continuum covariant derivative of the Dirac Hamiltonian.  \begingroup Here and below, \(f_{ij}\) denotes the geodesic Wilson-line phase \(f^{\rm W}_{ij}\) of Eq.~\eqref{eq:exact-wilson-line}, not the generic finite-isometry phase \(f_{\gamma}\) appearing in Eq.~\eqref{eq:infinitesimal-Dirac-transf}. \endgroup

Consider two neighboring sites \(i\) and \(j\), with coordinates \(x_i\) and
\begin{equation}
  x_j=x_i+\delta x_{ij},
\end{equation}
where \(\delta x_{ij}^{\mu}\) is the coordinate displacement along the link \(i\to j\). In the lattice construction, the spinor transformation associated with this link is encoded in the matrix
\begin{equation}
  \mathcal U_{ij}
=
\exp\!\left[
\frac{i}{2}f_{ij}\sigma_z
\right],
\qquad
f_{ij}\equiv f^{\rm W}_{ij},
\end{equation}
with \(x_j=\gamma^{-1}x_i\).  \begingroup This factor is the geodesic parallel-transport operator chosen for the oriented link; it is the finite-link discretization implemented in the Hamiltonian~\ref{eq:Hsc-TB}, which is used in the numerical analysis.
\endgroup

In the continuum theory, the corresponding spinor parallel transport from \(x_i\) to \(x_j\) is given by the Wilson line of the spin connection,
\begin{equation}
  \mathcal U_{ij}
=
\mathcal P\exp\!\left[
\int_{x_i}^{x_j} dx^\mu\,\Omega_\mu(x)
\right],
\end{equation}
where
\begin{equation}
  \Omega_\mu(x)
=
\frac{1}{2}
\omega_\mu{}^{ab}(x)\Sigma_{ab}.
\end{equation}
Here \(\mathcal P\) denotes path ordering. In the present two-dimensional case the spin connection is proportional to a single \(SO(2)\) generator, represented in the chosen spin basis by \(\sigma_z/2\), so the path ordering is trivial. We nevertheless keep \(\mathcal P\) to emphasize the Wilson-line structure and its extension to more general non-Abelian settings.

\begingroup
We now derive the closed form in Eq.~\eqref{eq:closed-form-geodesic-phase}. By introducing the disk automorphism
\begin{equation}
 h_i(z)=l\,\frac{z-z_i}{l^2-\bar z_i z},
\end{equation}
which sends \(z_i\) to the origin and maps the geodesic \(C_{ij}\) joining \(z_i\) and \(z_j\) to a radial geodesic. For the complex orthonormal vielbein \(e^+=e^1+i e^2\), its pullback is
\begin{equation}
 h_i^*e^+=e^{i\alpha_i(z)}e^+,
 \qquad
 e^{i\alpha_i(z)}=\frac{l^2-z_i\bar z}{l^2-\bar z_i z}.
\end{equation}
The spin connection correspondingly transforms as
\[
h_i^*\omega=\omega+\de\alpha_i,
\,\, {\rm with} \,\,
\de\alpha_i=\partial_\mu\alpha_i\,\de x^\mu,
\]
where \(\de\) denotes the exterior derivative. Since \(\omega\) vanishes on geodesics defined in Eq.~\eqref{geodesic1} and \(\alpha_i(z_i)=0\),
\begin{align}
 0
 &=\int_{h_i(C_{ij})}\omega
 =\int_{C_{ij}}h_i^*\omega \\
 &=f^{\rm W}_{ij}+\alpha_i(z_j),
\end{align}
which yields
\begin{equation}
f_{ij}^W=-i\log\left(\frac{l^2-z_i\bar{z_j}}{l^2-\bar{z_i}z_j}\right)
\end{equation}
and Eq.~\eqref{eq:closed-form-geodesic-phase-exp}. This also gives \(f^{\rm W}_{ji}=-f^{\rm W}_{ij}\) and \(\mathcal U_{ji}=\mathcal U_{ij}^{\dagger}\) exactly.
\endgroup

For a sufficiently short link, whose length is small compared with the curvature radius, the Wilson line can be expanded as
\begin{equation}
  \mathcal U_{ij}
=
1+\delta x_{ij}^{\mu}\Omega_\mu(x_i)
+
O(\delta x_{ij}^2).
\end{equation}
The spinor field at the neighboring site similarly admits the expansion
\begin{equation}
  \Psi(x_j)
=
\Psi(x_i)
+
\delta x_{ij}^{\mu}\partial_\mu\Psi(x_i)
+
O(\delta x_{ij}^2).
\end{equation}
Combining these two expansions gives
\begin{align}
\mathcal U_{ij}\Psi(x_j)
&=
\bigl[1+\delta x_{ij}^{\mu}\Omega_\mu(x_i)\bigr]
\bigl[\Psi(x_i)
+\delta x_{ij}^{\nu}\partial_\nu\Psi(x_i)\bigr]
\nonumber\\
&\quad
+O(\delta x_{ij}^2).
\end{align}
and therefore
\begin{align}
  \mathcal U_{ij}\Psi(x_j)
=&
\Psi(x_i)
+
\delta x_{ij}^{\mu}
\left(
\partial_\mu+\Omega_\mu
\right)\Psi(x_i) \nonumber\\
&+
O(\delta x_{ij}^2).
\end{align}
It follows that the lattice finite difference dressed by the discrete spin connection satisfies
\begin{equation}
 \frac{\mathcal U_{ij}\Psi(x_j)-\Psi(x_i)}{|\delta x_{ij}|}
\longrightarrow
\hat e_{ij}^{\mu}
\left(
\partial_\mu+\Omega_\mu
\right)\Psi(x_i),
\end{equation}
where
\begin{equation}
  \hat e_{ij}^{\mu}
=
\frac{\delta x_{ij}^{\mu}}{|\delta x_{ij}|}
\end{equation}
is the unit tangent vector along the link. Therefore, in the long-wavelength limit, the discrete spin-connection dressing promotes the ordinary derivative to the continuum covariant derivative,
\begin{equation}
  \partial_\mu
\longrightarrow
D_\mu
=
\partial_\mu+\Omega_\mu
=
\partial_\mu
+
\frac{1}{2}
\omega_\mu{}^{ab}\Sigma_{ab},
\end{equation}
thus establishing the connection with the continuum form of the covariant derivative.
}

\begin{figure*}[t!]
\centering
\includegraphics[width=0.95\linewidth]{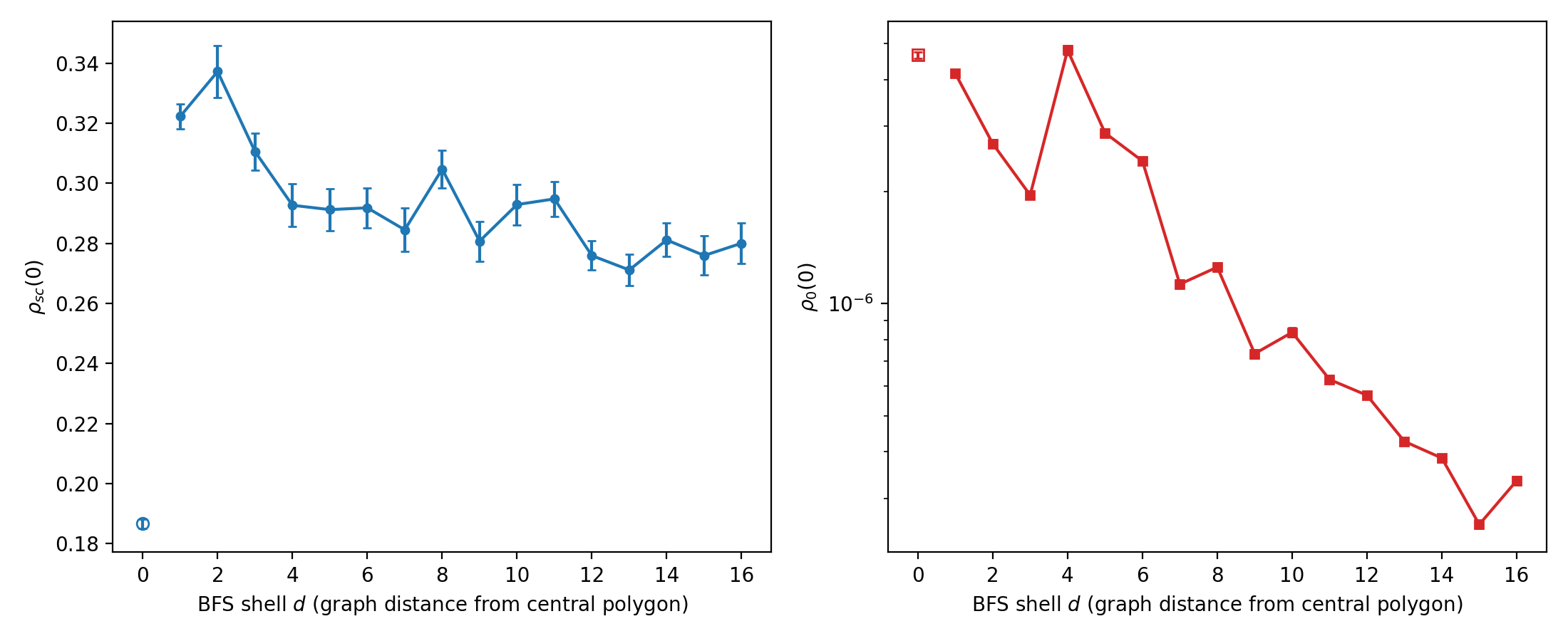}
\caption{Shell-resolved zero-energy DOS versus breadth-first-search (BFS) graph distance $d$ from the central polygon. The $\{10,3\}$ lattice contains $N=16810$ sites, $l=1$, and $M=4096$ Chebyshev moments (Table~\ref{tab:shell-resolved}). The open marker denotes $d=0$, which is treated as a central-shell outlier. Left: $\rho_{\rm sc}(0)$ remains finite from the inner shells to the true boundary at $d=16$. Right: $\rho_0(0)$ on a logarithmic scale remains five to six orders of magnitude below $\rho_{\rm sc}(0)$. Error bars show stochastic standard errors obtained from the independent trace estimates at each shell ($R=200$ for $d=0,1$ and $R=60$ for $d=2,\ldots,16$). They do not include finite-size, finite-$M$, Jackson-broadening, or spectral-rescaling systematics.}
\label{fig:shell-resolved}
\end{figure*}

\subsection{Closed-loop holonomy of the geodesic Wilson line}
\label{app:exact-holonomy}

The short-link expansion above establishes that $f_{ij}$ reproduces the continuum covariant derivative only to leading order in $\delta x_{ij}$. Here we show that the geodesic Wilson-line definition of Eq.~\eqref{eq:exact-wilson-line} used in Sec.~\ref{LatticeDiracFermions}, $f_{ij}=\int_{{\rm geodesic},x_i\to x_j}\de x^\mu\omega_\mu(x)$, satisfies the discrete Gauss--Bonnet holonomy condition exactly, at any bond length, in contrast. In the numerical code this integral is evaluated through the closed form in Eq.~\eqref{eq:closed-form-geodesic-phase}, so the implementation contains no path-discretization error.

Let $\Omega$ be an elementary $p$-gon plaquette of the $\{p,q\}$ tessellation with vertices $x_1,\dots,x_p$ ($x_{p+1}\equiv x_1$) joined by geodesic edges. Line integrals of a one-form are additive under composition of oriented curves, with no closedness or exactness required; since the geodesic edges of $\Omega$ are joined consecutively vertex-to-vertex, their composition is exactly the closed boundary curve $\partial\Omega$, so
\begin{equation}
  \Delta\theta\equiv\sum_{k=1}^{p}f_{k,k+1}=\sum_{k=1}^p\int_{x_k}^{x_{k+1}}\omega=\oint_{\partial\Omega}\omega.
\end{equation}
By Stokes' theorem, valid for any smooth one-form on a region with piecewise-smooth boundary,
\begin{equation}
  \oint_{\partial\Omega}\omega=\int_\Omega d\omega=\int_\Omega K\,\de A,
\end{equation}
using $\de\omega=K\,\de A$ (Cartan's second structure equation in two dimensions) and the constant Gaussian curvature $K=R/2=-4/l^2$ of the metric~\eqref{Poincare_metric}. Since every elementary face of a regular $\{p,q\}$ tessellation is isometric to every other face (the isometries preserving the tessellation act transitively on its faces), $\mathrm{Area}(\Omega)$ is the same for every plaquette, giving
\begin{equation}
|\Delta\theta|=|K|\cdot\mathrm{Area}(\Omega)
\end{equation}
identically, independent of bond length or position in the lattice. This is the discrete Gauss--Bonnet target used to benchmark the construction in Sec.~\ref{LatticeDiracFermions}, and it holds exactly because the  line integral leaves no residual of any order in $\delta x_{ij}$, guaranteeing that Eq.~\eqref{eq:exact-wilson-line} reproduces the correct enclosed curvature.

For a regular $\{p,q\}$ tessellation, exactly $q$ polygons meet at each vertex, so the interior angle of every elementary $p$-gon plaquette is $2\pi/q$. The hyperbolic Gauss--Bonnet theorem for geodesic
polygons~\cite{Ratcliffe2019} then fixes the angle defect
directly, without requiring a separate computation of
$\mathrm{Area}(\Omega)$ or the curvature radius $l$:
\begin{equation}
|\Delta\theta|=|K|\cdot\mathrm{Area}(\Omega)=(p-2)\pi-p\cdot\frac{2\pi}{q}.
\label{eq:DeltaTheta-pq}
\end{equation}
Specializing to $q=3$, the coordination number used for the $\{p,3\}$ tessellations, the angle defect is given by
\begin{equation}
\label{eq:plaquette-holonomy-p3}
  |\Delta\theta|=(p-2)\pi-\frac{2p\pi}{3}=\frac{(p-6)\pi}{3}.
\end{equation}
Since the plaquette holonomy enters the spin-connection factor as the $SU(2)$ rotation $\exp[i\Delta\theta\,\sigma_z/2]$, and $\sigma_z$ has eigenvalues $\pm1$, this rotation is exactly the identity whenever $\Delta\theta$ is an integer multiple of $4\pi$, i.e. whenever $(p-6)\pi/3=4\pi n$ for integer $n$, equivalently $p\equiv6\pmod{12}$. For these special values of $p$ (e.g. $p=18,30,\dots$), the elementary plaquette holonomies are trivial. On the simply connected open patches considered here, this makes the link field gauge equivalent to the identity and therefore removes any spectral effect, even though the lattice curvature remains nonzero. On closed quotients or multiply connected domains, additional global Wilson loops may remain. This is the origin of the selection rule quoted in Sec.~\ref{ConclusionsAndOutlook}. The $\{10,3\}$ lattice used throughout the numerical results below has $p=10\not=6\pmod{12}$ and is therefore unaffected by this condition.
Notice that the flat-space limit is recovered exactly from Eq.~\eqref{eq:DeltaTheta-pq}, since
\(\Delta\theta=0\) implies \(pq-2p-2q=0\), or equivalently \(1/p+1/q=1/2\), which is precisely the Euclidean tiling condition. Accordingly, \(\Delta\theta=0\) for the honeycomb lattice \(\{6,3\}\), for the square lattice \(\{4,4\}\), and  for the triangular lattice \(\{3,6\}\).

\section{Spatial resolution of the zero-energy density of states: bulk versus boundary}
\label{App:SpatialBoundaryCheck}

Because hyperbolic tessellations grow geometrically with generation, a finite open $\{p,q\}$ lattice does not have a vanishing boundary-to-bulk ratio the way a finite Euclidean lattice does. For the $N=16810$ ($\{10,3\}$) lattice used in Figs.~\ref{fig:DOS_spin_connection_generations}--\ref{fig:saturation-vs-N}, the outermost generation alone accounts for $82.9\%$ of all sites ($16810-2880$ of $16810$), and $71\%$ of all sites have coordination number below $q=3$. Since the KPM trace in the main text is taken over the complete open graph, it is natural to ask whether the finite zero-energy DOS, $\rho_{\rm sc}(0)$, is a genuine bulk property of the curved lattice or is instead dominated by under-coordinated sites near the outer edge. This appendix answers that question directly by resolving $\rho(0)$ spatially.

\subsection{Masked stochastic trace estimator}

We label every site by its graph distance $d$ from the central polygon, obtained by breadth-first search on the bond graph starting from the $p$ sites of generation $0$. This graph distance, rather than the Euclidean radius, provides the appropriate radial coordinate here: the Poincar\'e-disk embedding strongly compresses coordinates near the boundary, so that the Euclidean radius alone barely distinguishes the outer generations. For the $N=16810$ lattice, this procedure partitions the sites into $17$ concentric shells, $d=0,\dots,16$, containing between $10$ and $3010$ sites each. The shell populations first increase and then decrease toward the boundary, as shown in Table~\ref{tab:shell-resolved}, because the outermost one or two shells consist predominantly of the most under-coordinated sites.

\begin{table}
\begin{center}
\begin{tabular}{|c|c|c|c|}
\hline
$d$ & sites & $\rho_0(0)$ & $\rho_{\mathrm{sc}}(0)$ \\
\hline
0  & 10  &  $(4.660\pm0.089)\times10^{-6}$ &  $0.1867\pm0.0015^{*}$ \\
\hline
1  & 10  &  $(4.157\pm0.052)\times10^{-6}$ &  $0.3223\pm0.0042$ \\
\hline
2  & 20  &  $(2.693\pm0.064)\times10^{-6}$ &  $0.3372\pm0.0087$ \\
\hline
3  & 40  &  $(1.959\pm0.043)\times10^{-6}$ &  $0.3106\pm0.0062$ \\
\hline
4  & 80  &  $(4.80\pm0.14)\times10^{-6}$ &  $0.2927\pm0.0072$ \\
\hline
5  & 140  &  $(2.878\pm0.072)\times10^{-6}$ &  $0.2912\pm0.0069$ \\
\hline
6  & 270  &  $(2.419\pm0.055)\times10^{-6}$ &  $0.2918\pm0.0067$ \\
\hline
7  & 510  &  $(1.129\pm0.029)\times10^{-6}$ &  $0.2845\pm0.0072$ \\
\hline
8  & 960  &  $(1.254\pm0.033)\times10^{-6}$ &  $0.3047\pm0.0063$ \\
\hline
9  & 1640 &  $(7.34\pm0.20)\times10^{-7}$ &  $0.2807\pm0.0067$ \\
\hline
10 & 2440 &  $(8.38\pm0.23)\times10^{-7}$ &  $0.2929\pm0.0068$ \\
\hline
11 & 3000 &  $(6.26\pm0.15)\times10^{-7}$ &  $0.2948\pm0.0058$ \\
\hline
12 & 3010 &  $(5.68\pm0.10)\times10^{-7}$ &  $0.2759\pm0.0049$ \\
\hline
13 & 2400 &  $(4.266\pm0.095)\times10^{-7}$ &  $0.2711\pm0.0053$ \\
\hline
14 & 1480 &  $(3.849\pm0.092)\times10^{-7}$ &  $0.2811\pm0.0056$ \\
\hline
15 & 640  &  $(2.551\pm0.057)\times10^{-7}$ &  $0.2759\pm0.0065$ \\
\hline
16 & 160  &  $(3.346\pm0.084)\times10^{-7}$ &  $0.2800\pm0.0068$ \\
\hline
\end{tabular}
\end{center}
\caption{Shell-resolved zero-energy density of states (DOS) for the
$\{10,3\}$ lattice ($\ell=1$, $M=4096$ Chebyshev moments), $d=$ graph
distance from the central polygon ($d=16$ is the true open boundary).
$^*$The central shell $d=0$ is a statistically stable local outlier. For all shells with $d\geq1$, $\rho_{\rm sc}(0)$ remains finite within the range $0.27$--$0.34$, with statistically resolved shell-to-shell variations but no systematic trend toward the true open boundary. The quoted standard errors quantify only stochastic-trace sampling and do not include finite-size, finite-$M$, Jackson-broadening, or spectral-rescaling systematics.}
\label{tab:shell-resolved}
\end{table}

To probe the spatial distribution of the low-energy density of states, we compute the DOS associated with a shell $S$ of fixed graph distance. The stochastic trace estimator introduced in Sec.~\ref{LatticeDiracFermions} is modified by restricting the support of each random-phase vector to the sites belonging to $S$,
\begin{equation}
  r_i=
  \begin{cases}
  e^{i\phi_i}/\sqrt{|S|}, & i\in S,\\
  0, & i\notin S,
  \end{cases}
\end{equation}
where the phases $\phi_i$ are independent and uniformly distributed on $[0,2\pi)$. This procedure yields the shell-averaged local density of states,
\begin{equation}
  \rho_S(E)=\frac{1}{|S|}\sum_{i\in S}\rho_i(E),
\end{equation}
which is normalized in the same way as the full-lattice DOS $\rho(E)$ used in the main text and can therefore be compared directly in magnitude. For the shell-resolved calculation we evaluate one spin block, $H_+$. Since $H_-=H_+^*$, the two blocks have identical spectra, and the resulting shell DOS equals the per-state-normalized DOS of the full two-component Hamiltonian. Importantly, the Chebyshev recursion is still performed with the complete Hamiltonian $H$ (respectively $H_{\rm sc}$) at every step. The states contributing to $\rho_S(E)$ are therefore allowed to hybridize with the entire lattice, including the true open boundary; only the final stochastic trace evaluation is restricted to the shell $S$. We use  $M=4096$ Chebyshev moments, matching the resolution used in the main text, and $R=60$ random-phase vectors per shell. For  the two smallest shells, $d=0,1$, we increase the number to $R=200$ as an additional stability check, as discussed below.

\subsection{Results}

Table~\ref{tab:shell-resolved} lists $\rho_0(0)$ and $\rho_{\rm sc}(0)$ for all $17$ shells, and Fig.~\ref{fig:shell-resolved} plots the same data.  At $M=4096$, $\rho_{\rm sc}(0)$ remains finite throughout all shells with $d\geq1$, taking values in the range $0.27$--$0.34$. Although statistically resolved shell-to-shell variations are present, no systematic increase or decrease develops toward the true boundary at $d=16$. The outermost shells remain finite and show no enhancement associated with proximity to the open boundary. Over the same range, $\rho_0(0)$ remains five to six orders of magnitude smaller and likewise shows no boundary enhancement.

  The central polygon, $d=0$, yields the lower value $\rho_{\rm sc}(0)\simeq0.19$, whereas all shells with $d\geq1$ retain a finite enhanced DOS with shell-to-shell variations and no systematic boundary trend.
Increasing the stochastic sampling to $R=200$ confirms that
the central-shell deviation is not a trace-estimation artifact.
It should therefore be regarded as a localized central-region
effect rather than evidence for boundary domination. The
boundary conclusion is unchanged: no systematic enhancement
or suppression develops toward the outer shells.

\bibliography{reference}

\begin{thebibliography}{87}%
\makeatletter
\providecommand \@ifxundefined [1]{%
 \@ifx{#1\undefined}
}%
\providecommand \@ifnum [1]{%
 \ifnum #1\expandafter \@firstoftwo
 \else \expandafter \@secondoftwo
 \fi
}%
\providecommand \@ifx [1]{%
 \ifx #1\expandafter \@firstoftwo
 \else \expandafter \@secondoftwo
 \fi
}%
\providecommand \natexlab [1]{#1}%
\providecommand \enquote  [1]{``#1''}%
\providecommand \bibnamefont  [1]{#1}%
\providecommand \bibfnamefont [1]{#1}%
\providecommand \citenamefont [1]{#1}%
\providecommand \href@noop [0]{\@secondoftwo}%
\providecommand \href [0]{\begingroup \@sanitize@url \@href}%
\providecommand \@href[1]{\@@startlink{#1}\@@href}%
\providecommand \@@href[1]{\endgroup#1\@@endlink}%
\providecommand \@sanitize@url [0]{\catcode `\\12\catcode `\$12\catcode `\&12\catcode `\#12\catcode `\^12\catcode `\_12\catcode `\%12\relax}%
\providecommand \@@startlink[1]{}%
\providecommand \@@endlink[0]{}%
\providecommand \url  [0]{\begingroup\@sanitize@url \@url }%
\providecommand \@url [1]{\endgroup\@href {#1}{\urlprefix }}%
\providecommand \urlprefix  [0]{URL }%
\providecommand \Eprint [0]{\href }%
\providecommand \doibase [0]{https://doi.org/}%
\providecommand \selectlanguage [0]{\@gobble}%
\providecommand \bibinfo  [0]{\@secondoftwo}%
\providecommand \bibfield  [0]{\@secondoftwo}%
\providecommand \translation [1]{[#1]}%
\providecommand \BibitemOpen [0]{}%
\providecommand \bibitemStop [0]{}%
\providecommand \bibitemNoStop [0]{.\EOS\space}%
\providecommand \EOS [0]{\spacefactor3000\relax}%
\providecommand \BibitemShut  [1]{\csname bibitem#1\endcsname}%
\let\auto@bib@innerbib\@empty
\bibitem [{\citenamefont {Koll\'{a}r}\ \emph {et~al.}(2019{\natexlab{a}})\citenamefont {Koll\'{a}r}, \citenamefont {Fitzpatrick},\ and\ \citenamefont {Houck}}]{NatureExperiment}%
  \BibitemOpen
  \bibfield  {author} {\bibinfo {author} {\bibfnamefont {A.~J.}\ \bibnamefont {Koll\'{a}r}}, \bibinfo {author} {\bibfnamefont {M.}~\bibnamefont {Fitzpatrick}},\ and\ \bibinfo {author} {\bibfnamefont {A.~A.}\ \bibnamefont {Houck}},\ }\bibfield  {title} {\bibinfo {title} {Hyperbolic lattices in circuit quantum electrodynamics},\ }\href {https://doi.org/10.1038/s41586-019-1348-3} {\bibfield  {journal} {\bibinfo  {journal} {Nature}\ }\textbf {\bibinfo {volume} {571}},\ \bibinfo {pages} {45} (\bibinfo {year} {2019}{\natexlab{a}})}\BibitemShut {NoStop}%
\bibitem [{\citenamefont {Boettcher}\ \emph {et~al.}(2020)\citenamefont {Boettcher}, \citenamefont {Bienias}, \citenamefont {Belyansky}, \citenamefont {Koll\'ar},\ and\ \citenamefont {Gorshkov}}]{QuantumSimulationOfHyperbolicSpace}%
  \BibitemOpen
  \bibfield  {author} {\bibinfo {author} {\bibfnamefont {I.}~\bibnamefont {Boettcher}}, \bibinfo {author} {\bibfnamefont {P.}~\bibnamefont {Bienias}}, \bibinfo {author} {\bibfnamefont {R.}~\bibnamefont {Belyansky}}, \bibinfo {author} {\bibfnamefont {A.~J.}\ \bibnamefont {Koll\'ar}},\ and\ \bibinfo {author} {\bibfnamefont {A.~V.}\ \bibnamefont {Gorshkov}},\ }\bibfield  {title} {\bibinfo {title} {Quantum simulation of hyperbolic space with circuit quantum electrodynamics: From graphs to geometry},\ }\href {https://doi.org/10.1103/PhysRevA.102.032208} {\bibfield  {journal} {\bibinfo  {journal} {Phys. Rev. A}\ }\textbf {\bibinfo {volume} {102}},\ \bibinfo {pages} {032208} (\bibinfo {year} {2020})}\BibitemShut {NoStop}%
\bibitem [{\citenamefont {Bekenstein}\ \emph {et~al.}(2017)\citenamefont {Bekenstein}, \citenamefont {Kabessa}, \citenamefont {Sharabi}, \citenamefont {Tal}, \citenamefont {Engheta}, \citenamefont {Eisenstein}, \citenamefont {Agranat},\ and\ \citenamefont {Segev}}]{DielectricConstant5}%
  \BibitemOpen
  \bibfield  {author} {\bibinfo {author} {\bibfnamefont {R.}~\bibnamefont {Bekenstein}}, \bibinfo {author} {\bibfnamefont {Y.}~\bibnamefont {Kabessa}}, \bibinfo {author} {\bibfnamefont {Y.}~\bibnamefont {Sharabi}}, \bibinfo {author} {\bibfnamefont {O.}~\bibnamefont {Tal}}, \bibinfo {author} {\bibfnamefont {N.}~\bibnamefont {Engheta}}, \bibinfo {author} {\bibfnamefont {G.}~\bibnamefont {Eisenstein}}, \bibinfo {author} {\bibfnamefont {A.}~\bibnamefont {Agranat}},\ and\ \bibinfo {author} {\bibfnamefont {M.}~\bibnamefont {Segev}},\ }\bibfield  {title} {\bibinfo {title} {Control of light by curved space in nanophotonic structures},\ }\href {https://doi.org/10.1038/s41566-017-0008-0} {\bibfield  {journal} {\bibinfo  {journal} {Nature Photonics}\ }\textbf {\bibinfo {volume} {11}} (\bibinfo {year} {2017})}\BibitemShut {NoStop}%
\bibitem [{\citenamefont {Bekenstein}\ \emph {et~al.}(2015)\citenamefont {Bekenstein}, \citenamefont {Schley}, \citenamefont {Mutzafi}, \citenamefont {Rotschild},\ and\ \citenamefont {Segev}}]{DielectricConstant6}%
  \BibitemOpen
  \bibfield  {author} {\bibinfo {author} {\bibfnamefont {R.}~\bibnamefont {Bekenstein}}, \bibinfo {author} {\bibfnamefont {R.}~\bibnamefont {Schley}}, \bibinfo {author} {\bibfnamefont {M.}~\bibnamefont {Mutzafi}}, \bibinfo {author} {\bibfnamefont {C.}~\bibnamefont {Rotschild}},\ and\ \bibinfo {author} {\bibfnamefont {M.}~\bibnamefont {Segev}},\ }\bibfield  {title} {\bibinfo {title} {Optical simulations of gravitational effects in the newton–schrödinger system},\ }\href {https://doi.org/10.1038/nphys3451} {\bibfield  {journal} {\bibinfo  {journal} {Nature Physics}\ }\textbf {\bibinfo {volume} {11}} (\bibinfo {year} {2015})}\BibitemShut {NoStop}%
\bibitem [{\citenamefont {Boettcher}\ \emph {et~al.}(2022)\citenamefont {Boettcher}, \citenamefont {Gorshkov}, \citenamefont {Koll\'{a}r}, \citenamefont {Maciejko}, \citenamefont {Rayan},\ and\ \citenamefont {Thomale}}]{CrystallographyOfHyperbolicLattices}%
  \BibitemOpen
  \bibfield  {author} {\bibinfo {author} {\bibfnamefont {I.}~\bibnamefont {Boettcher}}, \bibinfo {author} {\bibfnamefont {A.~V.}\ \bibnamefont {Gorshkov}}, \bibinfo {author} {\bibfnamefont {A.~J.}\ \bibnamefont {Koll\'{a}r}}, \bibinfo {author} {\bibfnamefont {J.}~\bibnamefont {Maciejko}}, \bibinfo {author} {\bibfnamefont {S.}~\bibnamefont {Rayan}},\ and\ \bibinfo {author} {\bibfnamefont {R.}~\bibnamefont {Thomale}},\ }\bibfield  {title} {\bibinfo {title} {Crystallography of hyperbolic lattices},\ }\href {https://doi.org/10.1103/PhysRevB.105.125118} {\bibfield  {journal} {\bibinfo  {journal} {Phys. Rev. B}\ }\textbf {\bibinfo {volume} {105}},\ \bibinfo {pages} {125118} (\bibinfo {year} {2022})}\BibitemShut {NoStop}%
\bibitem [{\citenamefont {Maldacena}(1999)}]{Maldacena1997}%
  \BibitemOpen
  \bibfield  {author} {\bibinfo {author} {\bibfnamefont {J.~M.}\ \bibnamefont {Maldacena}},\ }\bibfield  {title} {\bibinfo {title} {{The large N limit of superconformal field theories and supergravity}},\ }\href {https://doi.org/10.1023/A:1026654312961} {\bibfield  {journal} {\bibinfo  {journal} {Int. J. Theor. Phys.}\ }\textbf {\bibinfo {volume} {38}},\ \bibinfo {pages} {1113} (\bibinfo {year} {1999})}\BibitemShut {NoStop}%
\bibitem [{\citenamefont {Witten}(1998)}]{Witten1998}%
  \BibitemOpen
  \bibfield  {author} {\bibinfo {author} {\bibfnamefont {E.}~\bibnamefont {Witten}},\ }\bibfield  {title} {\bibinfo {title} {{Anti-de Sitter space and holography}},\ }\href {https://doi.org/10.4310/ATMP.1998.v2.n2.a2} {\bibfield  {journal} {\bibinfo  {journal} {Adv. Theor. Math. Phys.}\ }\textbf {\bibinfo {volume} {2}},\ \bibinfo {pages} {253} (\bibinfo {year} {1998})}\BibitemShut {NoStop}%
\bibitem [{\citenamefont {Gubser}\ \emph {et~al.}(1998)\citenamefont {Gubser}, \citenamefont {Klebanov},\ and\ \citenamefont {Polyakov}}]{Gubser1998}%
  \BibitemOpen
  \bibfield  {author} {\bibinfo {author} {\bibfnamefont {S.~S.}\ \bibnamefont {Gubser}}, \bibinfo {author} {\bibfnamefont {I.~R.}\ \bibnamefont {Klebanov}},\ and\ \bibinfo {author} {\bibfnamefont {A.~M.}\ \bibnamefont {Polyakov}},\ }\bibfield  {title} {\bibinfo {title} {{Gauge theory correlators from noncritical string theory}},\ }\href {https://doi.org/10.1016/S0370-2693(98)00377-3} {\bibfield  {journal} {\bibinfo  {journal} {Phys. Lett. B}\ }\textbf {\bibinfo {volume} {428}},\ \bibinfo {pages} {105} (\bibinfo {year} {1998})}\BibitemShut {NoStop}%
\bibitem [{\citenamefont {Ryu}\ and\ \citenamefont {Takayanagi}(2006{\natexlab{a}})}]{Ryu_PRL2006}%
  \BibitemOpen
  \bibfield  {author} {\bibinfo {author} {\bibfnamefont {S.}~\bibnamefont {Ryu}}\ and\ \bibinfo {author} {\bibfnamefont {T.}~\bibnamefont {Takayanagi}},\ }\bibfield  {title} {\bibinfo {title} {Holographic derivation of entanglement entropy from the anti--de sitter space/conformal field theory correspondence},\ }\href {https://doi.org/10.1103/PhysRevLett.96.181602} {\bibfield  {journal} {\bibinfo  {journal} {Phys. Rev. Lett.}\ }\textbf {\bibinfo {volume} {96}},\ \bibinfo {pages} {181602} (\bibinfo {year} {2006}{\natexlab{a}})}\BibitemShut {NoStop}%
\bibitem [{\citenamefont {Ryu}\ and\ \citenamefont {Takayanagi}(2006{\natexlab{b}})}]{Ryu_JHEP2006}%
  \BibitemOpen
  \bibfield  {author} {\bibinfo {author} {\bibfnamefont {S.}~\bibnamefont {Ryu}}\ and\ \bibinfo {author} {\bibfnamefont {T.}~\bibnamefont {Takayanagi}},\ }\bibfield  {title} {\bibinfo {title} {Aspects of holographic entanglement entropy},\ }\href {https://doi.org/10.1088/1126-6708/2006/08/045} {\bibfield  {journal} {\bibinfo  {journal} {Journal of High Energy Physics}\ }\textbf {\bibinfo {volume} {2006}},\ \bibinfo {pages} {045} (\bibinfo {year} {2006}{\natexlab{b}})}\BibitemShut {NoStop}%
\bibitem [{\citenamefont {Swingle}(2012)}]{Swingle2012}%
  \BibitemOpen
  \bibfield  {author} {\bibinfo {author} {\bibfnamefont {B.}~\bibnamefont {Swingle}},\ }\bibfield  {title} {\bibinfo {title} {Entanglement renormalization and holography},\ }\href {https://doi.org/10.1103/PhysRevD.86.065007} {\bibfield  {journal} {\bibinfo  {journal} {Phys. Rev. D}\ }\textbf {\bibinfo {volume} {86}},\ \bibinfo {pages} {065007} (\bibinfo {year} {2012})}\BibitemShut {NoStop}%
\bibitem [{\citenamefont {Czech}\ \emph {et~al.}(2015)\citenamefont {Czech}, \citenamefont {Lamprou}, \citenamefont {McCandlish},\ and\ \citenamefont {Sully}}]{Czech2015}%
  \BibitemOpen
  \bibfield  {author} {\bibinfo {author} {\bibfnamefont {B.}~\bibnamefont {Czech}}, \bibinfo {author} {\bibfnamefont {L.}~\bibnamefont {Lamprou}}, \bibinfo {author} {\bibfnamefont {S.}~\bibnamefont {McCandlish}},\ and\ \bibinfo {author} {\bibfnamefont {J.}~\bibnamefont {Sully}},\ }\bibfield  {title} {\bibinfo {title} {Integral geometry and holography},\ }\href {https://doi.org/10.1007/JHEP10(2015)175} {\bibfield  {journal} {\bibinfo  {journal} {Journal of High Energy Physics}\ }\textbf {\bibinfo {volume} {2015}},\ \bibinfo {pages} {175} (\bibinfo {year} {2015})}\BibitemShut {NoStop}%
\bibitem [{\citenamefont {Nishioka}(2018)}]{Nishioka2018}%
  \BibitemOpen
  \bibfield  {author} {\bibinfo {author} {\bibfnamefont {T.}~\bibnamefont {Nishioka}},\ }\bibfield  {title} {\bibinfo {title} {Entanglement entropy: Holography and renormalization group},\ }\href {https://doi.org/10.1103/RevModPhys.90.035007} {\bibfield  {journal} {\bibinfo  {journal} {Rev. Mod. Phys.}\ }\textbf {\bibinfo {volume} {90}},\ \bibinfo {pages} {035007} (\bibinfo {year} {2018})}\BibitemShut {NoStop}%
\bibitem [{\citenamefont {Leonhardt}\ and\ \citenamefont {Philbin}(2006)}]{Leonhardt_2006}%
  \BibitemOpen
  \bibfield  {author} {\bibinfo {author} {\bibfnamefont {U.}~\bibnamefont {Leonhardt}}\ and\ \bibinfo {author} {\bibfnamefont {T.~G.}\ \bibnamefont {Philbin}},\ }\bibfield  {title} {\bibinfo {title} {General relativity in electrical engineering},\ }\href {https://doi.org/10.1088/1367-2630/8/10/247} {\bibfield  {journal} {\bibinfo  {journal} {New Journal of Physics}\ }\textbf {\bibinfo {volume} {8}},\ \bibinfo {pages} {247–247} (\bibinfo {year} {2006})}\BibitemShut {NoStop}%
\bibitem [{\citenamefont {Batz}\ and\ \citenamefont {Peschel}(2008)}]{PhysRevA.78.043821}%
  \BibitemOpen
  \bibfield  {author} {\bibinfo {author} {\bibfnamefont {S.}~\bibnamefont {Batz}}\ and\ \bibinfo {author} {\bibfnamefont {U.}~\bibnamefont {Peschel}},\ }\bibfield  {title} {\bibinfo {title} {Linear and nonlinear optics in curved space},\ }\href {https://doi.org/10.1103/PhysRevA.78.043821} {\bibfield  {journal} {\bibinfo  {journal} {Phys. Rev. A}\ }\textbf {\bibinfo {volume} {78}},\ \bibinfo {pages} {043821} (\bibinfo {year} {2008})}\BibitemShut {NoStop}%
\bibitem [{\citenamefont {Smolyaninov}\ and\ \citenamefont {Narimanov}(2010)}]{PhysRevLett.105.067402}%
  \BibitemOpen
  \bibfield  {author} {\bibinfo {author} {\bibfnamefont {I.~I.}\ \bibnamefont {Smolyaninov}}\ and\ \bibinfo {author} {\bibfnamefont {E.~E.}\ \bibnamefont {Narimanov}},\ }\bibfield  {title} {\bibinfo {title} {Metric signature transitions in optical metamaterials},\ }\href {https://doi.org/10.1103/PhysRevLett.105.067402} {\bibfield  {journal} {\bibinfo  {journal} {Phys. Rev. Lett.}\ }\textbf {\bibinfo {volume} {105}},\ \bibinfo {pages} {067402} (\bibinfo {year} {2010})}\BibitemShut {NoStop}%
\bibitem [{\citenamefont {Genov}\ \emph {et~al.}(2009)\citenamefont {Genov}, \citenamefont {Zhang},\ and\ \citenamefont {Zhang}}]{DielectricConstant3}%
  \BibitemOpen
  \bibfield  {author} {\bibinfo {author} {\bibfnamefont {D.}~\bibnamefont {Genov}}, \bibinfo {author} {\bibfnamefont {S.}~\bibnamefont {Zhang}},\ and\ \bibinfo {author} {\bibfnamefont {X.}~\bibnamefont {Zhang}},\ }\bibfield  {title} {\bibinfo {title} {Mimicking celestial mechanics in metamaterials},\ }\href {https://doi.org/10.1038/nphys1338} {\bibfield  {journal} {\bibinfo  {journal} {Nature Physics}\ }\textbf {\bibinfo {volume} {5}} (\bibinfo {year} {2009})}\BibitemShut {NoStop}%
\bibitem [{\citenamefont {Chen}\ \emph {et~al.}(2010)\citenamefont {Chen}, \citenamefont {Miao},\ and\ \citenamefont {Li}}]{Chen_2010}%
  \BibitemOpen
  \bibfield  {author} {\bibinfo {author} {\bibfnamefont {H.}~\bibnamefont {Chen}}, \bibinfo {author} {\bibfnamefont {R.-X.}\ \bibnamefont {Miao}},\ and\ \bibinfo {author} {\bibfnamefont {M.}~\bibnamefont {Li}},\ }\bibfield  {title} {\bibinfo {title} {Transformation optics that mimics the system outside a schwarzschild black hole},\ }\href {https://doi.org/10.1364/oe.18.015183} {\bibfield  {journal} {\bibinfo  {journal} {Optics Express}\ }\textbf {\bibinfo {volume} {18}},\ \bibinfo {pages} {15183} (\bibinfo {year} {2010})}\BibitemShut {NoStop}%
\bibitem [{\citenamefont {Sab{\'\i}n}(2017)}]{Sab_n_2017}%
  \BibitemOpen
  \bibfield  {author} {\bibinfo {author} {\bibfnamefont {C.}~\bibnamefont {Sab{\'\i}n}},\ }\bibfield  {title} {\bibinfo {title} {Mapping curved spacetimes into dirac spinors},\ }\href {https://doi.org/10.1038/srep40346} {\bibfield  {journal} {\bibinfo  {journal} {Scientific Reports}\ }\textbf {\bibinfo {volume} {7}},\ \bibinfo {pages} {40346} (\bibinfo {year} {2017})}\BibitemShut {NoStop}%
\bibitem [{\citenamefont {Pedernales}\ \emph {et~al.}(2018)\citenamefont {Pedernales}, \citenamefont {Beau}, \citenamefont {Pittman}, \citenamefont {Egusquiza}, \citenamefont {Lamata}, \citenamefont {Solano},\ and\ \citenamefont {del Campo}}]{Pedernales_2018}%
  \BibitemOpen
  \bibfield  {author} {\bibinfo {author} {\bibfnamefont {J.~S.}\ \bibnamefont {Pedernales}}, \bibinfo {author} {\bibfnamefont {M.}~\bibnamefont {Beau}}, \bibinfo {author} {\bibfnamefont {S.~M.}\ \bibnamefont {Pittman}}, \bibinfo {author} {\bibfnamefont {I.~L.}\ \bibnamefont {Egusquiza}}, \bibinfo {author} {\bibfnamefont {L.}~\bibnamefont {Lamata}}, \bibinfo {author} {\bibfnamefont {E.}~\bibnamefont {Solano}},\ and\ \bibinfo {author} {\bibfnamefont {A.}~\bibnamefont {del Campo}},\ }\bibfield  {title} {\bibinfo {title} {Dirac equation in ($1+1$)-dimensional curved spacetime and the multiphoton quantum rabi model},\ }\href {https://doi.org/10.1103/PhysRevLett.120.160403} {\bibfield  {journal} {\bibinfo  {journal} {Phys. Rev. Lett.}\ }\textbf {\bibinfo {volume} {120}},\ \bibinfo {pages} {160403} (\bibinfo {year} {2018})}\BibitemShut {NoStop}%
\bibitem [{\citenamefont {Koke}\ \emph {et~al.}(2016)\citenamefont {Koke}, \citenamefont {Noh},\ and\ \citenamefont {Angelakis}}]{Koke_2016}%
  \BibitemOpen
  \bibfield  {author} {\bibinfo {author} {\bibfnamefont {C.}~\bibnamefont {Koke}}, \bibinfo {author} {\bibfnamefont {C.}~\bibnamefont {Noh}},\ and\ \bibinfo {author} {\bibfnamefont {D.~G.}\ \bibnamefont {Angelakis}},\ }\bibfield  {title} {\bibinfo {title} {Dirac equation in 2-dimensional curved spacetime, particle creation, and coupled waveguide arrays},\ }\href {https://doi.org/10.1016/j.aop.2016.08.013} {\bibfield  {journal} {\bibinfo  {journal} {Annals of Physics}\ }\textbf {\bibinfo {volume} {374}},\ \bibinfo {pages} {162–178} (\bibinfo {year} {2016})}\BibitemShut {NoStop}%
\bibitem [{\citenamefont {Boada}\ \emph {et~al.}(2011)\citenamefont {Boada}, \citenamefont {Celi}, \citenamefont {Latorre},\ and\ \citenamefont {Lewenstein}}]{Boada_2011}%
  \BibitemOpen
  \bibfield  {author} {\bibinfo {author} {\bibfnamefont {O.}~\bibnamefont {Boada}}, \bibinfo {author} {\bibfnamefont {A.}~\bibnamefont {Celi}}, \bibinfo {author} {\bibfnamefont {J.~I.}\ \bibnamefont {Latorre}},\ and\ \bibinfo {author} {\bibfnamefont {M.}~\bibnamefont {Lewenstein}},\ }\bibfield  {title} {\bibinfo {title} {Dirac equation for cold atoms in artificial curved spacetimes},\ }\href {https://doi.org/10.1088/1367-2630/13/3/035002} {\bibfield  {journal} {\bibinfo  {journal} {New Journal of Physics}\ }\textbf {\bibinfo {volume} {13}},\ \bibinfo {pages} {035002} (\bibinfo {year} {2011})}\BibitemShut {NoStop}%
\bibitem [{\citenamefont {Zhang}\ \emph {et~al.}(2022{\natexlab{a}})\citenamefont {Zhang}, \citenamefont {Yuan}, \citenamefont {Sun}, \citenamefont {Sun},\ and\ \citenamefont {Zhang}}]{Zhang2022}%
  \BibitemOpen
  \bibfield  {author} {\bibinfo {author} {\bibfnamefont {W.}~\bibnamefont {Zhang}}, \bibinfo {author} {\bibfnamefont {H.}~\bibnamefont {Yuan}}, \bibinfo {author} {\bibfnamefont {N.}~\bibnamefont {Sun}}, \bibinfo {author} {\bibfnamefont {H.}~\bibnamefont {Sun}},\ and\ \bibinfo {author} {\bibfnamefont {X.}~\bibnamefont {Zhang}},\ }\bibfield  {title} {\bibinfo {title} {Observation of novel topological states in hyperbolic lattices},\ }\href {https://doi.org/10.1038/s41467-022-30631-x} {\bibfield  {journal} {\bibinfo  {journal} {Nature Communications}\ }\textbf {\bibinfo {volume} {13}},\ \bibinfo {pages} {2937} (\bibinfo {year} {2022}{\natexlab{a}})}\BibitemShut {NoStop}%
\bibitem [{\citenamefont {Lenggenhager}\ \emph {et~al.}(2022)\citenamefont {Lenggenhager}, \citenamefont {Stegmaier}, \citenamefont {Upreti}, \citenamefont {Hofmann}, \citenamefont {Helbig}, \citenamefont {Vollhardt}, \citenamefont {Greiter}, \citenamefont {Lee}, \citenamefont {Imhof}, \citenamefont {Brand}, \citenamefont {Kie{\ss}ling}, \citenamefont {Boettcher}, \citenamefont {Neupert}, \citenamefont {Thomale},\ and\ \citenamefont {Bzdu{\v{s}}ek}}]{Lenggenhager2022}%
  \BibitemOpen
  \bibfield  {author} {\bibinfo {author} {\bibfnamefont {P.~M.}\ \bibnamefont {Lenggenhager}}, \bibinfo {author} {\bibfnamefont {A.}~\bibnamefont {Stegmaier}}, \bibinfo {author} {\bibfnamefont {L.~K.}\ \bibnamefont {Upreti}}, \bibinfo {author} {\bibfnamefont {T.}~\bibnamefont {Hofmann}}, \bibinfo {author} {\bibfnamefont {T.}~\bibnamefont {Helbig}}, \bibinfo {author} {\bibfnamefont {A.}~\bibnamefont {Vollhardt}}, \bibinfo {author} {\bibfnamefont {M.}~\bibnamefont {Greiter}}, \bibinfo {author} {\bibfnamefont {C.~H.}\ \bibnamefont {Lee}}, \bibinfo {author} {\bibfnamefont {S.}~\bibnamefont {Imhof}}, \bibinfo {author} {\bibfnamefont {H.}~\bibnamefont {Brand}}, \bibinfo {author} {\bibfnamefont {T.}~\bibnamefont {Kie{\ss}ling}}, \bibinfo {author} {\bibfnamefont {I.}~\bibnamefont {Boettcher}}, \bibinfo {author} {\bibfnamefont {T.}~\bibnamefont {Neupert}}, \bibinfo {author} {\bibfnamefont {R.}~\bibnamefont {Thomale}},\ and\ \bibinfo {author} {\bibfnamefont {T.}~\bibnamefont {Bzdu{\v{s}}ek}},\ }\bibfield  {title}
  {\bibinfo {title} {Simulating hyperbolic space on a circuit board},\ }\href {https://doi.org/10.1038/s41467-022-32042-4} {\bibfield  {journal} {\bibinfo  {journal} {Nature Communications}\ }\textbf {\bibinfo {volume} {13}},\ \bibinfo {pages} {4373} (\bibinfo {year} {2022})}\BibitemShut {NoStop}%
\bibitem [{\citenamefont {Zhang}\ \emph {et~al.}(2023)\citenamefont {Zhang}, \citenamefont {Di}, \citenamefont {Zheng}, \citenamefont {Sun},\ and\ \citenamefont {Zhang}}]{Zhang2023natcomm}%
  \BibitemOpen
  \bibfield  {author} {\bibinfo {author} {\bibfnamefont {W.}~\bibnamefont {Zhang}}, \bibinfo {author} {\bibfnamefont {F.}~\bibnamefont {Di}}, \bibinfo {author} {\bibfnamefont {X.}~\bibnamefont {Zheng}}, \bibinfo {author} {\bibfnamefont {H.}~\bibnamefont {Sun}},\ and\ \bibinfo {author} {\bibfnamefont {X.}~\bibnamefont {Zhang}},\ }\bibfield  {title} {\bibinfo {title} {Hyperbolic band topology with non-trivial second chern numbers},\ }\href {https://doi.org/10.1038/s41467-023-36767-8} {\bibfield  {journal} {\bibinfo  {journal} {Nature Communications}\ }\textbf {\bibinfo {volume} {14}},\ \bibinfo {pages} {1083} (\bibinfo {year} {2023})}\BibitemShut {NoStop}%
\bibitem [{\citenamefont {Chen}\ \emph {et~al.}(2023)\citenamefont {Chen}, \citenamefont {Brand}, \citenamefont {Helbig}, \citenamefont {Hofmann}, \citenamefont {Imhof}, \citenamefont {Fritzsche}, \citenamefont {Kie{\ss}ling}, \citenamefont {Stegmaier}, \citenamefont {Upreti}, \citenamefont {Neupert}, \citenamefont {Bzdu{\v{s}}ek}, \citenamefont {Greiter}, \citenamefont {Thomale},\ and\ \citenamefont {Boettcher}}]{Chen2023}%
  \BibitemOpen
  \bibfield  {author} {\bibinfo {author} {\bibfnamefont {A.}~\bibnamefont {Chen}}, \bibinfo {author} {\bibfnamefont {H.}~\bibnamefont {Brand}}, \bibinfo {author} {\bibfnamefont {T.}~\bibnamefont {Helbig}}, \bibinfo {author} {\bibfnamefont {T.}~\bibnamefont {Hofmann}}, \bibinfo {author} {\bibfnamefont {S.}~\bibnamefont {Imhof}}, \bibinfo {author} {\bibfnamefont {A.}~\bibnamefont {Fritzsche}}, \bibinfo {author} {\bibfnamefont {T.}~\bibnamefont {Kie{\ss}ling}}, \bibinfo {author} {\bibfnamefont {A.}~\bibnamefont {Stegmaier}}, \bibinfo {author} {\bibfnamefont {L.~K.}\ \bibnamefont {Upreti}}, \bibinfo {author} {\bibfnamefont {T.}~\bibnamefont {Neupert}}, \bibinfo {author} {\bibfnamefont {T.}~\bibnamefont {Bzdu{\v{s}}ek}}, \bibinfo {author} {\bibfnamefont {M.}~\bibnamefont {Greiter}}, \bibinfo {author} {\bibfnamefont {R.}~\bibnamefont {Thomale}},\ and\ \bibinfo {author} {\bibfnamefont {I.}~\bibnamefont {Boettcher}},\ }\bibfield  {title} {\bibinfo {title} {Hyperbolic matter in electrical circuits with tunable complex
  phases},\ }\href {https://doi.org/10.1038/s41467-023-36359-6} {\bibfield  {journal} {\bibinfo  {journal} {Nature Communications}\ }\textbf {\bibinfo {volume} {14}},\ \bibinfo {pages} {622} (\bibinfo {year} {2023})}\BibitemShut {NoStop}%
\bibitem [{\citenamefont {Maciejko}\ and\ \citenamefont {Rayan}(2021)}]{Maciejko2021}%
  \BibitemOpen
  \bibfield  {author} {\bibinfo {author} {\bibfnamefont {J.}~\bibnamefont {Maciejko}}\ and\ \bibinfo {author} {\bibfnamefont {S.}~\bibnamefont {Rayan}},\ }\bibfield  {title} {\bibinfo {title} {Hyperbolic band theory},\ }\href {https://doi.org/10.1126/sciadv.abe9170} {\bibfield  {journal} {\bibinfo  {journal} {Sci. Adv.}\ }\textbf {\bibinfo {volume} {7}},\ \bibinfo {pages} {eabe9170} (\bibinfo {year} {2021})}\BibitemShut {NoStop}%
\bibitem [{\citenamefont {Cheng}\ \emph {et~al.}(2022)\citenamefont {Cheng}, \citenamefont {Serafin}, \citenamefont {McInerney}, \citenamefont {Rocklin}, \citenamefont {Sun},\ and\ \citenamefont {Mao}}]{Cheng2022}%
  \BibitemOpen
  \bibfield  {author} {\bibinfo {author} {\bibfnamefont {N.}~\bibnamefont {Cheng}}, \bibinfo {author} {\bibfnamefont {F.}~\bibnamefont {Serafin}}, \bibinfo {author} {\bibfnamefont {J.}~\bibnamefont {McInerney}}, \bibinfo {author} {\bibfnamefont {Z.}~\bibnamefont {Rocklin}}, \bibinfo {author} {\bibfnamefont {K.}~\bibnamefont {Sun}},\ and\ \bibinfo {author} {\bibfnamefont {X.}~\bibnamefont {Mao}},\ }\bibfield  {title} {\bibinfo {title} {Band theory and boundary modes of high-dimensional representations of infinite hyperbolic lattices},\ }\href {https://doi.org/10.1103/PhysRevLett.129.088002} {\bibfield  {journal} {\bibinfo  {journal} {Phys. Rev. Lett.}\ }\textbf {\bibinfo {volume} {129}},\ \bibinfo {pages} {088002} (\bibinfo {year} {2022})}\BibitemShut {NoStop}%
\bibitem [{\citenamefont {Kienzle}\ and\ \citenamefont {Rayan}(2022)}]{Kienzle2022}%
  \BibitemOpen
  \bibfield  {author} {\bibinfo {author} {\bibfnamefont {E.}~\bibnamefont {Kienzle}}\ and\ \bibinfo {author} {\bibfnamefont {S.}~\bibnamefont {Rayan}},\ }\bibfield  {title} {\bibinfo {title} {Hyperbolic band theory through higgs bundles},\ }\href {https://doi.org/10.1016/j.aim.2022.108664} {\bibfield  {journal} {\bibinfo  {journal} {Adv. Math.}\ }\textbf {\bibinfo {volume} {409}},\ \bibinfo {pages} {108664} (\bibinfo {year} {2022})}\BibitemShut {NoStop}%
\bibitem [{\citenamefont {Maciejko}\ and\ \citenamefont {Rayan}(2022)}]{Maciejko2022}%
  \BibitemOpen
  \bibfield  {author} {\bibinfo {author} {\bibfnamefont {J.}~\bibnamefont {Maciejko}}\ and\ \bibinfo {author} {\bibfnamefont {S.}~\bibnamefont {Rayan}},\ }\bibfield  {title} {\bibinfo {title} {Automorphic bloch theorems for hyperbolic lattices},\ }\href {https://doi.org/10.1073/pnas.2116869119} {\bibfield  {journal} {\bibinfo  {journal} {Proc. Natl. Acad. Sci. USA}\ }\textbf {\bibinfo {volume} {119}},\ \bibinfo {pages} {e2116869119} (\bibinfo {year} {2022})}\BibitemShut {NoStop}%
\bibitem [{\citenamefont {Attar}\ and\ \citenamefont {Boettcher}(2022)}]{Attar2022}%
  \BibitemOpen
  \bibfield  {author} {\bibinfo {author} {\bibfnamefont {A.}~\bibnamefont {Attar}}\ and\ \bibinfo {author} {\bibfnamefont {I.}~\bibnamefont {Boettcher}},\ }\bibfield  {title} {\bibinfo {title} {Selberg trace formula in hyperbolic band theory},\ }\href {https://doi.org/10.1103/PhysRevE.106.034114} {\bibfield  {journal} {\bibinfo  {journal} {Phys. Rev. E}\ }\textbf {\bibinfo {volume} {106}},\ \bibinfo {pages} {034114} (\bibinfo {year} {2022})}\BibitemShut {NoStop}%
\bibitem [{\citenamefont {Bzdušek}\ and\ \citenamefont {Maciejko}(2022)}]{Bzdusek2022}%
  \BibitemOpen
  \bibfield  {author} {\bibinfo {author} {\bibfnamefont {T.}~\bibnamefont {Bzdušek}}\ and\ \bibinfo {author} {\bibfnamefont {J.}~\bibnamefont {Maciejko}},\ }\bibfield  {title} {\bibinfo {title} {Flat bands and band-touching from real-space topology in hyperbolic lattices},\ }\href {https://doi.org/10.1103/PhysRevB.106.155146} {\bibfield  {journal} {\bibinfo  {journal} {Phys. Rev. B}\ }\textbf {\bibinfo {volume} {106}},\ \bibinfo {pages} {155146} (\bibinfo {year} {2022})}\BibitemShut {NoStop}%
\bibitem [{\citenamefont {Yu}\ \emph {et~al.}(2020)\citenamefont {Yu}, \citenamefont {Piao},\ and\ \citenamefont {Park}}]{THL-PRL2020}%
  \BibitemOpen
  \bibfield  {author} {\bibinfo {author} {\bibfnamefont {S.}~\bibnamefont {Yu}}, \bibinfo {author} {\bibfnamefont {X.}~\bibnamefont {Piao}},\ and\ \bibinfo {author} {\bibfnamefont {N.}~\bibnamefont {Park}},\ }\bibfield  {title} {\bibinfo {title} {Topological hyperbolic lattices},\ }\href {https://doi.org/10.1103/PhysRevLett.125.053901} {\bibfield  {journal} {\bibinfo  {journal} {Phys. Rev. Lett.}\ }\textbf {\bibinfo {volume} {125}},\ \bibinfo {pages} {053901} (\bibinfo {year} {2020})}\BibitemShut {NoStop}%
\bibitem [{\citenamefont {Urwyler}\ \emph {et~al.}(2022)\citenamefont {Urwyler}, \citenamefont {Lenggenhager}, \citenamefont {Boettcher}, \citenamefont {Thomale}, \citenamefont {Neupert},\ and\ \citenamefont {Bzdu\ifmmode~\check{s}\else \v{s}\fi{}ek}}]{HPTB-PRL2022}%
  \BibitemOpen
  \bibfield  {author} {\bibinfo {author} {\bibfnamefont {D.~M.}\ \bibnamefont {Urwyler}}, \bibinfo {author} {\bibfnamefont {P.~M.}\ \bibnamefont {Lenggenhager}}, \bibinfo {author} {\bibfnamefont {I.}~\bibnamefont {Boettcher}}, \bibinfo {author} {\bibfnamefont {R.}~\bibnamefont {Thomale}}, \bibinfo {author} {\bibfnamefont {T.}~\bibnamefont {Neupert}},\ and\ \bibinfo {author} {\bibfnamefont {T.~c.~v.}\ \bibnamefont {Bzdu\ifmmode~\check{s}\else \v{s}\fi{}ek}},\ }\bibfield  {title} {\bibinfo {title} {Hyperbolic topological band insulators},\ }\href {https://doi.org/10.1103/PhysRevLett.129.246402} {\bibfield  {journal} {\bibinfo  {journal} {Phys. Rev. Lett.}\ }\textbf {\bibinfo {volume} {129}},\ \bibinfo {pages} {246402} (\bibinfo {year} {2022})}\BibitemShut {NoStop}%
\bibitem [{\citenamefont {Liu}\ \emph {et~al.}(2022)\citenamefont {Liu}, \citenamefont {Hua}, \citenamefont {Peng},\ and\ \citenamefont {Zhou}}]{CIHL-PRB2022}%
  \BibitemOpen
  \bibfield  {author} {\bibinfo {author} {\bibfnamefont {Z.-R.}\ \bibnamefont {Liu}}, \bibinfo {author} {\bibfnamefont {C.-B.}\ \bibnamefont {Hua}}, \bibinfo {author} {\bibfnamefont {T.}~\bibnamefont {Peng}},\ and\ \bibinfo {author} {\bibfnamefont {B.}~\bibnamefont {Zhou}},\ }\bibfield  {title} {\bibinfo {title} {Chern insulator in a hyperbolic lattice},\ }\href {https://doi.org/10.1103/PhysRevB.105.245301} {\bibfield  {journal} {\bibinfo  {journal} {Phys. Rev. B}\ }\textbf {\bibinfo {volume} {105}},\ \bibinfo {pages} {245301} (\bibinfo {year} {2022})}\BibitemShut {NoStop}%
\bibitem [{\citenamefont {Zhang}\ \emph {et~al.}(2022{\natexlab{b}})\citenamefont {Zhang}, \citenamefont {Yuan}, \citenamefont {Sun}, \citenamefont {Sun},\ and\ \citenamefont {Zhang}}]{Zhang2022NatComm}%
  \BibitemOpen
  \bibfield  {author} {\bibinfo {author} {\bibfnamefont {W.}~\bibnamefont {Zhang}}, \bibinfo {author} {\bibfnamefont {H.}~\bibnamefont {Yuan}}, \bibinfo {author} {\bibfnamefont {N.}~\bibnamefont {Sun}}, \bibinfo {author} {\bibfnamefont {H.}~\bibnamefont {Sun}},\ and\ \bibinfo {author} {\bibfnamefont {X.}~\bibnamefont {Zhang}},\ }\bibfield  {title} {\bibinfo {title} {Observation of novel topological states in hyperbolic lattices},\ }\href {https://doi.org/10.1038/s41467-022-30631-x} {\bibfield  {journal} {\bibinfo  {journal} {Nature Communications}\ }\textbf {\bibinfo {volume} {13}},\ \bibinfo {pages} {2937} (\bibinfo {year} {2022}{\natexlab{b}})}\BibitemShut {NoStop}%
\bibitem [{\citenamefont {Huang}\ \emph {et~al.}(2024)\citenamefont {Huang}, \citenamefont {He}, \citenamefont {Zhang}, \citenamefont {Zhang}, \citenamefont {Liu}, \citenamefont {Feng}, \citenamefont {Liu}, \citenamefont {Cui}, \citenamefont {Huang}, \citenamefont {Zhang},\ and\ \citenamefont {Zhang}}]{Huang2024-NatComm}%
  \BibitemOpen
  \bibfield  {author} {\bibinfo {author} {\bibfnamefont {L.}~\bibnamefont {Huang}}, \bibinfo {author} {\bibfnamefont {L.}~\bibnamefont {He}}, \bibinfo {author} {\bibfnamefont {W.}~\bibnamefont {Zhang}}, \bibinfo {author} {\bibfnamefont {H.}~\bibnamefont {Zhang}}, \bibinfo {author} {\bibfnamefont {D.}~\bibnamefont {Liu}}, \bibinfo {author} {\bibfnamefont {X.}~\bibnamefont {Feng}}, \bibinfo {author} {\bibfnamefont {F.}~\bibnamefont {Liu}}, \bibinfo {author} {\bibfnamefont {K.}~\bibnamefont {Cui}}, \bibinfo {author} {\bibfnamefont {Y.}~\bibnamefont {Huang}}, \bibinfo {author} {\bibfnamefont {W.}~\bibnamefont {Zhang}},\ and\ \bibinfo {author} {\bibfnamefont {X.}~\bibnamefont {Zhang}},\ }\bibfield  {title} {\bibinfo {title} {Hyperbolic photonic topological insulators},\ }\href {https://doi.org/10.1038/s41467-024-46035-y} {\bibfield  {journal} {\bibinfo  {journal} {Nature Communications}\ }\textbf {\bibinfo {volume} {15}},\ \bibinfo {pages} {1647} (\bibinfo {year} {2024})}\BibitemShut {NoStop}%
\bibitem [{\citenamefont {Chen}\ \emph {et~al.}(2024)\citenamefont {Chen}, \citenamefont {Zhang}, \citenamefont {Qin}, \citenamefont {Bossart}, \citenamefont {Yang}, \citenamefont {Chen},\ and\ \citenamefont {Fleury}}]{Chen2024}%
  \BibitemOpen
  \bibfield  {author} {\bibinfo {author} {\bibfnamefont {Q.}~\bibnamefont {Chen}}, \bibinfo {author} {\bibfnamefont {Z.}~\bibnamefont {Zhang}}, \bibinfo {author} {\bibfnamefont {H.}~\bibnamefont {Qin}}, \bibinfo {author} {\bibfnamefont {A.}~\bibnamefont {Bossart}}, \bibinfo {author} {\bibfnamefont {Y.}~\bibnamefont {Yang}}, \bibinfo {author} {\bibfnamefont {H.}~\bibnamefont {Chen}},\ and\ \bibinfo {author} {\bibfnamefont {R.}~\bibnamefont {Fleury}},\ }\bibfield  {title} {\bibinfo {title} {Anomalous and chern topological waves in hyperbolic networks},\ }\href {https://doi.org/10.1038/s41467-024-46551-x} {\bibfield  {journal} {\bibinfo  {journal} {Nature Communications}\ }\textbf {\bibinfo {volume} {15}},\ \bibinfo {pages} {2293} (\bibinfo {year} {2024})}\BibitemShut {NoStop}%
\bibitem [{\citenamefont {Sun}\ \emph {et~al.}(2024)\citenamefont {Sun}, \citenamefont {Chen}, \citenamefont {Bzdušek},\ and\ \citenamefont {Maciejko}}]{Sun-SciPost2024}%
  \BibitemOpen
  \bibfield  {author} {\bibinfo {author} {\bibfnamefont {C.}~\bibnamefont {Sun}}, \bibinfo {author} {\bibfnamefont {A.}~\bibnamefont {Chen}}, \bibinfo {author} {\bibfnamefont {T.}~\bibnamefont {Bzdušek}},\ and\ \bibinfo {author} {\bibfnamefont {J.}~\bibnamefont {Maciejko}},\ }\bibfield  {title} {\bibinfo {title} {{Topological linear response of hyperbolic Chern insulators}},\ }\href {https://doi.org/10.21468/SciPostPhys.17.5.124} {\bibfield  {journal} {\bibinfo  {journal} {SciPost Phys.}\ }\textbf {\bibinfo {volume} {17}},\ \bibinfo {pages} {124} (\bibinfo {year} {2024})}\BibitemShut {NoStop}%
\bibitem [{\citenamefont {Tao}\ and\ \citenamefont {Xu}(2023)}]{HOTopology2}%
  \BibitemOpen
  \bibfield  {author} {\bibinfo {author} {\bibfnamefont {Y.-L.}\ \bibnamefont {Tao}}\ and\ \bibinfo {author} {\bibfnamefont {Y.}~\bibnamefont {Xu}},\ }\bibfield  {title} {\bibinfo {title} {Higher-order topological hyperbolic lattices},\ }\href {https://doi.org/10.1103/PhysRevB.107.184201} {\bibfield  {journal} {\bibinfo  {journal} {Phys. Rev. B}\ }\textbf {\bibinfo {volume} {107}},\ \bibinfo {pages} {184201} (\bibinfo {year} {2023})}\BibitemShut {NoStop}%
\bibitem [{\citenamefont {Sun}\ \emph {et~al.}(2023)\citenamefont {Sun}, \citenamefont {Li}, \citenamefont {Feng},\ and\ \citenamefont {Guo}}]{HOTopology3}%
  \BibitemOpen
  \bibfield  {author} {\bibinfo {author} {\bibfnamefont {J.}~\bibnamefont {Sun}}, \bibinfo {author} {\bibfnamefont {C.-A.}\ \bibnamefont {Li}}, \bibinfo {author} {\bibfnamefont {S.}~\bibnamefont {Feng}},\ and\ \bibinfo {author} {\bibfnamefont {H.}~\bibnamefont {Guo}},\ }\bibfield  {title} {\bibinfo {title} {Hybrid higher-order skin-topological effect in hyperbolic lattices},\ }\href {https://doi.org/10.1103/PhysRevB.108.075122} {\bibfield  {journal} {\bibinfo  {journal} {Phys. Rev. B}\ }\textbf {\bibinfo {volume} {108}},\ \bibinfo {pages} {075122} (\bibinfo {year} {2023})}\BibitemShut {NoStop}%
\bibitem [{\citenamefont {Tummuru}\ \emph {et~al.}(2024)\citenamefont {Tummuru}, \citenamefont {Chen}, \citenamefont {Lenggenhager}, \citenamefont {Neupert}, \citenamefont {Maciejko},\ and\ \citenamefont {Bzdu\v{s}ek}}]{Tummuru:2023wuq}%
  \BibitemOpen
  \bibfield  {author} {\bibinfo {author} {\bibfnamefont {T.}~\bibnamefont {Tummuru}}, \bibinfo {author} {\bibfnamefont {A.}~\bibnamefont {Chen}}, \bibinfo {author} {\bibfnamefont {P.~M.}\ \bibnamefont {Lenggenhager}}, \bibinfo {author} {\bibfnamefont {T.}~\bibnamefont {Neupert}}, \bibinfo {author} {\bibfnamefont {J.}~\bibnamefont {Maciejko}},\ and\ \bibinfo {author} {\bibfnamefont {T.}~\bibnamefont {Bzdu\v{s}ek}},\ }\bibfield  {title} {\bibinfo {title} {{Hyperbolic Non-Abelian Semimetal}},\ }\href {https://doi.org/10.1103/PhysRevLett.132.206601} {\bibfield  {journal} {\bibinfo  {journal} {Phys. Rev. Lett.}\ }\textbf {\bibinfo {volume} {132}},\ \bibinfo {pages} {206601} (\bibinfo {year} {2024})}\BibitemShut {NoStop}%
\bibitem [{\citenamefont {Roy}(2024)}]{Roy_2024}%
  \BibitemOpen
  \bibfield  {author} {\bibinfo {author} {\bibfnamefont {B.}~\bibnamefont {Roy}},\ }\bibfield  {title} {\bibinfo {title} {Magnetic catalysis in weakly interacting hyperbolic dirac materials},\ }\href {https://doi.org/10.1103/PhysRevB.110.245117} {\bibfield  {journal} {\bibinfo  {journal} {Phys. Rev. B}\ }\textbf {\bibinfo {volume} {110}},\ \bibinfo {pages} {245117} (\bibinfo {year} {2024})}\BibitemShut {NoStop}%
\bibitem [{\citenamefont {Ikeda}\ \emph {et~al.}(2021)\citenamefont {Ikeda}, \citenamefont {Aoki},\ and\ \citenamefont {Matsuki}}]{Hofstadter1}%
  \BibitemOpen
  \bibfield  {author} {\bibinfo {author} {\bibfnamefont {K.}~\bibnamefont {Ikeda}}, \bibinfo {author} {\bibfnamefont {S.}~\bibnamefont {Aoki}},\ and\ \bibinfo {author} {\bibfnamefont {Y.}~\bibnamefont {Matsuki}},\ }\bibfield  {title} {\bibinfo {title} {Hyperbolic band theory under magnetic field and dirac cones on a higher genus surface},\ }\href {https://doi.org/10.1088/1361-648x/ac24c4} {\bibfield  {journal} {\bibinfo  {journal} {Journal of Physics: Condensed Matter}\ }\textbf {\bibinfo {volume} {33}},\ \bibinfo {pages} {485602} (\bibinfo {year} {2021})}\BibitemShut {NoStop}%
\bibitem [{\citenamefont {Stegmaier}\ \emph {et~al.}(2022)\citenamefont {Stegmaier}, \citenamefont {Upreti}, \citenamefont {Thomale},\ and\ \citenamefont {Boettcher}}]{Hofstadter2}%
  \BibitemOpen
  \bibfield  {author} {\bibinfo {author} {\bibfnamefont {A.}~\bibnamefont {Stegmaier}}, \bibinfo {author} {\bibfnamefont {L.~K.}\ \bibnamefont {Upreti}}, \bibinfo {author} {\bibfnamefont {R.}~\bibnamefont {Thomale}},\ and\ \bibinfo {author} {\bibfnamefont {I.}~\bibnamefont {Boettcher}},\ }\bibfield  {title} {\bibinfo {title} {Universality of hofstadter butterflies on hyperbolic lattices},\ }\bibfield  {journal} {\bibinfo  {journal} {Physical Review Letters}\ }\textbf {\bibinfo {volume} {128}},\ \href {https://doi.org/10.1103/physrevlett.128.166402} {10.1103/physrevlett.128.166402} (\bibinfo {year} {2022})\BibitemShut {NoStop}%
\bibitem [{\citenamefont {Koll\'{a}r}\ \emph {et~al.}(2019{\natexlab{b}})\citenamefont {Koll\'{a}r}, \citenamefont {Fitzpatrick}, \citenamefont {Sarnak},\ and\ \citenamefont {Houck}}]{FlatBands1}%
  \BibitemOpen
  \bibfield  {author} {\bibinfo {author} {\bibfnamefont {A.~J.}\ \bibnamefont {Koll\'{a}r}}, \bibinfo {author} {\bibfnamefont {M.}~\bibnamefont {Fitzpatrick}}, \bibinfo {author} {\bibfnamefont {P.}~\bibnamefont {Sarnak}},\ and\ \bibinfo {author} {\bibfnamefont {A.~A.}\ \bibnamefont {Houck}},\ }\bibfield  {title} {\bibinfo {title} {Line-graph lattices: Euclidean and non-euclidean flat bands, and implementations in circuit quantum electrodynamics},\ }\href {https://doi.org/10.1007/s00220-019-03645-8} {\bibfield  {journal} {\bibinfo  {journal} {Communications in Mathematical Physics}\ }\textbf {\bibinfo {volume} {376}},\ \bibinfo {pages} {1909–1956} (\bibinfo {year} {2019}{\natexlab{b}})}\BibitemShut {NoStop}%
\bibitem [{\citenamefont {Saa}\ \emph {et~al.}(2021)\citenamefont {Saa}, \citenamefont {Miranda},\ and\ \citenamefont {Rouxinol}}]{FlatBands2}%
  \BibitemOpen
  \bibfield  {author} {\bibinfo {author} {\bibfnamefont {A.}~\bibnamefont {Saa}}, \bibinfo {author} {\bibfnamefont {E.}~\bibnamefont {Miranda}},\ and\ \bibinfo {author} {\bibfnamefont {F.}~\bibnamefont {Rouxinol}},\ }\href {https://arxiv.org/abs/2108.08854} {\bibinfo {title} {Higher-dimensional euclidean and non-euclidean structures in planar circuit quantum electrodynamics}} (\bibinfo {year} {2021}),\ \Eprint {https://arxiv.org/abs/2108.08854} {arXiv:2108.08854 [quant-ph]} \BibitemShut {NoStop}%
\bibitem [{\citenamefont {Bzdu\ifmmode~\check{s}\else \v{s}\fi{}ek}\ and\ \citenamefont {Maciejko}(2022)}]{FlatBands3}%
  \BibitemOpen
  \bibfield  {author} {\bibinfo {author} {\bibfnamefont {T.~c.~v.}\ \bibnamefont {Bzdu\ifmmode~\check{s}\else \v{s}\fi{}ek}}\ and\ \bibinfo {author} {\bibfnamefont {J.}~\bibnamefont {Maciejko}},\ }\bibfield  {title} {\bibinfo {title} {Flat bands and band-touching from real-space topology in hyperbolic lattices},\ }\href {https://doi.org/10.1103/PhysRevB.106.155146} {\bibfield  {journal} {\bibinfo  {journal} {Phys. Rev. B}\ }\textbf {\bibinfo {volume} {106}},\ \bibinfo {pages} {155146} (\bibinfo {year} {2022})}\BibitemShut {NoStop}%
\bibitem [{\citenamefont {Mosseri}\ \emph {et~al.}(2022)\citenamefont {Mosseri}, \citenamefont {Vogeler},\ and\ \citenamefont {Vidal}}]{FlatBands4}%
  \BibitemOpen
  \bibfield  {author} {\bibinfo {author} {\bibfnamefont {R.}~\bibnamefont {Mosseri}}, \bibinfo {author} {\bibfnamefont {R.}~\bibnamefont {Vogeler}},\ and\ \bibinfo {author} {\bibfnamefont {J.}~\bibnamefont {Vidal}},\ }\bibfield  {title} {\bibinfo {title} {Aharonov-bohm cages, flat bands, and gap labeling in hyperbolic tilings},\ }\href {https://doi.org/10.1103/PhysRevB.106.155120} {\bibfield  {journal} {\bibinfo  {journal} {Phys. Rev. B}\ }\textbf {\bibinfo {volume} {106}},\ \bibinfo {pages} {155120} (\bibinfo {year} {2022})}\BibitemShut {NoStop}%
\bibitem [{\citenamefont {Zhu}\ \emph {et~al.}(2021)\citenamefont {Zhu}, \citenamefont {Guo}, \citenamefont {Breuckmann}, \citenamefont {Guo},\ and\ \citenamefont {Feng}}]{StrongCorrelations1}%
  \BibitemOpen
  \bibfield  {author} {\bibinfo {author} {\bibfnamefont {X.}~\bibnamefont {Zhu}}, \bibinfo {author} {\bibfnamefont {J.}~\bibnamefont {Guo}}, \bibinfo {author} {\bibfnamefont {N.~P.}\ \bibnamefont {Breuckmann}}, \bibinfo {author} {\bibfnamefont {H.}~\bibnamefont {Guo}},\ and\ \bibinfo {author} {\bibfnamefont {S.}~\bibnamefont {Feng}},\ }\bibfield  {title} {\bibinfo {title} {Quantum phase transitions of interacting bosons on hyperbolic lattices},\ }\href {https://doi.org/10.1088/1361-648x/ac0a1a} {\bibfield  {journal} {\bibinfo  {journal} {Journal of Physics: Condensed Matter}\ }\textbf {\bibinfo {volume} {33}},\ \bibinfo {pages} {335602} (\bibinfo {year} {2021})}\BibitemShut {NoStop}%
\bibitem [{\citenamefont {Bienias}\ \emph {et~al.}(2022)\citenamefont {Bienias}, \citenamefont {Boettcher}, \citenamefont {Belyansky}, \citenamefont {Koll\'{a}r},\ and\ \citenamefont {Gorshkov}}]{StrongCorrelations2}%
  \BibitemOpen
  \bibfield  {author} {\bibinfo {author} {\bibfnamefont {P.}~\bibnamefont {Bienias}}, \bibinfo {author} {\bibfnamefont {I.}~\bibnamefont {Boettcher}}, \bibinfo {author} {\bibfnamefont {R.}~\bibnamefont {Belyansky}}, \bibinfo {author} {\bibfnamefont {A.~J.}\ \bibnamefont {Koll\'{a}r}},\ and\ \bibinfo {author} {\bibfnamefont {A.~V.}\ \bibnamefont {Gorshkov}},\ }\bibfield  {title} {\bibinfo {title} {Circuit quantum electrodynamics in hyperbolic space: From photon bound states to frustrated spin models},\ }\href {https://doi.org/10.1103/PhysRevLett.128.013601} {\bibfield  {journal} {\bibinfo  {journal} {Phys. Rev. Lett.}\ }\textbf {\bibinfo {volume} {128}},\ \bibinfo {pages} {013601} (\bibinfo {year} {2022})}\BibitemShut {NoStop}%
\bibitem [{\citenamefont {Gluscevich}\ \emph {et~al.}(2025)\citenamefont {Gluscevich}, \citenamefont {Samanta}, \citenamefont {Manna},\ and\ \citenamefont {Roy}}]{StrongCorrelations3}%
  \BibitemOpen
  \bibfield  {author} {\bibinfo {author} {\bibfnamefont {N.}~\bibnamefont {Gluscevich}}, \bibinfo {author} {\bibfnamefont {A.}~\bibnamefont {Samanta}}, \bibinfo {author} {\bibfnamefont {S.}~\bibnamefont {Manna}},\ and\ \bibinfo {author} {\bibfnamefont {B.}~\bibnamefont {Roy}},\ }\bibfield  {title} {\bibinfo {title} {Dynamic mass generation on two-dimensional electronic hyperbolic lattices},\ }\href {https://doi.org/10.1103/PhysRevB.111.L121108} {\bibfield  {journal} {\bibinfo  {journal} {Phys. Rev. B}\ }\textbf {\bibinfo {volume} {111}},\ \bibinfo {pages} {L121108} (\bibinfo {year} {2025})}\BibitemShut {NoStop}%
\bibitem [{\citenamefont {Leong}\ and\ \citenamefont {Roy}(2026)}]{Leong-Roy-PRB2026}%
  \BibitemOpen
  \bibfield  {author} {\bibinfo {author} {\bibfnamefont {C.~A.}\ \bibnamefont {Leong}}\ and\ \bibinfo {author} {\bibfnamefont {B.}~\bibnamefont {Roy}},\ }\bibfield  {title} {\bibinfo {title} {Non-hermitian catalysis of spontaneous symmetry breaking on euclidean and hyperbolic lattices},\ }\href {https://doi.org/10.1103/3l8x-ffmn} {\bibfield  {journal} {\bibinfo  {journal} {Phys. Rev. B}\ }\textbf {\bibinfo {volume} {113}},\ \bibinfo {pages} {155152} (\bibinfo {year} {2026})}\BibitemShut {NoStop}%
\bibitem [{\citenamefont {Lv}\ \emph {et~al.}(2022)\citenamefont {Lv}, \citenamefont {Zhang}, \citenamefont {Zhai},\ and\ \citenamefont {Zhou}}]{Lv2022}%
  \BibitemOpen
  \bibfield  {author} {\bibinfo {author} {\bibfnamefont {C.}~\bibnamefont {Lv}}, \bibinfo {author} {\bibfnamefont {R.}~\bibnamefont {Zhang}}, \bibinfo {author} {\bibfnamefont {Z.}~\bibnamefont {Zhai}},\ and\ \bibinfo {author} {\bibfnamefont {Q.}~\bibnamefont {Zhou}},\ }\bibfield  {title} {\bibinfo {title} {Curving the space by non-hermiticity},\ }\href {https://doi.org/10.1038/s41467-022-29774-8} {\bibfield  {journal} {\bibinfo  {journal} {Nature Communications}\ }\textbf {\bibinfo {volume} {13}},\ \bibinfo {pages} {2184} (\bibinfo {year} {2022})}\BibitemShut {NoStop}%
\bibitem [{\citenamefont {Hu}\ \emph {et~al.}(2024)\citenamefont {Hu}, \citenamefont {Lin},\ and\ \citenamefont {Ding}}]{hu2024-NHhyper}%
  \BibitemOpen
  \bibfield  {author} {\bibinfo {author} {\bibfnamefont {M.}~\bibnamefont {Hu}}, \bibinfo {author} {\bibfnamefont {J.}~\bibnamefont {Lin}},\ and\ \bibinfo {author} {\bibfnamefont {K.}~\bibnamefont {Ding}},\ }\href {https://arxiv.org/abs/2412.05607} {\bibinfo {title} {Unveiling non-hermitian spectral topology in hyperbolic lattices with non-abelian translation symmetry}} (\bibinfo {year} {2024}),\ \Eprint {https://arxiv.org/abs/2412.05607} {arXiv:2412.05607 [cond-mat.mes-hall]} \BibitemShut {NoStop}%
\bibitem [{\citenamefont {Shen}\ \emph {et~al.}(2025)\citenamefont {Shen}, \citenamefont {Chan},\ and\ \citenamefont {Lee}}]{Shen-PRB2025}%
  \BibitemOpen
  \bibfield  {author} {\bibinfo {author} {\bibfnamefont {R.}~\bibnamefont {Shen}}, \bibinfo {author} {\bibfnamefont {W.~J.}\ \bibnamefont {Chan}},\ and\ \bibinfo {author} {\bibfnamefont {C.~H.}\ \bibnamefont {Lee}},\ }\bibfield  {title} {\bibinfo {title} {Non-hermitian skin effect along hyperbolic geodesics},\ }\href {https://doi.org/10.1103/PhysRevB.111.045420} {\bibfield  {journal} {\bibinfo  {journal} {Phys. Rev. B}\ }\textbf {\bibinfo {volume} {111}},\ \bibinfo {pages} {045420} (\bibinfo {year} {2025})}\BibitemShut {NoStop}%
\bibitem [{\citenamefont {Yan}(2019)}]{Fractons1}%
  \BibitemOpen
  \bibfield  {author} {\bibinfo {author} {\bibfnamefont {H.}~\bibnamefont {Yan}},\ }\bibfield  {title} {\bibinfo {title} {Hyperbolic fracton model, subsystem symmetry, and holography},\ }\href {https://doi.org/10.1103/PhysRevB.99.155126} {\bibfield  {journal} {\bibinfo  {journal} {Phys. Rev. B}\ }\textbf {\bibinfo {volume} {99}},\ \bibinfo {pages} {155126} (\bibinfo {year} {2019})}\BibitemShut {NoStop}%
\bibitem [{\citenamefont {Yan}\ \emph {et~al.}(2022)\citenamefont {Yan}, \citenamefont {Slagle},\ and\ \citenamefont {Nevidomskyy}}]{Fractons2}%
  \BibitemOpen
  \bibfield  {author} {\bibinfo {author} {\bibfnamefont {H.}~\bibnamefont {Yan}}, \bibinfo {author} {\bibfnamefont {K.}~\bibnamefont {Slagle}},\ and\ \bibinfo {author} {\bibfnamefont {A.~H.}\ \bibnamefont {Nevidomskyy}},\ }\href {https://arxiv.org/abs/2211.15829} {\bibinfo {title} {Y-cube model and fractal structure of subdimensional particles on hyperbolic lattices}} (\bibinfo {year} {2022}),\ \Eprint {https://arxiv.org/abs/2211.15829} {arXiv:2211.15829 [quant-ph]} \BibitemShut {NoStop}%
\bibitem [{\citenamefont {Brill}\ and\ \citenamefont {Wheeler}(1957)}]{Brill-Wheeler-1957}%
  \BibitemOpen
  \bibfield  {author} {\bibinfo {author} {\bibfnamefont {D.~R.}\ \bibnamefont {Brill}}\ and\ \bibinfo {author} {\bibfnamefont {J.~A.}\ \bibnamefont {Wheeler}},\ }\bibfield  {title} {\bibinfo {title} {Interaction of neutrinos and gravitational fields},\ }\href {https://doi.org/10.1103/RevModPhys.29.465} {\bibfield  {journal} {\bibinfo  {journal} {Rev. Mod. Phys.}\ }\textbf {\bibinfo {volume} {29}},\ \bibinfo {pages} {465} (\bibinfo {year} {1957})}\BibitemShut {NoStop}%
\bibitem [{\citenamefont {Obukhov}\ \emph {et~al.}(2011)\citenamefont {Obukhov}, \citenamefont {Silenko},\ and\ \citenamefont {Teryaev}}]{ObukhovDiracfermions}%
  \BibitemOpen
  \bibfield  {author} {\bibinfo {author} {\bibfnamefont {Y.~N.}\ \bibnamefont {Obukhov}}, \bibinfo {author} {\bibfnamefont {A.~J.}\ \bibnamefont {Silenko}},\ and\ \bibinfo {author} {\bibfnamefont {O.~V.}\ \bibnamefont {Teryaev}},\ }\bibfield  {title} {\bibinfo {title} {Dirac fermions in strong gravitational fields},\ }\href {https://doi.org/10.1103/PhysRevD.84.024025} {\bibfield  {journal} {\bibinfo  {journal} {Phys. Rev. D}\ }\textbf {\bibinfo {volume} {84}},\ \bibinfo {pages} {024025} (\bibinfo {year} {2011})}\BibitemShut {NoStop}%
\bibitem [{\citenamefont {Falcone}\ and\ \citenamefont {Conti}(2023)}]{PhysRevD.107.045012}%
  \BibitemOpen
  \bibfield  {author} {\bibinfo {author} {\bibfnamefont {R.}~\bibnamefont {Falcone}}\ and\ \bibinfo {author} {\bibfnamefont {C.}~\bibnamefont {Conti}},\ }\bibfield  {title} {\bibinfo {title} {Nonrelativistic limit of scalar and dirac fields in curved spacetime},\ }\href {https://doi.org/10.1103/PhysRevD.107.045012} {\bibfield  {journal} {\bibinfo  {journal} {Phys. Rev. D}\ }\textbf {\bibinfo {volume} {107}},\ \bibinfo {pages} {045012} (\bibinfo {year} {2023})}\BibitemShut {NoStop}%
\bibitem [{\citenamefont {Vozmediano}\ \emph {et~al.}(2010)\citenamefont {Vozmediano}, \citenamefont {Katsnelson},\ and\ \citenamefont {Guinea}}]{VOZMEDIANO2010109}%
  \BibitemOpen
  \bibfield  {author} {\bibinfo {author} {\bibfnamefont {M.}~\bibnamefont {Vozmediano}}, \bibinfo {author} {\bibfnamefont {M.}~\bibnamefont {Katsnelson}},\ and\ \bibinfo {author} {\bibfnamefont {F.}~\bibnamefont {Guinea}},\ }\bibfield  {title} {\bibinfo {title} {Gauge fields in graphene},\ }\href {https://doi.org/https://doi.org/10.1016/j.physrep.2010.07.003} {\bibfield  {journal} {\bibinfo  {journal} {Physics Reports}\ }\textbf {\bibinfo {volume} {496}},\ \bibinfo {pages} {109} (\bibinfo {year} {2010})}\BibitemShut {NoStop}%
\bibitem [{\citenamefont {{de Juan}}\ \emph {et~al.}(2010)\citenamefont {{de Juan}}, \citenamefont {Cortijo},\ and\ \citenamefont {Vozmediano}}]{DEJUAN2010625}%
  \BibitemOpen
  \bibfield  {author} {\bibinfo {author} {\bibfnamefont {F.}~\bibnamefont {{de Juan}}}, \bibinfo {author} {\bibfnamefont {A.}~\bibnamefont {Cortijo}},\ and\ \bibinfo {author} {\bibfnamefont {M.~A.}\ \bibnamefont {Vozmediano}},\ }\bibfield  {title} {\bibinfo {title} {Dislocations and torsion in graphene and related systems},\ }\href {https://doi.org/https://doi.org/10.1016/j.nuclphysb.2009.11.012} {\bibfield  {journal} {\bibinfo  {journal} {Nuclear Physics B}\ }\textbf {\bibinfo {volume} {828}},\ \bibinfo {pages} {625} (\bibinfo {year} {2010})}\BibitemShut {NoStop}%
\bibitem [{\citenamefont {Son}(2025)}]{Son_2025}%
  \BibitemOpen
  \bibfield  {author} {\bibinfo {author} {\bibfnamefont {D.~T.}\ \bibnamefont {Son}},\ }\bibfield  {title} {\bibinfo {title} {Newton-cartan geometry and the quantum hall effect},\ }\href {https://doi.org/10.1063/10.0035975} {\bibfield  {journal} {\bibinfo  {journal} {Low Temperature Physics}\ }\textbf {\bibinfo {volume} {51}},\ \bibinfo {pages} {384–391} (\bibinfo {year} {2025})}\BibitemShut {NoStop}%
\bibitem [{\citenamefont {Andringa}\ \emph {et~al.}(2011)\citenamefont {Andringa}, \citenamefont {Bergshoeff}, \citenamefont {Panda},\ and\ \citenamefont {de~Roo}}]{Andringa_2011}%
  \BibitemOpen
  \bibfield  {author} {\bibinfo {author} {\bibfnamefont {R.}~\bibnamefont {Andringa}}, \bibinfo {author} {\bibfnamefont {E.}~\bibnamefont {Bergshoeff}}, \bibinfo {author} {\bibfnamefont {S.}~\bibnamefont {Panda}},\ and\ \bibinfo {author} {\bibfnamefont {M.}~\bibnamefont {de~Roo}},\ }\bibfield  {title} {\bibinfo {title} {Newtonian gravity and the bargmann algebra},\ }\href {https://doi.org/10.1088/0264-9381/28/10/105011} {\bibfield  {journal} {\bibinfo  {journal} {Classical and Quantum Gravity}\ }\textbf {\bibinfo {volume} {28}},\ \bibinfo {pages} {105011} (\bibinfo {year} {2011})}\BibitemShut {NoStop}%
\bibitem [{\citenamefont {Geracie}\ \emph {et~al.}(2015)\citenamefont {Geracie}, \citenamefont {Son}, \citenamefont {Wu},\ and\ \citenamefont {Wu}}]{PhysRevD.91.045030}%
  \BibitemOpen
  \bibfield  {author} {\bibinfo {author} {\bibfnamefont {M.}~\bibnamefont {Geracie}}, \bibinfo {author} {\bibfnamefont {D.~T.}\ \bibnamefont {Son}}, \bibinfo {author} {\bibfnamefont {C.}~\bibnamefont {Wu}},\ and\ \bibinfo {author} {\bibfnamefont {S.-F.}\ \bibnamefont {Wu}},\ }\bibfield  {title} {\bibinfo {title} {Spacetime symmetries of the quantum hall effect},\ }\href {https://doi.org/10.1103/PhysRevD.91.045030} {\bibfield  {journal} {\bibinfo  {journal} {Phys. Rev. D}\ }\textbf {\bibinfo {volume} {91}},\ \bibinfo {pages} {045030} (\bibinfo {year} {2015})}\BibitemShut {NoStop}%
\bibitem [{\citenamefont {Kobayashi}(1972)}]{Kobayashi1972}%
  \BibitemOpen
  \bibfield  {author} {\bibinfo {author} {\bibfnamefont {S.}~\bibnamefont {Kobayashi}},\ }\href {https://doi.org/10.1007/978-3-642-61983-1} {\emph {\bibinfo {title} {Transformation Groups in Differential Geometry}}},\ Classics in Mathematics\ (\bibinfo  {publisher} {Springer-Verlag},\ \bibinfo {address} {Berlin},\ \bibinfo {year} {1972})\BibitemShut {NoStop}%
\bibitem [{\citenamefont {Helgason}(2001)}]{Helgason2001}%
  \BibitemOpen
  \bibfield  {author} {\bibinfo {author} {\bibfnamefont {S.}~\bibnamefont {Helgason}},\ }\href@noop {} {\emph {\bibinfo {title} {Differential Geometry, Lie Groups, and Symmetric Spaces}}},\ \bibinfo {series} {Graduate Studies in Mathematics}, Vol.~\bibinfo {volume} {34}\ (\bibinfo  {publisher} {American Mathematical Society},\ \bibinfo {address} {Providence, RI},\ \bibinfo {year} {2001})\BibitemShut {NoStop}%
\bibitem [{\citenamefont {Ratcliffe}(2019)}]{Ratcliffe2019}%
  \BibitemOpen
  \bibfield  {author} {\bibinfo {author} {\bibfnamefont {J.~G.}\ \bibnamefont {Ratcliffe}},\ }\href {https://doi.org/10.1007/978-3-030-31532-9} {\emph {\bibinfo {title} {Foundations of Hyperbolic Manifolds}}},\ \bibinfo {edition} {3rd}\ ed.,\ \bibinfo {series} {Graduate Texts in Mathematics}, Vol.\ \bibinfo {volume} {149}\ (\bibinfo  {publisher} {Springer},\ \bibinfo {address} {New York},\ \bibinfo {year} {2019})\BibitemShut {NoStop}%
\bibitem [{\citenamefont {Inagaki}\ \emph {et~al.}(1997)\citenamefont {Inagaki}, \citenamefont {Muta},\ and\ \citenamefont {Odintsov}}]{Inagaki-review-1997}%
  \BibitemOpen
  \bibfield  {author} {\bibinfo {author} {\bibfnamefont {T.}~\bibnamefont {Inagaki}}, \bibinfo {author} {\bibfnamefont {T.}~\bibnamefont {Muta}},\ and\ \bibinfo {author} {\bibfnamefont {S.~D.}\ \bibnamefont {Odintsov}},\ }\bibfield  {title} {\bibinfo {title} {Dynamical symmetry breaking in curved spacetime},\ }\href {https://doi.org/10.1143/PTP.127.93} {\bibfield  {journal} {\bibinfo  {journal} {Progress of Theoretical Physics Supplement}\ }\textbf {\bibinfo {volume} {127}},\ \bibinfo {pages} {93} (\bibinfo {year} {1997})},\ \Eprint {https://arxiv.org/abs/https://academic.oup.com/ptps/article-pdf/doi/10.1143/PTP.127.93/5160221/127-93.pdf} {https://academic.oup.com/ptps/article-pdf/doi/10.1143/PTP.127.93/5160221/127-93.pdf} \BibitemShut {NoStop}%
\bibitem [{\citenamefont {Blagojevic}(2001)}]{MBlagojevic:2001}%
  \BibitemOpen
  \bibfield  {author} {\bibinfo {author} {\bibfnamefont {M.}~\bibnamefont {Blagojevic}},\ }\href {https://doi.org/10.1201/9781420034264} {\emph {\bibinfo {title} {Gravitation and Gauge Symmetries}}}\ (\bibinfo  {publisher} {CRC Press},\ \bibinfo {year} {2001})\BibitemShut {NoStop}%
\bibitem [{\citenamefont {Kitaev}(2018)}]{kitaev2018}%
  \BibitemOpen
  \bibfield  {author} {\bibinfo {author} {\bibfnamefont {A.}~\bibnamefont {Kitaev}},\ }\href {https://arxiv.org/abs/1711.08169} {\bibinfo {title} {Notes on $\widetilde{\mathrm{sl}}(2,\mathbb{R})$ representations}} (\bibinfo {year} {2018}),\ \Eprint {https://arxiv.org/abs/1711.08169} {arXiv:1711.08169 [hep-th]} \BibitemShut {NoStop}%
\bibitem [{\citenamefont {Bargmann}(1947)}]{bargmann1947}%
  \BibitemOpen
  \bibfield  {author} {\bibinfo {author} {\bibfnamefont {V.}~\bibnamefont {Bargmann}},\ }\bibfield  {title} {\bibinfo {title} {Irreducible unitary representations of the lorentz group},\ }\href {http://www.jstor.org/stable/1969129} {\bibfield  {journal} {\bibinfo  {journal} {Annals of Mathematics}\ }\textbf {\bibinfo {volume} {48}},\ \bibinfo {pages} {568} (\bibinfo {year} {1947})}\BibitemShut {NoStop}%
\bibitem [{\citenamefont {Camporesi}(1992)}]{Camporesi:1992tm}%
  \BibitemOpen
  \bibfield  {author} {\bibinfo {author} {\bibfnamefont {R.}~\bibnamefont {Camporesi}},\ }\bibfield  {title} {\bibinfo {title} {{The Spinor heat kernel in maximally symmetric spaces}},\ }\href {https://doi.org/10.1007/BF02100862} {\bibfield  {journal} {\bibinfo  {journal} {Commun. Math. Phys.}\ }\textbf {\bibinfo {volume} {148}},\ \bibinfo {pages} {283} (\bibinfo {year} {1992})}\BibitemShut {NoStop}%
\bibitem [{\citenamefont {Gorbar}\ and\ \citenamefont {Gusynin}(2008)}]{gorbar2008gap}%
  \BibitemOpen
  \bibfield  {author} {\bibinfo {author} {\bibfnamefont {E.}~\bibnamefont {Gorbar}}\ and\ \bibinfo {author} {\bibfnamefont {V.}~\bibnamefont {Gusynin}},\ }\bibfield  {title} {\bibinfo {title} {Gap generation for dirac fermions on lobachevsky plane in a magnetic field},\ }\href {https://doi.org/doi.org/10.1016/j.aop.2007.11.005} {\bibfield  {journal} {\bibinfo  {journal} {Annals of Physics}\ }\textbf {\bibinfo {volume} {323}},\ \bibinfo {pages} {2132} (\bibinfo {year} {2008})}\BibitemShut {NoStop}%
\bibitem [{\citenamefont {Wei\ss{}e}\ \emph {et~al.}(2006)\citenamefont {Wei\ss{}e}, \citenamefont {Wellein}, \citenamefont {Alvermann},\ and\ \citenamefont {Fehske}}]{KPM}%
  \BibitemOpen
  \bibfield  {author} {\bibinfo {author} {\bibfnamefont {A.}~\bibnamefont {Wei\ss{}e}}, \bibinfo {author} {\bibfnamefont {G.}~\bibnamefont {Wellein}}, \bibinfo {author} {\bibfnamefont {A.}~\bibnamefont {Alvermann}},\ and\ \bibinfo {author} {\bibfnamefont {H.}~\bibnamefont {Fehske}},\ }\bibfield  {title} {\bibinfo {title} {The kernel polynomial method},\ }\href {https://doi.org/10.1103/RevModPhys.78.275} {\bibfield  {journal} {\bibinfo  {journal} {Rev. Mod. Phys.}\ }\textbf {\bibinfo {volume} {78}},\ \bibinfo {pages} {275} (\bibinfo {year} {2006})}\BibitemShut {NoStop}%
\bibitem [{\citenamefont {Sorella}\ and\ \citenamefont {Tosatti}(1992)}]{Sorella_1992}%
  \BibitemOpen
  \bibfield  {author} {\bibinfo {author} {\bibfnamefont {S.}~\bibnamefont {Sorella}}\ and\ \bibinfo {author} {\bibfnamefont {E.}~\bibnamefont {Tosatti}},\ }\bibfield  {title} {\bibinfo {title} {Semi-metal-insulator transition of the hubbard model in the honeycomb lattice},\ }\href {https://doi.org/10.1209/0295-5075/19/8/007} {\bibfield  {journal} {\bibinfo  {journal} {Europhysics Letters}\ }\textbf {\bibinfo {volume} {19}},\ \bibinfo {pages} {699} (\bibinfo {year} {1992})}\BibitemShut {NoStop}%
\bibitem [{\citenamefont {Assaad}\ and\ \citenamefont {Herbut}(2013)}]{Assaad-Herbut-2013}%
  \BibitemOpen
  \bibfield  {author} {\bibinfo {author} {\bibfnamefont {F.~F.}\ \bibnamefont {Assaad}}\ and\ \bibinfo {author} {\bibfnamefont {I.~F.}\ \bibnamefont {Herbut}},\ }\bibfield  {title} {\bibinfo {title} {Pinning the order: The nature of quantum criticality in the hubbard model on honeycomb lattice},\ }\href {https://doi.org/10.1103/PhysRevX.3.031010} {\bibfield  {journal} {\bibinfo  {journal} {Phys. Rev. X}\ }\textbf {\bibinfo {volume} {3}},\ \bibinfo {pages} {031010} (\bibinfo {year} {2013})}\BibitemShut {NoStop}%
\bibitem [{\citenamefont {Otsuka}\ \emph {et~al.}(2016)\citenamefont {Otsuka}, \citenamefont {Yunoki},\ and\ \citenamefont {Sorella}}]{Sorella-2016}%
  \BibitemOpen
  \bibfield  {author} {\bibinfo {author} {\bibfnamefont {Y.}~\bibnamefont {Otsuka}}, \bibinfo {author} {\bibfnamefont {S.}~\bibnamefont {Yunoki}},\ and\ \bibinfo {author} {\bibfnamefont {S.}~\bibnamefont {Sorella}},\ }\bibfield  {title} {\bibinfo {title} {Universal quantum criticality in the metal-insulator transition of two-dimensional interacting dirac electrons},\ }\href {https://doi.org/10.1103/PhysRevX.6.011029} {\bibfield  {journal} {\bibinfo  {journal} {Phys. Rev. X}\ }\textbf {\bibinfo {volume} {6}},\ \bibinfo {pages} {011029} (\bibinfo {year} {2016})}\BibitemShut {NoStop}%
\bibitem [{\citenamefont {Kane}\ and\ \citenamefont {Mele}(2005)}]{kane-mele-2005}%
  \BibitemOpen
  \bibfield  {author} {\bibinfo {author} {\bibfnamefont {C.~L.}\ \bibnamefont {Kane}}\ and\ \bibinfo {author} {\bibfnamefont {E.~J.}\ \bibnamefont {Mele}},\ }\bibfield  {title} {\bibinfo {title} {Quantum spin hall effect in graphene},\ }\href {https://doi.org/10.1103/PhysRevLett.95.226801} {\bibfield  {journal} {\bibinfo  {journal} {Phys. Rev. Lett.}\ }\textbf {\bibinfo {volume} {95}},\ \bibinfo {pages} {226801} (\bibinfo {year} {2005})}\BibitemShut {NoStop}%
\bibitem [{\citenamefont {Bernevig}\ \emph {et~al.}(2006)\citenamefont {Bernevig}, \citenamefont {Hughes},\ and\ \citenamefont {Zhang}}]{BHZ-2006}%
  \BibitemOpen
  \bibfield  {author} {\bibinfo {author} {\bibfnamefont {B.~A.}\ \bibnamefont {Bernevig}}, \bibinfo {author} {\bibfnamefont {T.~L.}\ \bibnamefont {Hughes}},\ and\ \bibinfo {author} {\bibfnamefont {S.-C.}\ \bibnamefont {Zhang}},\ }\bibfield  {title} {\bibinfo {title} {Quantum spin hall effect and topological phase transition in hgte quantum wells},\ }\href {https://doi.org/10.1126/science.1133734} {\bibfield  {journal} {\bibinfo  {journal} {Science}\ }\textbf {\bibinfo {volume} {314}},\ \bibinfo {pages} {1757} (\bibinfo {year} {2006})}\BibitemShut {NoStop}%
\bibitem [{\citenamefont {Ba\~nados}\ \emph {et~al.}(1992)\citenamefont {Ba\~nados}, \citenamefont {Teitelboim},\ and\ \citenamefont {Zanelli}}]{BTZ}%
  \BibitemOpen
  \bibfield  {author} {\bibinfo {author} {\bibfnamefont {M.}~\bibnamefont {Ba\~nados}}, \bibinfo {author} {\bibfnamefont {C.}~\bibnamefont {Teitelboim}},\ and\ \bibinfo {author} {\bibfnamefont {J.}~\bibnamefont {Zanelli}},\ }\bibfield  {title} {\bibinfo {title} {Black hole in three-dimensional spacetime},\ }\href {https://doi.org/10.1103/PhysRevLett.69.1849} {\bibfield  {journal} {\bibinfo  {journal} {Phys. Rev. Lett.}\ }\textbf {\bibinfo {volume} {69}},\ \bibinfo {pages} {1849} (\bibinfo {year} {1992})}\BibitemShut {NoStop}%
\bibitem [{\citenamefont {Cheng}\ \emph {et~al.}(2023)\citenamefont {Cheng}, \citenamefont {Wang},\ and\ \citenamefont {Fan}}]{Cheng2023_NonAbelianSU2Photons}%
  \BibitemOpen
  \bibfield  {author} {\bibinfo {author} {\bibfnamefont {D.}~\bibnamefont {Cheng}}, \bibinfo {author} {\bibfnamefont {K.}~\bibnamefont {Wang}},\ and\ \bibinfo {author} {\bibfnamefont {S.}~\bibnamefont {Fan}},\ }\bibfield  {title} {\bibinfo {title} {Artificial non-{A}belian lattice gauge fields for photons in the synthetic frequency dimension},\ }\href {https://doi.org/10.1103/PhysRevLett.130.083601} {\bibfield  {journal} {\bibinfo  {journal} {Physical Review Letters}\ }\textbf {\bibinfo {volume} {130}},\ \bibinfo {pages} {083601} (\bibinfo {year} {2023})}\BibitemShut {NoStop}%
\bibitem [{\citenamefont {Yang}\ \emph {et~al.}(2024)\citenamefont {Yang}, \citenamefont {Yang}, \citenamefont {Ma}, \citenamefont {Li}, \citenamefont {Zhang},\ and\ \citenamefont {Chan}}]{Yang2024_NonAbelianLightSound}%
  \BibitemOpen
  \bibfield  {author} {\bibinfo {author} {\bibfnamefont {Y.}~\bibnamefont {Yang}}, \bibinfo {author} {\bibfnamefont {B.}~\bibnamefont {Yang}}, \bibinfo {author} {\bibfnamefont {G.}~\bibnamefont {Ma}}, \bibinfo {author} {\bibfnamefont {J.}~\bibnamefont {Li}}, \bibinfo {author} {\bibfnamefont {S.}~\bibnamefont {Zhang}},\ and\ \bibinfo {author} {\bibfnamefont {C.~T.}\ \bibnamefont {Chan}},\ }\bibfield  {title} {\bibinfo {title} {Non-{A}belian physics in light and sound},\ }\href {https://doi.org/10.1126/science.adf9621} {\bibfield  {journal} {\bibinfo  {journal} {Science}\ }\textbf {\bibinfo {volume} {383}},\ \bibinfo {pages} {eadf9621} (\bibinfo {year} {2024})}\BibitemShut {NoStop}%
\bibitem [{\citenamefont {Đorđevic}(2026)}]{GitHub-Code}%
  \BibitemOpen
  \bibfield  {author} {\bibinfo {author} {\bibfnamefont {A.}~\bibnamefont {Đorđevic}},\ }\href {https://github.com/ana-djordjevic-98/Hyperbolic-lattice-TB-model-spin-connection} {\bibinfo {title} {Hyperbolic crystals and tb model with spin connection}} (\bibinfo {year} {2026})\BibitemShut {NoStop}%
\bibitem [{\citenamefont {Bishop~C.}(1993)}]{ActaMath:1993}%
  \BibitemOpen
  \bibfield  {author} {\bibinfo {author} {\bibfnamefont {S.~T.}\ \bibnamefont {Bishop~C.}},\ }\bibfield  {title} {\bibinfo {title} {{Representation theoretic rigidity in PSL (2,R)}},\ }\href {https://doi.org/10.1007/BF02392456} {\bibfield  {journal} {\bibinfo  {journal} {Acta Math.}\ }\textbf {\bibinfo {volume} {170}},\ \bibinfo {pages} {121} (\bibinfo {year} {1993})}\BibitemShut {NoStop}%
\bibitem [{\citenamefont {Camporesi}\ and\ \citenamefont {Higuchi}(1996)}]{HyperbolicEigenFunction_Camporesi}%
  \BibitemOpen
  \bibfield  {author} {\bibinfo {author} {\bibfnamefont {R.}~\bibnamefont {Camporesi}}\ and\ \bibinfo {author} {\bibfnamefont {A.}~\bibnamefont {Higuchi}},\ }\bibfield  {title} {\bibinfo {title} {On the eigenfunctions of the dirac operator on spheres and real hyperbolic spaces},\ }\href {https://doi.org/10.1016/0393-0440(95)00042-9} {\bibfield  {journal} {\bibinfo  {journal} {Journal of Geometry and Physics}\ }\textbf {\bibinfo {volume} {20}},\ \bibinfo {pages} {1} (\bibinfo {year} {1996})}\BibitemShut {NoStop}%
\end{thebibliography}%
\end{document}